\documentclass[10pt,oneside,a4paper,onecolumn]{elsarticle}

\journal{Journal of Computers \& Fluids}

\usepackage{graphicx}
\usepackage{dcolumn}
\usepackage{bm}

\usepackage[utf8]{inputenc}
\usepackage[T1]{fontenc}
\usepackage{mathptmx}

\usepackage{graphicx}
\usepackage{epstopdf, epsfig}

\usepackage{color,soul}
\usepackage{amssymb}
\usepackage{amsmath}
\usepackage{bm}
\usepackage{multirow}
\usepackage{xcolor}
\usepackage{tikz}
\usepackage{subfig}
\usepackage{subfloat}
\usepackage{multirow}
\usepackage{rotating}
\usepackage{booktabs}
\usepackage{relsize}
\usepackage{nomencl}
\usepackage{etoolbox}
\usepackage{float}
 \usepackage{url}
 \usepackage{hyperref}
 \usepackage{algorithm2e}
\usepackage[export]{adjustbox}
\usepackage{upgreek}
\usepackage{geometry}
\usepackage{booktabs}
\usepackage{tabularx}
\usepackage{changepage}                 
\usepackage{lipsum}
\usepackage{setspace}
\usepackage{array}
\usepackage{hyperref}
\usepackage[framemethod=tikz]{mdframed}
\newcolumntype{P}[1]{>{\centering\arraybackslash}p{#1}}

\makeatletter
\newcommand{\labeltext}[3][]{%
    \@bsphack%
    \csname phantomsection\endcsname
    \def\tst{#1}%
    \def\labelmarkup{}
    \def\refmarkup{}%
    \ifx\tst\empty\def\@currentlabel{\refmarkup{#2}}{\label{#3}}%
    \else\def\@currentlabel{\refmarkup{#1}}{\label{#3}}\fi%
    \@esphack%
    \labelmarkup{#2}
}
\makeatother

\newlength{\offsetpage}
\newenvironment{widepage}{\begin{adjustwidth}{-\offsetpage}{-\offsetpage}%
    \addtolength{\textwidth}{2\offsetpage}}%
{\end{adjustwidth}}

\newcommand{\buf}{\textbf{u}^\mathrm{f}}
\newcommand{\bus}{\textbf{u}^\mathrm{s}}
\newcommand{\bbf}{\textbf{b}^\mathrm{f}}
\newcommand{\bbs}{\textbf{b}^\mathrm{s}}
\newcommand{\bFs}{\textbf{F}^\mathrm{s}}
\newcommand{\stf}{\bm{\sigma}^{\mathrm{f}}}
\newcommand{\stnf}{\bm{\sigma}}

\newcommand{\df}{\Omega^\mathrm{f}(t)}
\newcommand{\dfnt}{\Omega^\mathrm{f}}

\newcommand{\dsnt}{\Omega^\mathrm{s}}
\newcommand{\ifs}{\Gamma^\mathrm{fs}(t)}
\newcommand{\ifsnt}{\Gamma^\mathrm{fs}}
\newcommand{\vf}{\mu^\mathrm{f}}

\usepackage{amsthm}
\newtheorem{remark}{Remark}

\begin{document}

\begin{frontmatter}

\title{A hybrid {\color{black} partitioned} deep learning methodology for moving interface and fluid-structure interaction}

\author[mymainaddress]{Rachit Gupta}
\ead{rachit.gupta@ubc.ca}

\author[mymainaddress]{Rajeev Jaiman\corref{mycorrespondingauthor}}
\cortext[mycorrespondingauthor]{Corresponding author}
\ead{rjaiman@mech.ubc.ca}

\address[mymainaddress]{Department of Mechanical Engineering, University of British Columbia, Vancouver, BC Canada V6T 1Z4}

\begin{abstract}

In this work, we present a hybrid partitioned deep learning framework for the reduced-order modeling of moving interfaces and predicting fluid-structure interaction. Using the discretized Navier-Stokes in the arbitrary Lagrangian-Eulerian reference frame, we generate the full-order flow snapshots and point cloud displacements as target physical data for the learning and inference of coupled fluid-structure dynamics. {\color{black} The hybrid operation of this methodology comes by combining two separate data-driven models for fluid and solid subdomains via deep learning-based reduced-order models (DL-ROMs).
The proposed multi-level framework comprises the partitioned data-driven drivers for unsteady flow and the moving point cloud displacements.} At the fluid-structure interface, the force information is exchanged synchronously between the two partitioned subdomain solvers. The first component of our proposed framework relies on the proper orthogonal decomposition-based recurrent neural network (POD-RNN) as a DL-ROM procedure to infer the point cloud with a moving interface. This model utilizes the POD basis modes to reduce dimensionality and evolve them in time via long short-term memory-based recurrent neural networks (LSTM-RNNs). The second component employs the convolution-based recurrent autoencoder network (CRAN) as a self-supervised DL-ROM procedure to infer the nonlinear flow dynamics at static Eulerian probes. 
We introduce these probes as spatially structured query nodes in the moving point cloud to treat the Lagrangian-to-Eulerian conflict together with convenience in training the CRAN driver. To determine these Eulerian probes, we construct a novel snapshot-field transfer and load recovery algorithm. They are chosen in such a way that the two components (i.e., POD-RNN and CRAN) are constrained at the interface to recover the bulk force quantities. These DL-ROM-based data-driven drivers rely on the LSTM-RNNs to evolve the low-dimensional states. {\color{black} A popular prototypical fluid-structure interaction problem of flow past a freely oscillating cylinder is considered to assess the efficacy of the proposed methodology for a different set of reduced velocities that lead to vortex-induced vibrations. The proposed framework tracks the interface description with acceptable accuracy and predicts the nonlinear wake dynamics over the chosen test data range.}
The proposed framework aligns with the development of partitioned digital twin of engineering systems, especially those involving moving boundaries and fluid-structure interactions. 
\end{abstract}


\end{frontmatter}

\section{Introduction}
Fluid-structure interaction (FSI) is a coupled physical phenomenon that involves a mutual interplay and bidirectional interaction of fluid flow with structural dynamics. This phenomenon is ubiquitous in nature and engineering systems  such as fluttering flags \cite{gurugubelli2015self}, flying bats \cite{joshi2020variational}, offshore platforms and pipelines \cite{jaiman2016partitioned,joshi20183d}, oscillating hydrofoils with cavitation \cite{kashyap2021}, two-phase flow in a flexible pipeline \cite{joshi2019hybrid} and among others. For example, the two-way coupling between the fluid and solid exhibits rich flow dynamics such as the wake-body interaction and vortex-induced vibrations \cite{li2016vortex}, which are important to understand from an engineering design or a decision-making standpoint. Due to the complex characteristics of the fluid-structure coupling, frequent but reliable practice is to model these complex interplay and underlying dynamics by solving numerically a coupled set of partial differential equations (PDEs) describing the physical laws. Of particular interest in these numerical techniques for unsteady fluid-structure interaction is to accurately simulate the wake-body interaction in terms of vortex-induced loads and structural displacements, which represent the prominent coupled dynamical effects.  

Fluid-structure systems can undergo highly nonlinear interactions involving complicated interface dynamics and a wide range of spatial and temporal scales.
Moreover, these complex spatial-temporal
characteristics are very sensitive to physical parameters and geometric variations.
Numerical techniques such as the arbitrary Lagrangian-Eulerian \cite{donaea1983arbitrary, jaiman2016partitioned}, level-set \cite{sethian1999level}, immersed boundary \cite{peskin2002immersed}, fictitious domain \cite{yu2005dlm} and phase-field modeling \cite{mokbel2018, joshi2019hybrid} can provide high-fidelity PDE-based numerical solutions and physical insights of the underlying FSI phenomena. Using the state-of-the-art discretization techniques such the finite element method,  accurate solutions have been possible by solving millions of fluid and structural variables using the full-order models (FOMs) and incorporating proper treatment of the fluid-structure interface \cite{joshi2020variational}. 
As the complexity of an FSI system elevates, the accurate treatment of the interface is directly linked with increasing fidelity near the fluid-structure interface, which implies solving more and more unknown variables numerically \cite{joshi2019hybrid}. 
{\color{black} At higher resolutions of space and time, the equation-based forward simulations can come at the expense of prohibitively large computing time and high-dimensionality, rendering them almost ineffective in real-time predictions or control required by digital twin development \cite{tuegel2011}.}
%

{\color{black} The applicability of deep learning has emerged as a promising alternative for constructing data-driven prediction models for fluid flow \cite{miyanawala2017efficient, miyanawala2018novel} and nonlinear dynamical systems e.g., fluid-structure interaction \cite{miyanawala2018low}. Deep learning is a subset of machine learning that refers to the use of multilayered neural networks to classify spatial-temporal datasets and make inference from a set of training data \cite{goodfellow2016deep,lecun2015deep,schmidhuber2014}}. By learning the underlying data-driven input-output model via deep neural networks, the goal is to make predictions to unseen input data.
The construction of heavily over-parametrized functions by deep neural networks rely on the foundations of
the Kolmogorov–Arnold representation theorem \cite{schmidtHieber2021} 
and the universal approximation of functions via neural networks \cite{cybenko1989, chen1995universal, hornik1990universal}.
%
Using deep neural networks along with a collection of algorithms, one can find useful features or embedding functions in a low-dimensional space from the input-output relation in datasets. 
Deep neural networks have the ability to separate spatial-temporal scales and automatically extract functional relations from high-dimensional data with hierarchical importance.
%
%
Bestowed by the state-of-the-art back-propagation and stochastic gradient descent techniques for estimating weights and biases adaptively, deep neural networks can provide efficient low-dimensional representation in a flexible way while learning multiple levels of hierarchy in data \cite{goodfellow2016deep}.
 While deep neural networks are heavily overparametrized, they have an inherent bias to induce and make inferences to unseen data which is termed as inductive bias \cite{bronstein2021geometric}.
 For example, convolutional neural nets possess an implicit inductive bias via convolutional filters with
shared weights (i.e., translational symmetry) and pooling to exploit
scale separation \cite{bronstein2021geometric}.
Nevertheless, these black-box deep learning techniques do not account for prior domain knowledge that can be important for interpretability, data efficiency and generalization.

In the last few years, there is a growing interest to exploit the inherent inductive bias in deep learning and to infuse explicit bias or domain knowledge in the network architectures for efficient predictions and interpretability \cite{goyal2021}. In that direction, there have been many promising approaches established in the research community for a synergistic coupling of the deep learning and physical-based models \cite{bukka2019data,bukka2021assessment}.
%
These models are often trained to represent a full or partial parametrization of a forward physical process while emulating the governing equations. These coarse-grained inference models and surrogate representations aim to reduce the high computational costs in forecasting dynamics and to model the quantities of interest \cite{bukka2021assessment}. 
For example, one can use the trained parameters to carry out {the state-to-state time advancements} \cite{ham2019deep, brown2008neural,san2018machine} and {inverse modeling} \cite{chen2017low, lunz2018adversarial,parish2016paradigm}. 
Here, we refer to the state-to-state time advancements as inferring dependent physical variables from the previous states. At the same time, inverse modeling identifies the physical system parameters from the output state information. One can also exploit the approximation properties and inductive bias of neural calculus to solve the underlying differential equations within a deep learning framework \cite{han2018solving, lagaris1998artificial, rudy2017data, lu2019deeponet}. {\color{black} Without appropriate modifications of the state-of-the-art DL techniques, the approximate solution of the PDEs via black-box DL approaches may lead to slow training, reduced accuracy, and a lack of generality to multiphysics and multiscale systems involving complex geometries and boundary conditions \cite{willard2020integrating, wang2021physics, pfaff2020learning, miyanawala2018novel, wang2020towards}.} 

For increasing generality, we can broadly classify the physics-based machine learning into three categories: {\color{black} (a) the modification of the objective or loss function by adding a regularizer, (b) the adjustment of neural architecture designs, and (c) the hybridization of deep learning with projection-based model reduction.}
%
In the first category, one can infuse the physical information by applying a regularizer to the standard loss function. Such regularizers are generally based on physical conservation principles or the governing equations of the underlying problem. Many researchers have explicitly enforced such regularizers to boost generalizability; for instance Karpatne \emph{et al.} \cite{karpatne2017physics}, Raissi \emph{et al.} \cite{raissi2019physics}, Zhu \emph{et al.} \cite{zhu2019physics}, Erichson \emph{et al.} \cite{erichson2019physics}, Geneva \emph{et al.} \cite{geneva2020modeling}, Mallik \emph{et al.} \cite{mallik2021}, and among others. For physically constrained ML architectures, as the second category, Daw \emph{et al.} \cite{daw2020physics} and Chang \emph{et al.} \cite{chang2019antisymmetricrnn} modified the RNN cell structure by intermediate constraints while Muralidhar \emph{et al.} \cite{muralidhar2018incorporating} and Ruthotto and Haber \cite{ruthotto2019deep} took a similar route for the convolutional neural networks (CNNs). Recently, Li \emph{et al.} \cite{li2020fourier} developed the Fourier neural operators by applying a Fourier transform and linear weight operator on the input before the activation function in a neural network. 
{\color{black} The third category involves the development of the hybrid DL-ROMs that take into account the spatial and temporal domain knowledge in neural networks. The spatial knowledge is incorporated by effectively building the low-dimensional states using the projection-based techniques which inherit physical interpretation as discussed in Miyanawala \& Jaiman \citep{miyanawala2019hybrid}. The temporal knowledge comes from evolving these low-dimensional states in time. We refer to such DL-ROM frameworks as physics-based because they incorporate physical interpretability via proper orthogonal decomposition and its variants.} These DL-ROMs can operate in synchronization with the FOMs to boost predictive abilities. 

In this paper, we are interested in developing a hybrid DL-ROM framework to address the curse of spatial dimensionality while utilizing the desired properties of projection-based model reduction and deep neural network. 
Our intent is to construct the {state-to-state time advancements} of the flow fields while handling the moving interface description in a coupled data-driven system of fluid-structure interaction.
%
%
%
In particular, we address the issue of high-dimensionality by combining the projection-based ROMs with deep learning techniques such as the convolutional autoencoder and recurrent neural networks based on long short-term memory.  
These models can project the high-dimensional data on the low-dimensional spaces linearly or nonlinearly for data compression \cite{berkooz1993proper} and feature extraction \cite{miyanawala2017efficient}. The projected spaces are formed such that the loss of information is minimum or recoverable. More recently, these spaces have been combined with deep learning models to enhance the predictive abilities and are termed as projection-based DL models. This combination of projection and deep learning helps in reducing dimensionality and boosting real-time predictive abilities. There exist many variants of such hybrid projection-based DL models in the literature such as the POD-CNN by Miyanawala \& Jaiman \cite{miyanawala2019hybrid}, the POD-RNN from Bukka \emph{et al.} \cite{reddy2019data}, the CNN-RNN by Gonzalez \emph{et al.} \cite{gonzalez2018deep}.
Notably, for the first time,  Miyanawala \& Jaiman \cite{miyanawala2019hybrid} proposed a hybrid partitioned DL-ROM technique for fluid-structure interaction to  combine the POD and CNN to take the advantage of the optimal low-dimensional representation given by POD and the multiscale feature extraction by the CNN.
{\color{black} Besides the prediction of unsteady flow and wake-body interaction, such hybrid DL-ROM models have also been explored for other problems including turbulent flow control application \cite{mohan2018deep}, bifurcating flow phenomenon \cite{pichi2021artificial}, combustion \cite{wang2019non} and parametric steady-state solutions of the PDEs \cite{hesthaven2018non, chen2021physics}. These works reported substantial speed-ups using the projection-based DL models during the online predictions compared with their full-order counterpart. }

 The primary objective of this paper is to develop a partitioned and modular data-driven framework that can simulate a coupled fluid-structure interaction system with efficient predictive abilities. The present work builds upon the previous works \cite{bukka2021assessment, gupta2020assessment}, wherein the authors presented an assessment of two DL-ROMs, the so-called POD-RNN and the CRAN, for the flow past single and side-by-side stationary cylinder configurations. It was demonstrated that the CRAN framework achieves a longer time series prediction due to its nonlinear projection space compared with the POD-RNN, but the former relies on a uniform grid. This uniform grid can lose the interface description when projecting and interpolating the fluid variables from the full-order grid. Although the POD-RNN has comparatively lesser flow predictive abilities for complex flows, this model recovers a high-dimensional interface during the reconstruction process.
Therefore the question arises is there a way to learn and predict structural motion and fluid dynamics in a modular data-driven framework? How can we develop an efficient data-driven model that predicts the flow fields and the moving interface description simultaneously? 

In this paper, we develop a deep learning based reduced-order modeling framework to address the aforementioned questions. %
We consider a partitioned data-driven approach to couple the fluid and solid subdomains via deep neural networks.
Specifically, the present work unifies the attributes of the POD-RNN and the CRAN frameworks for reducing the coupled dynamics and inferring the fluid flow interacting with a freely vibrating structure. Of particular interest is to predict the flow fields via the CRAN framework, while handling the moving interfaces via the POD-RNN.
Using high-fidelity time series data, we first project the dataset to a low-dimensional subspace using POD and then the time-dependent coefficients of the POD are iteratively trained using the recurrent neural network. 
We model the flow variables in a uniform with a static pixeled grid via convolutional architecture. This simplified procedure addresses the Lagrangian-to-Eulerian conflict by embedding the full-order fluid-structure data on the uniform grid while providing convenience in the training of the CRAN framework. 
Based on an unstructured irregular grid for the full-order simulation, we propose a strategy termed as the snapshot-field transfer and load recovery (snapshot-FTLR) to select the structured grid for the fluid domain. Together with the CRAN framework, this strategy maintains the accuracy of the full-order interface forces and provides a reduced learning model for fluid-structure interaction.
The proposed end-to-end partitioned DL-ROM framework is modular and fully data-driven; hence, it aligns with the development of a digital twin involving FSI effects.

The article is organized as follows: Section \ref{fom_vs_rom_v1} describes the full-order governing physical model and a reduced-order representation of the fluid-structure interaction. Section \ref{two_level} presents the hybrid partitioned DL methodology and the DL-ROM components for the solid and fluid dynamics. Section \ref{interface-load-recovery} introduces the snapshot-FTLR technique and the DL-ROM grid search. The article ends with an application to the VIV motion of a freely oscillating cylinder {\color{black} for different set of reduced velocities} in section \ref{VIV_application}  and the main conclusions are provided in section \ref{conclusions}.

\section{Full-order vs reduced-order modeling}\label{fom_vs_rom_v1}
{ \color{black} This section starts by describing the full-order equations of a coupled fluid-structure interaction, followed by a brief description of reduced-order modeling. 
\subsection{Full-order modeling}\label{fom_ns_ale}
A coupled fluid-solid system in an arbitrary Lagrangian-Eulerian (ALE) reference frame is modeled by the isothermal fluid flow interacting with a rigid body structure: 
\begin{align}
\rho^\mathrm{f}\frac{\partial  \buf}{\partial t}+ \rho^\mathrm{f}(\buf - \mathbf{w})\cdot\bm{\nabla}\buf = \bm{\nabla}\cdot\stf+\bbf \quad \text{on} \quad \df, \label{eq1}\\ 
\frac{\partial \rho^\mathrm{f} }{\partial t} + \bm{\nabla}\cdot (\rho^\mathrm{f} \buf) = 0 \quad \text{on}\quad \df, \label{eq2} \\ 
\boldsymbol{m}^{\mathrm{s}} \frac{\partial \bus}{\partial t} +\boldsymbol{c}^{\mathrm{s}}  \bus +\boldsymbol{k}^{\mathrm{s}}  \left(\varphi^{\mathrm{s}}\left(\boldsymbol{z}_{0}, t\right)-\boldsymbol{z}_{0}\right) = \bFs +\bbs \quad \text { on } \dsnt, \label{eq3}
\end{align}}
where superscripts $\mathrm{f}$ and $\mathrm{s}$ denotes the fluid and structural variables, respectively. In the fluid domain $\dfnt$,  $\buf$ and $\mathbf{w}$ represent the fluid and mesh velocities, respectively and $\bbf$ is the body force, and $\stf$ is the Cauchy stress tensor for a Newtonian fluid written as
$
\stf = -\textit{p}^{\mathrm{f}}\textbf{I}+\vf\left(\bm{\nabla}\buf+(\bm{\nabla}\buf)^{T}\right) 
$. Here, $p^\mathrm{f}$ is the pressure in the fluid, $\textbf{I}$ is the identity tensor and $\vf$ is the fluid viscosity. 
Any arbitrary submerged rigid body $\dsnt$ experiences transient vortex-induced loads and, as a result, may undergo large structural motion if mounted elastically. The rigid body motion along the two Cartesian axes is modeled by Eq. (\ref{eq3}), where $\boldsymbol{m}^{\mathrm{s}}, \boldsymbol{c}^{\mathrm{s}}$ and $\boldsymbol{k}^{\mathrm{s}}$ denote the mass, damping and stiffness matrices, respectively. $\bus$ represents the rigid body motion at time $t$ with $\bFs$ and $\bbs$ as the fluid traction and body forces acting on it, respectively. Here, $\mathbf{\varphi}^{\mathrm{s}}$ denotes the position vector that transforms the initial position $z_{0}$ of the rigid body to time $t$.  This coupled system must satisfy the no-slip and traction continuity conditions at the fluid-body interface $\ifs$ as follows:
\begin{align} 
\buf \left( t\right)= \bus\left(t\right), \label{eq4} \\ 
\int_{\ifs} \stf \cdot \mathbf{n} \mathrm{d} \Gamma+ \bFs  = \mathbf{0},  \label{eq5}
\end{align}
where $\mathbf{n}$ and $\mathrm{d} \Gamma$ denote the outward normal and the differential surface area of the fluid-solid interface, respectively. While Eq. (\ref{eq4}) enforces the velocity continuity on the moving interface, the balance of the net force exerted by the fluid on the rigid body is given by Eq. (\ref{eq5}). 
Due to the body-conforming Eulerian-Lagrangian treatment, the interface conditions provide accurate modeling of the boundary layer and the vorticity generation over a moving body.
The coupled differential equations in Eqs. (\ref{eq1})-(\ref{eq3}) are numerically solved using the Petrov–Galerkin finite element and the semi-discrete time stepping \cite{jaiman2016partitioned}. The weak form of the incompressible Navier-Stokes equations is solved in space using equal-order isoparametric finite elements for the fluid velocity and pressure. We employ the nonlinear partitioned staggered procedure for the stable and robust coupling of the fluid-structure interaction \cite{jaiman2016partitioned}. This completes the description of the coupled full-order FSI system. 

In relation to the model-order reduction, the coupled equations (Eqs. (\ref{eq1})-(\ref{eq3})) can be written in an abstract state-space form as
\begin{equation}
\begin{split}
\frac{d \mathbf{z}}{d t} = \textbf{F} (\textbf{z}), 
\end{split}
\end{equation}
{\color{black} where $\mathbf{z} \in \mathbb{R}^{M}$ is the state vector for a coupled FSI domain with a total of $M$ variables in the system. 
For our fluid-body system in the current study, the state vector involves the fluid velocity and the pressure as $\mathbf{z} = \{\mathbf{u}^\mathrm{f}, p^\mathrm{f}  \}$ and the structural velocity includes the three translational degrees-of-freedom.
Note that the pressure $p^{f}$ can be written as some function of density $\rho^{f}$ via the state law of a fluid flow. The right hand side term $\textbf{F}$ represents a dynamic model and can be associated with a vector-valued differential operator describing the spatially discretized PDEs in Eqs. (\ref{eq1})-(\ref{eq3}).} The temporal term $ d \mathbf{z} / d t$ is the system dynamics which determines the instantaneous physics of a fluid-structure system and is useful for creating a low-order representation. The resultant spatial-temporal dynamics of the fluid-structure interaction are driven by the inputs such as the modeling parameters and boundary conditions. Next, we briefly review a data-driven reduced-order modeling based on traditional projection and convolutional autoencoder techniques.

{ \color{black}
\subsection{Reduced-order modeling}
From the perspective of a data-driven approach, the idea is to build the differential operator $\textbf{F}$ by projecting onto a set of low-dimensional trial subspace. In that sense, one can decompose $\textbf{F}(\textbf{z})$ to encapsulate the constant term $\mathbf{C}$, a linear $\mathbf{B} \mathbf{z}$ and nonlinear $\mathbf{F}^{\prime}(\mathbf{z})$ dynamical components as 
\begin{equation}
\mathbf{F}(\mathbf{z})=\mathbf{C}+\mathbf{B} \mathbf{z}+\mathbf{F}^{\prime}(\mathbf{z}).
\end{equation} 
Using the Galerkin-based ROMs, we represent the state vector $\mathbf{z} \in \mathbb{R}^{M}$ via a subspace spanned by the column vectors of
the low-dimensional modes $\mathcal{V} \in \mathbb{R}^{M \times K}$, with $K << M$. 
The matrix $\mathcal{V}$ is referred to as the reduced basis spanning the subspace onto which the dynamics is projected.
This allows to approximate the state vector $\mathbf{z}$ as $\mathcal{V}\mathbf{\tilde{z}}$ with $\mathbf{\tilde{z}}\in \mathbb{R}^{K}$ and thereby reducing the system dynamics as  
\begin{equation} \label{low_order_dynamics}
\frac{d \mathbf{\tilde{z}}}{d t}=\mathcal{V}^{T} \mathbf{C}+\mathcal{V}^{T} \mathbf{B} \mathcal{V} \mathbf{\tilde{z}}+\mathcal{V}^{T} \mathbf{F}^{\prime}(\mathcal{V} \mathbf{\tilde{z}}).
\end{equation}
{ \color{black} Here, the reduced-order space is constructed by defining a choice of modes $\mathcal{V} \in \mathbb{R}^{M \times K}$ using the snapshot matrix $\textbf{Z} = \{\mathbf{z}^{1} \; \mathbf{z}^{2} \; \ldots  \; \mathbf{z}^{S}\} \in \mathbb{R}^{M \times S} $, with $M$ number of variables and $S$ denotes the number of snapshots. The columns of the matrix $\mathcal{V}=\{\mathcal{V}^{1}\; \mathcal{V}^{2}\; \ldots\; \mathcal{V}^{K}\}$ form an orthonormal basis of $\mathbf{\tilde{Z}} = \{\mathbf{\tilde{z}}^{1} \; \mathbf{\tilde{z}}^{2} \; \ldots \; \mathbf{\tilde{z}}^{S}\} \in \mathbb{R}^{K \times S}$, a $K$ dimensional subspace of $\mathbb{R}^{M}$. The linear expansion of the state vector snapshot matrix $\textbf{Z} = \mathcal{V} \tilde{\textbf{Z}}$ allows the reduction to take place through special choices of subspace $\tilde{\textbf{Z}}$ for any dataset. For example, the well known POD derives this subspace to be such that the manifold variance is preserved as much as possible when projected to $\tilde{\textbf{Z}}$, given a fixed dimension constraint $K$. Using the singular value decomposition (SVD) decomposition, the snapshot matrix can be expressed as 
$
\mathbf{Z}=\mathcal{V} \boldsymbol{\Sigma} \mathcal{W}^{\mathrm{T}}=\sum_{J=1}^{K} \sigma_{J} \mathbf{v}_{J} \mathbf{w}_{J}^{T}, 
$
where $\mathbf{v}_{J}$ are the POD modes of the snapshot matrix $\mathbf{Z}$ and $\boldsymbol{\Sigma}  \in \mathbb{R}^{K \times K}$ is a diagonal matrix with diagonal entries $\sigma_{1} \geqslant \sigma_{2} \geqslant \ldots \geqslant \sigma_{K} \geqslant 0$. 
The total energy contained in each $\mathrm{POD}$ mode $\mathbf{v}_{J}$ is given by $\sigma_{J}^{2}$. $\mathcal{V}$ and $\mathcal{W}$ are the matrices with orthonormal columns that represent the eigenvectors of $\mathbf{Z} \mathbf{Z}^{T}$ and $\mathbf{Z}^{T} \mathbf{Z}$, respectively.}

It is well known that the above POD-Galerkin process is very effective for
dimensionality reduction of the linear term while the nonlinear term cannot be reconstructed properly for the unsteady wake flow dynamics. Therefore, the linear POD method may result in a similar order of computational expense to the full-order simulation.
For a low-order representation of such nonlinear terms, hyperreduction techniques such as the discrete empirical interpolation method \cite{chaturantabut2010} and  energy-conserving
sampling and weighting \cite{An2008ECSW} method can provide an additional level of approximation for the dimensionality reduction.
These hyperreduction strategies can reduce the required number of
modes, hence decreasing the computational cost while capturing the nonlinear regions properly \cite{miyanawala2019decomposition}.
All these techniques primarily aim to resolve the nonlinear term, i.e, $\mathcal{V}^{T} \mathbf{F}^{\prime}(\mathcal{V} \mathbf{\tilde{z}})$, where $\mathbf{\tilde{z}}$ is the reduced-order state of full-order state $\textbf{z}$. 
We note that these decomposition techniques are dependent on the choice of dataset, parameters and the initial condition of a nonlinear system. For instance, it has been shown that for the flow problems with small Kolmogorov $\mathrm{n}$-width, the POD modes optimally reduce the dimension of the nonlinear unsteady flows \cite{miyanawala2019hybrid, bukka2021assessment, bukka2019data}. However, in general, these empirical projection-based reduced-order models can come at the cost of large subspace dimensions for turbulence or convection-dominated problems characterized by large Kolmogorov $\mathrm{n}$-width. 

To address the drawbacks of the projection-based ROMs, the use of autoencoders provides a promising alternative for constructing the low-order projections of the state vector snapshot matrix $\textbf{Z}$. 
In an autoencoder, $\mathcal{F}$ is trained to output the same input data $\textbf{Z}$ such that $\textbf{Z} \approx \mathcal{F}(\textbf{Z} ; \boldsymbol{\theta}_{AE})$, where $\boldsymbol{\theta}_{AE}$ are the parameters of the end-to-end autoencoder model. The process is to train the parameters $\boldsymbol{\theta}_{AE}$ using an iterative minimization of an error function $E$
\begin{equation}
    \boldsymbol{\theta}_{AE}=\operatorname{argmin}_{\boldsymbol{\theta}_{AE}}[E(\textbf{Z}, \mathcal{F}(\boldsymbol{\textbf{Z}} ; \boldsymbol{\theta}_{AE}))].
\end{equation}
For the use of the autoencoder as a dimension compressor, the dimension of the low-order space called the latent space, say $\tilde{\textbf{H}}$, is smaller than that of the input or output data $\textbf{Z}$. When we obtain the output $\mathcal{F}(\textbf{Z})$ similar to the input such that $\textbf{Z} \approx \mathcal{F}(\textbf{Z})$, the latent space is a low-dimensional representation of its input which provides a low-rank embedding. In an autoencoder, the dimension compressor is called the encoder $\mathcal{F}_{e}$ and the counterpart is the decoder $\mathcal{F}_{d}$. Using the encoder-decoder architecture, the internal procedure of the autoencoder can be expressed as
$
\tilde{\textbf{H}}=\mathcal{F}_{e}(\textbf{Z}), \quad \textbf{Z}=\mathcal{F}_{d}(\tilde{\textbf{H}}).
$

For a general nonlinear system, one can construct the subspace projection as a self-supervised DL parametrization of the autoencoders without any SVD process of the state $\textbf{Z}$. Instead, the decomposition can be achieved by generating the trainable layers of the encoder and decoder space such that the error function is minimized.
By using a series of convolutional and nonlinear mapping process, the autoencoder can lead to an efficient construction of a compressed latent space to characterize the reduced dynamics of a given nonlinear system. The autoencoders can be interpreted as a
flexible and nonlinear generalization of POD \cite{reddy2020deep}.
Both projection-based POD and convolutional autoencoder help in reducing the dimensionality of the incoming data $\mathbf{Z}$ by encoding dominant patterns from the high-dimensional data.}
We next turn our attention to the proposed hybrid partitioned DL-ROM for solving fluid-structure interaction problems.
%

\section{Hybrid partitioned DL-ROM framework}\label{two_level}
%
Herein, we present a hybrid partitioned DL-ROM which is trained using the high-dimensional simulation generated by solving the full-order fluid-structure interaction model. 
Consistent with the high-dimensional FSI solver, we consider two separate data-driven solvers for fluid and solid subdomains with their own independent neural network in a partitioned manner. 
Once the high-dimensional data is effectively reduced via POD and convolutional autoencoder, the low-dimensional states can be used to infer the coupled behavior of a fluid-structure system. 
Together with POD and convolutional autoencoder for the dimensionality reduction, we use the well-known LSTM-RNNs for evolving the low-dimensional states in time.
The mathematical structure of the LSTM-RNN represents a nonlinear state-space form. This particular attribute makes them suitable for a nonlinear dynamical system of fluid-structure interaction.
When the dimensionality of the data is reduced using POD and evolved in time using RNN, the hybrid DL-ROM is referred to as the POD-RNN.  
On a similar note, when the low-dimensional states, obtained using a convolutional autoencoder, are similarly evolved in time, the hybrid framework is called the CRAN. 

For predicting the coupled dynamics of the fluid flow interacting with a freely vibrating structure, we unify the combined effects of the POD-RNN and the CRAN frameworks for the partitioned fluid and solid sub-domains.
The illustration of this hybrid partitioned DL-ROM approach is provided in Fig.~\ref{two_level_framework}. 
The first partition or level of our framework is the POD-RNN for learning the solid displacements, while the second partition is the CRAN for learning the flow fields in a uniform pixel grid. 
On a given training data, the POD-RNN model is employed to track the structural displacements. This tracking helps in constructing a level-set function. 
A level-set function is a distance function based on structural geometry. It is used for setting the level set boundaries on the computational grid based on the position of the structure with a binary condition as the input function. 
Miyanawala \& Jaiman \cite{miyanawala2017efficient, miyanawala2018novel} presented a CNN model combined with the level set  to predict the unsteady fluid forces  over stationary geometries.
Further details about the coupling of a level-set function with convolutional neural network can be found in \cite{miyanawala2017efficient,miyanawala2018novel,miyanawala2019hybrid}. 
An interface load recovery is constructed to select the pixel grid for the CRAN framework while extracting the interface information using the level-set and forces.

{ \color{black}
The solid and fluid equations are solved sequentially in a partitioned manner and constrained at the interface with traction and velocity continuity in a partitioned full-order solver for fluid-structure interaction. Analogous to the full-order partitioned solver, we apply the DL-ROMs sequentially for the solid and fluid variables and couple them via an interface recovery process. The choice of the so-called {physics-based DL drivers} can be different and dependent on the nature of the problem at hand. However, the interface load recovery is general data coupling for learning and modeling of fluid-structure interaction with pixel-based flow field information.
We first elaborate on the integration of POD-based low-dimensional approximation with the RNN process for predicting structural motion.
%
}
\begin{figure}
\centering
\includegraphics[width=1.00\textwidth]{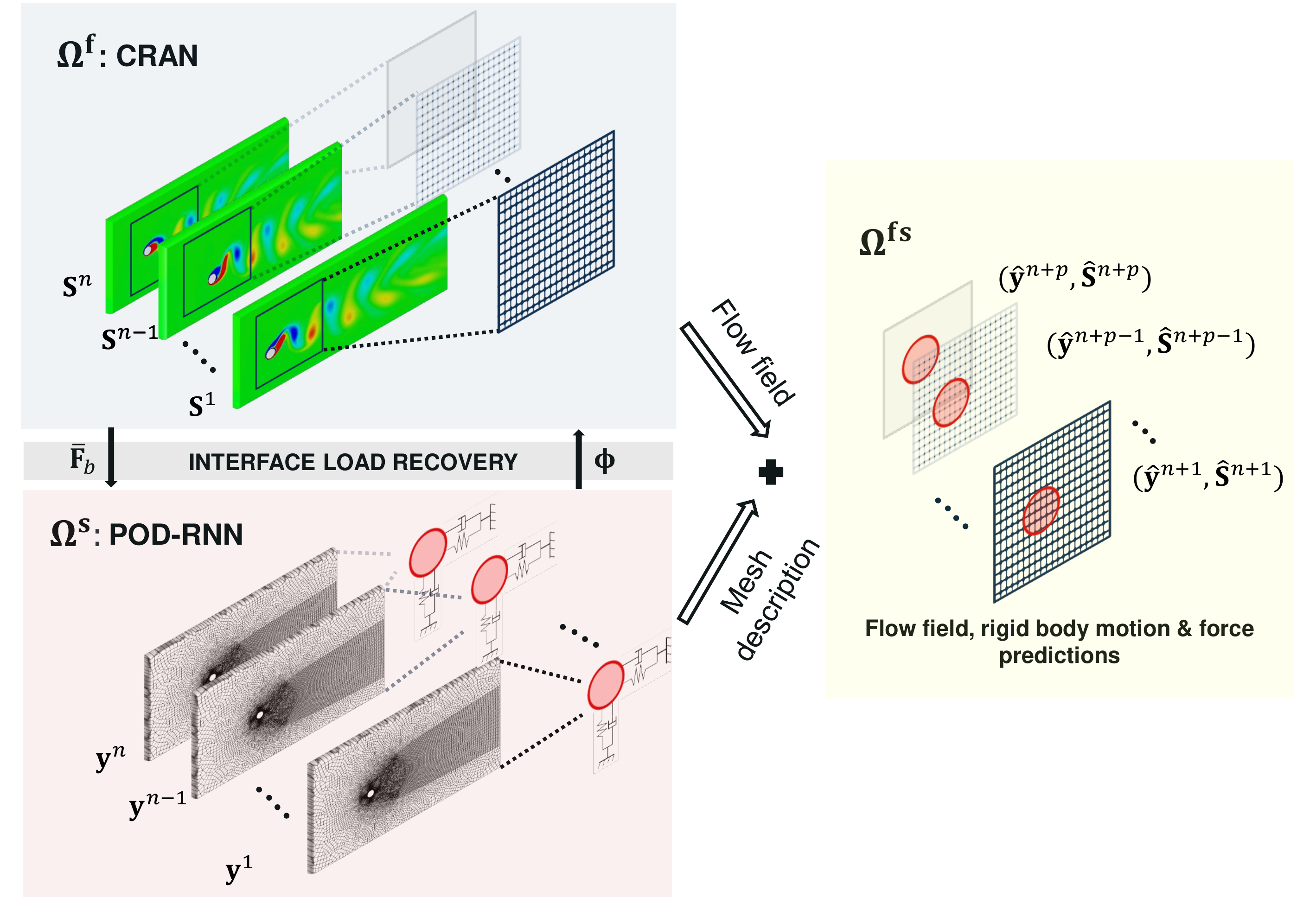}
\caption{Illustration of a hybrid partitioned  DL-ROM framework for fluid-structure interaction. The field variables $\df$ are learned and predicted on the uniform grid using the CRAN (blue box), while the interface information $\dsnt$ with moving point cloud is predicted via the POD-RNN (red box). These boxes exchange the interface information (grey box) that couples the system via level-set $\Phi$ and force signals $\bar{\textbf{F}}_{b}$. The yellow box demonstrates the synchronized predictions. As the CRAN learns the flow variables on the uniform grid (blue box), the level-set $\Phi$ is utilized in calculating the pixelated the force signals $\bar{\textbf{F}}_{b}$ on the interface. }
\label{two_level_framework}
\end{figure}

\subsection{The POD-RNN framework: structural motion } \label{pod_rnn}
This DL-ROM technique proposed by Bukka \emph{et al.} \cite{reddy2019data} utilizes high-fidelity time series of snapshot data obtained from the full-order simulations or experimental measurements. These high-dimensional data are decomposed into the spatial modes (dominant POD basis), the mean field, and the temporal coefficients as the reduced-order dynamics using the Galerkin projection of POD. The spatial modes and the mean field form the offline database, while the time coefficients are propagated and learned using variants of RNNs: closed-loop or encoder-decoder type. This projection and propagation technique is hence called the POD-RNN. In this work, we learn and infer the full-order ALE mesh displacements using the POD-RNN. {\color{black} The key assumptions adopted here are that (a) the complete position and shape of the exact fluid-solid interface are inherent in the point cloud displacements, and (b) the POD energy spectrum for these mesh displacements are fast decaying due to the finite degrees of freedom of the rigid body motion. 
This can allow us to compress the displacements using the POD modes and then propagate with RNNs.}    

Let $\textbf{Y}=\{\textbf{y}^1 \; \textbf{y}^2 \; ... \; \textbf{y}^n \} \in \mathbb{R}^{m\times n}$ be the ALE displacement snapshot matrix along a given cartesian direction from the initial position $\textbf{y}^0$. Here, $\textbf{y}^i \in \mathbb{R}^{m}$ is the displacement snapshot at time $t^i$ and $n$ represents the number of such snapshots. $m \gg n$ are the number of data probes, for instance, here the number mesh nodes in the full-order simulation.
The target is to predict the future mesh position: $\hat{\textbf{y}}^{n+1}, \hat{\textbf{y}}^{n+2}, ... $ using the training dataset $\textbf{Y}$. The POD-RNN framework can be constructed by the following three step process:\\ \\
\noindent
\emph{Step 1: Construct the POD modes} $\nu$ \emph{for the dataset} $\textbf{Y}$

Given the $n$ snapshots of the ALE displacements along any co-ordinate axis, determine the offline database: {\color{black} the temporal mean vector ($\overline{\textbf{y}}\in \mathbb{R}^m$) and the POD basis modes ($\mathbf{\nu}\in \mathbb{R}^{m\times k}$) using the snapshot matrix of the ALE displacement \textbf{Y}. Here $k < n \ll m$ and $k$ is selected based on the energy hierarchy of the POD modes.} This projection aims to reduce the order of the high-dimensional moving point cloud from $O(m)$ to $O(k)$. The low-dimensional states are obtained linearly using the POD reconstruction technique
\begin{equation} \label{eq_pod}
\textbf{y}^i \approx \overline{\textbf{y}} +\mathbf{\nu}\textbf{A}_{\nu}^i, \qquad i=1,2,...,n, 
\end{equation}
where $\textbf{A}_{\nu}^i=[\textbf{a}_{\nu}^{i1} \: \textbf{a}_{\nu}^{i2} \: ... \: \textbf{a}_{\nu}^{ik}]^{\mathrm{T}} \in \mathbb{R}^k$ are the time coefficients of the $k$ most energetic modes for the time instant $t^i$. The low-dimensional states $\textbf{A}_{\nu} = \{\textbf{A}_{\nu}^1\;\textbf{A}_{\nu}^2\;...\;\textbf{A}_{\nu}^{n}\} \in \mathbb{R}^{k \times n}$ are hence generated for the dataset $\textbf{Y}$. 
Once generated, the problem boils down to predict $\hat{\textbf{A}}_{\nu}^{n+1}, \hat{\textbf{A}}_{\nu}^{n+2}, ... $ using the reduced-order state $\textbf{A}_{\nu}^{n}$ in an iterative manner. For a detailed analysis on linear or nonlinear POD reconstruction, readers can refer to \cite{miyanawala2019decomposition, burkardt2006pod}. \\

\noindent
\emph{Step 2: Supervised learning and prediction of the low-dimensional states using LSTM-RNN}

Training consists of learning a dynamical operator $g_{\nu}$ that allows to recurrently predict finite time series of the low-dimensional states. We employ LSTM-RNN, which exploits both the long-term dependencies in the data and prevents the vanishing gradient problem during training \cite{hochreiter1997long}. This is employed as a one-to-one dynamical mapping between the low-dimensional states $\textbf{A}_{\nu}^{i-1}$ and $\textbf{A}_{\nu}^i$. 
\begin{equation}
\textbf{A}_{\nu}^i = f_{\nu}(\textbf{y}^i), \qquad \hat{\textbf{A}}_{\nu}^i = g_{\nu}(\textbf{A}_{\nu}^{i-1}; \theta_{\nu,evolver}),  \qquad i=1,2,...,n, 
\end{equation}
where $f_{\nu}$ is the known function relating the low-dimensional state and the full-order ALE data. {\color{black} The algebraic relation in Eq. (\ref{eq_pod}) can be re-written as  $\textbf{A}_{\nu}^i \approx \mathbf{\nu}^{T} (\textbf{y}^i - \overline{\textbf{y}}) \approx  f_{\nu}(\textbf{y}_{i})$. Hence, $f_{\nu}$ is the known mapping function utilized to obtain the POD time coefficients from a full-order snapshot.} 
The term $g_{\nu}$ is the trainable operator parametrized by $\theta_{\nu,evolver}$, which outputs a time advanced state $\hat{\textbf{A}}_{\nu}^i$ with an input $\textbf{A}_{\nu}^{i-1}$. 
To find $\theta_{\nu,evolver}$, one can set the LSTM-RNN such that the low-dimensional states $\{\textbf{A}_{\nu}^1\;\textbf{A}_{\nu}^2\;...\;\textbf{A}_{\nu}^{n-1}\}$ are mapped to the time advanced low-dimensional states $\{\hat{\textbf{A}}_{\nu}^2\; \hat{\textbf{A}}_{\nu}^3\;...\;\hat{\textbf{A}}_{\nu}^n\}$ in a way that $g_{\nu}$ mapping is one-to-one
\begin{equation}
\hat{\textbf{A}}_{\nu}^i = g_{\nu}(\textbf{A}_{\nu}^{i-1};\theta_{\nu,evolver}) , \qquad i=2,3,...,n, 
\end{equation}
where $\theta_{\nu,evolver}$ are the weight operators of the LSTM cell. Once trained, $\theta_{\nu,evolver}$ forms an offline parameter recurrent space which generates a finite amount of the predicted low-dimensional states. This closed-loop network predicts the output iteratively with the same trained weights, meaning that the generated output is fed as an input for the next prediction. \\ 

\noindent
\emph{Step 3: Reconstruction to the point cloud} $\hat{\textbf{y}}^{n+1}$

In Step 2, the reduced-order dynamics is predicted using the LSTM-RNN. Once $\hat{\textbf{A}}_{\nu}^{n+1}$ is generated, it is straightforward to reconstruct the full-order ALE displacement at $t^{n+1}$ as follows:
\begin{equation} \label{point_cloud}
\hat{\textbf{y}}^{n+1} \approx \overline{\textbf{y}} + \nu \hat{\textbf{A}}_{\nu}^{n+1}.
\end{equation}
Hence, to track the moving point cloud, the predicted low-dimensional states are linearly combined with the temporal mean vector ($\overline{\textbf{y}}\in \mathbb{R}^m$) and the POD basis modes ($\mathbf{\nu}\in \mathbb{R}^{m\times k}$). An illustration of the entire predictive process is shown in Fig.~\ref{pod_rnn_setup}. This technique of the POD-RNN on the ALE displacements can be capable of predicting the FSI motion for finite time steps $\hat{\textbf{y}}^{n+1}, \hat{\textbf{y}}^{n+2}, ...,\hat{\textbf{y}}^{n+p}$ autonomously provided that the error is within the acceptable range of accuracy.

\begin{figure}
\centering
\includegraphics[width=1.0\textwidth]{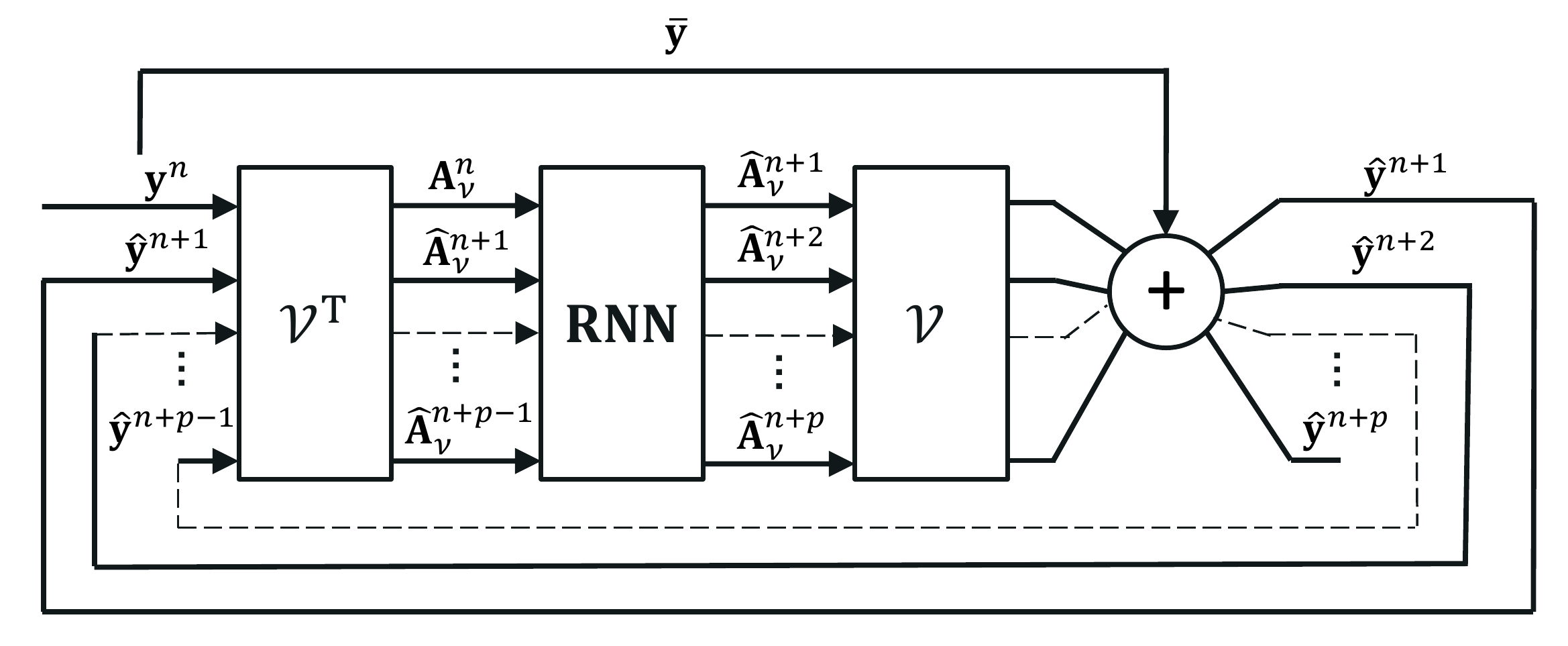}
\caption{Schematic of the POD-RNN framework with a block diagram for the iterative prediction. With one input driver $\textbf{y}^{n}$, the predictions $\hat{\textbf{y}}^{n+1}, \hat{\textbf{y}}^{n+2}, ... ,\hat{\textbf{y}}^{n+p}$ are achieved autonomously. The diagram illustrates the prediction of $p$ number of time steps from $\hat{\textbf{y}}^{n+1}$ to $\hat{\textbf{y}}^{n+p}$. {\color{black} The dashed lines imply an iterative process of generating all the predictions between $\hat{\textbf{y}}^{n+2}$ and $\hat{\textbf{y}}^{n+p}$ using the POD-RNN method.}}
\label{pod_rnn_setup}
\end{figure}

{ \color{black}
\textbf{Normalisation of dataset:} In the POD-RNN, we train and predict the time coefficients obtained from the POD analysis of the full-order dataset. Let $\textbf{A}_{\nu}=\{\textbf{A}_{\nu}^{1} \;\textbf{A}_{\nu}^{2} \; ... \; \textbf{A}_{\nu}^{n} \} \in \mathbb{R}^{k\times n}$ be the low-dimensional states obtained from the POD analysis of the ALE snapshot matrix $\textbf{Y}$. The scaling of $\mathbf{A}_{\nu}$ is conducted as 
\begin{equation}
\mathbf{A}_{\nu, s}^{i}=\frac{\mathbf{A}^{i}_{\nu} -\mathbf{A}_{\nu, \min}}  {\mathbf{A}_{\nu, \max}-\mathbf{A}_{\nu, \min}},  \qquad i=1,2,...,n,
\end{equation}
where $\mathbf{A}_{\nu, \max}$ and  $\mathbf{A}_{\nu, \min}$ are the maximum and minimum values in $\mathbf{A}_{\nu}$, respectively, and arranged as a vector. This ensures $\mathbf{A}_{\nu, s}^{i} \in [0,1]^{k}$. The obtained dataset matrix $\mathbf{A}_{\nu, s} = \left\{\mathbf{A}_{\nu, s}^{1}\; \mathbf{A}_{\nu, s}^{2}\; \ldots\; \mathbf{A}_{\nu, s}^{n} \right\} \in \mathbb{R}^{k \times n}$ consists of the normalised time coefficients.   \\
\textbf{Training:} The training consists of finding the evolver parameters of the LSTM-RNN $\theta_{\nu,evolver}$ such that the observable loss is minimized
\begin{equation}  \label{pod_rnn_loss}
\theta_{\nu, evolver} 
=
\mathrm{argmin}_{(\theta_{\nu, evolver})}\left[
\frac{1}{n-1} \sum_{i=2}^{n} \frac{\left\|\mathbf{A}^{i}_{\nu,s}- g_{\nu}(\mathbf{A}^{i-1}_{\nu,s}; \theta_{\nu, evolver})\right\|_{2}^{2}}{\left\|\mathbf{A}^{i}_{\nu,s}\right\|_{2}^{2}}
\right].
\end{equation}
This loss function is iteratively minimised over the entire training sequence using the adaptive moment optimization \cite{kingma2014adam}. 
}

\subsection{The CRAN framework: flow field} \label{cnn_rnn}
The POD-RNN method can provide an optimal low-dimensional space (encoding) in which the modes are not only computationally inexpensive to obtain but can be physically interpretable as well \cite{miyanawala2019decomposition}. However, there are several problems with the POD encoding for highly nonlinear flow physics: (a) the encoding scales linearly, which may cause significant loss of the flow physics characterized by a large Kolmogorov $\mathrm{n}$-width, (b) the POD reconstruction of the flow problems dominated by turbulence often results in the slow decay of the POD energy spectrum, implying that the number of POD modes can increase significantly and be difficult to learn, and (c) the POD basis can have additional orthogonality constraint on the low-dimensional space which can limit its flexibility in general. So an alternative approach is to utilize a much more flexible encoding that can address the above challenges. One method is to use CNNs instead of POD, which leads to a non-intrusive type of DL-ROM framework called the CNN-RNN or simply the CRAN. Furthermore, as depicted in our previous work \cite{bukka2021assessment}, CRAN outperforms the POD-RNN in complex flow predictions by nearly 25 times. Hence, we rely on the CRAN technique for learning nonlinear flow variables due to the aforementioned advantages.

While operating synchronously in a nonlinear neural space, this projection and propagation technique extracts the low-dimensional embedding whereby the flow variables are extracted via CNNs and the encoding evolved via LSTM-RNN. Since there is no knowledge of the mean or basis vectors here, we construct a decoding space of transpose convolution that up-samples the low-dimensional encoding back to the high-dimensional space. 
It is assumed that the solution space of the unsteady flow attracts a low-dimensional subspace, which allows the embedding in the high-dimensional space.
This end-to-end convolutional autoencoder architecture is illustrated in Fig.~\ref{cnn_rnn_setup_2}. For the sake of completeness, the CRAN architecture is further elaborated in this section. For more details about the CRAN architecture, readers can refer to our previous work \cite{bukka2021assessment}.

The CRAN framework needs a high-fidelity series of
snapshots of the flow field obtained by full-order simulations. 
Let $\textbf{S}=\{\textbf{S}^1 \; \textbf{S}^2 \; ... \; \textbf{S}^n \} \in \mathbb{R}^{N_{x}\times\ N_{y}\times n}$ denote the 2D snapshots of a field dataset (such as pressure or velocity data from a flow solver), where $\textbf{S}^i \in \mathbb{R}^{N_{x}\times N_{y}}$ is the field snapshot at time $t^i$ and $n$ represents the number of such snapshots. $N_{x}$ and $N_{y}$ are the number of data probes in the respective Cartesian axes. For instance, here, the number of fixed (Eulerian) query points introduced in the moving point cloud. These probes are structured as a spatially uniform space for the CRAN architecture as it relies on the field uniformity. These probes are optimally selected based on the field convergence and the interface load recovery using the snapshot-FTLR method as discussed in section \ref{interface-load-recovery}. The target of the CRAN-based end-to-end learning is to encode-propagate-decode the future values at the field probes: $\hat{\textbf{S}}^{n+1}, \hat{\textbf{S}}^{n+2}, ... $ using the training dataset $\textbf{S}$. The CRAN framework is constructed using the following process: \\

\noindent
\emph{Step 1: Find the nonlinear encoding feature space for the dataset} $\textbf{S}$\\
Given the $n$ snapshots of any field data, construct a trainable neural encoder space using layers of linear convolutional kernels with nonlinear activation (Conv2D) as shown in Fig.~\ref{cnn_rnn_setup_2}. The dimensionality of the field 2D probes is conveniently and gradually down-sampled using the convolutional filters and the plain feed-forward networks until a finite size of the low-dimensional state is reached, similar to the POD encoding as discussed earlier. The update of time coefficient can be expressed as
\begin{equation}
\textbf{A}_{c}^i = f_{c}(\textbf{\textbf{S}}^{i};\theta_{c,enc}),  \qquad i=1,2,...,n,
\end{equation}
where $f_{c}$ is the trainable encoder space that is parametrized by $\theta_{c,enc}$. The time coefficients $\textbf{A}_{c}^i=[\textbf{a}_{c}^{i1} \: \textbf{a}_{c}^{i2} \: ... \: \textbf{a}_{c}^{ih}]^{\mathrm{T}} \in \mathbb{R}^h$ represent $h$ encoded features at time $t^i$. The low-dimensional states $\textbf{A}_{c} =\{\textbf{A}_{c}^1\; \textbf{A}_{c}^2\; ...\; \textbf{A}_{c}^n\} \in \mathbb{R}^{h \times n}$ of the dataset $\textbf{S}$ are determined in a self-supervised fashion of the autoencoder, with no energy or orthogonality constraint unlike the POD. Here, $h < n \ll (N_{x} \times N_{y})$ and $h$ is the unknown hyperparamter for the optimal feature extraction based on the input dataset $\textbf{S}$. This projection reduces the order of high-dimensional field from $O(N_{x}\times N_{y})$ to $O(h)$.
After the decomposition, the problem boils down to predict $\hat{\textbf{A}}_{c}^{n+1}, \hat{\textbf{A}}_{c}^{n+2}, ...$ using the low-dimensional state  $\textbf{A}_{c}^{n}$ and spatially up-sample to $\hat{\textbf{S}}^{n+1}, \hat{\textbf{S}}^{n+2}, ...$ as generative outputs. \\ 

\noindent
\emph{Step 2: Supervised learning and prediction of the low-dimensional states using LSTM-RNN}

Similar to the POD-RNN, the training of the low-dimensional space consists of learning a one-to-one dynamical operator $g_{c}$ that allows to predict a certain series of the low-dimensional states. Without a loss of generality, we employ the LSTM-RNN in a closed-loop fashion. The LSTM-RNN is employed as a state transformation between the low-dimensional states $\textbf{A}_{c}^{i-1}$ and $\textbf{A}_{c}^i$. 
Instantiate the network such that the low-dimensional states $\{\textbf{A}_{c}^1\;\textbf{A}_{c}^2\;...\;\textbf{A}_{c}^{n-1}\}$ are mapped to the time advanced low-dimensional states $\{\hat{\textbf{A}}_{c}^2\; \hat{\textbf{A}}_{c}^3\;...\;\hat{\textbf{A}}_{c}^n\}$ so that dynamical operator $g_{c}$ is a one-to-one transformation
\begin{equation}
\hat{\textbf{A}}_{c}^i = g_{c}(\textbf{A}_{c}^{i-1};\theta_{c,evolver}), \qquad i=2,...,n,
\end{equation}
where $\theta_{c,evolver}$ are the weight operators of the LSTM cell. Likewise, once trained, $\theta_{c,evolver}$ also forms an offline recurrent parameter space which predicts finite time steps of the self-supervised features obtained from the neural encoder space. \\ 

\noindent
\emph{Step 3: Generating the high-dimensional time advanced state} $\hat{\textbf{S}}^{n+1}$

Once $\hat{\textbf{A}}_{c}^{n+1}$ is generated, there is a need to prolongate the time advanced features at $t^{n+1}$ using gradual reverse convolution (DeConv2D) as these features are often hard to interpret. As there is no knowledge of the mean or the POD basis vectors, we rely on decoding these low-dimensional features gradually using a decoder space which is a mirror of the encoder space. 
\begin{equation}
\hat{\textbf{S}}^{n+1} = f_{c}^{-1}(\hat{\textbf{A}}_{c}^{n+1};\theta_{c,dec}),  
\end{equation}
where $f_{c}^{-1}$ is the trainable decoder space that is parametrized by $\theta_{c,dec}$. The trained LSTM-RNN is employed to iteratively generate the desired future states $\hat{\textbf{S}}^{n+1}, \hat{\textbf{S}}^{n+2} ...., \hat{\textbf{S}}^{n+p}$ starting from $\textbf{S}^{n}$. This technique of the CRAN on the flow variables is self-supervised and is capable of predicting the flow variables at the fixed probes provided that the prediction error is within the acceptable range of accuracy. The complete predictive process is illustrated in Fig. \ref{cnn_rnn_setup_2}. 

\begin{figure}
\centering
\includegraphics[width=1.0\textwidth]{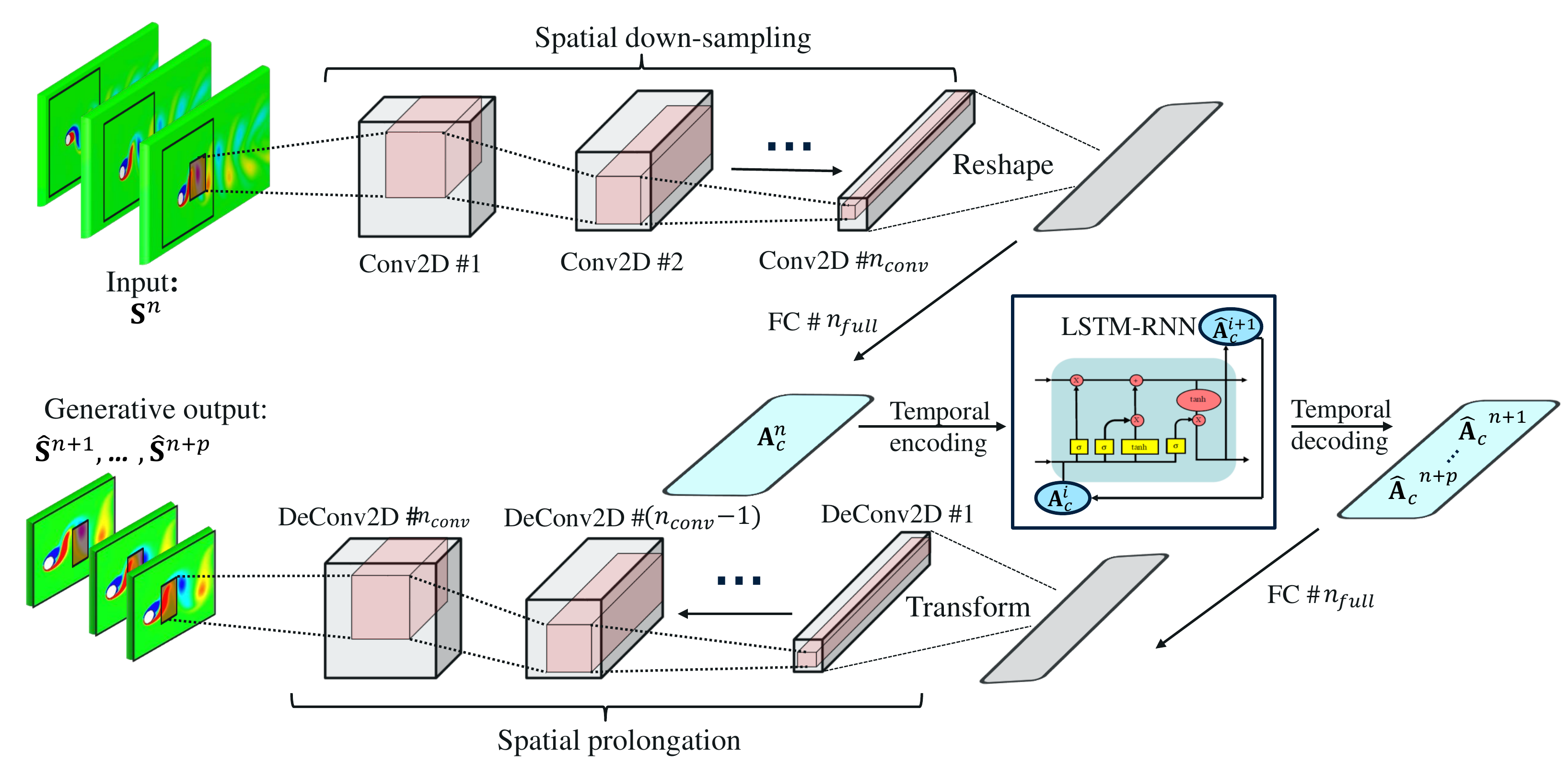}
\caption{Schematic of the CRAN framework for flow field prediction. The encoding is achieved by reducing the input dimension from $N_{x} \times N_{y}$ to $ \textbf{A}_{c}$ via CNNs (Conv2D) and the fully connected networks. The decoding is achieved using the fully connected layers and the transpose CNNs (DeConv2D). Between the encoding and decoding space, the LSTM-RNN evolves the low-dimensional state $\textbf{A}_c$.}
\label{cnn_rnn_setup_2}
\end{figure}

{ \color{black}
\textbf{Normalization of dataset:} We normalise the flow dataset snapshot matrix $\textbf{S}=\{\textbf{S}^1 \; \textbf{S}^2 \; ... \; \textbf{S}^n \} \in \mathbb{R}^{N_{x}\times\ N_{y}\times n}$ by first finding the fluctuation around the temporal mean using
\begin{equation}  \label{dn_cran_1}
\mathbf{S}^{\prime i}=\mathbf{S}^{i}-\overline{\mathbf{S}} ,  \qquad i=1,2,...,n,
\end{equation}
where $\overline{\mathbf{S}}=\frac{1}{n} \sum_{i=1}^{n} \mathbf{S}^{i}$ is the temporal average over the entire dataset and $\mathbf{S}^{\prime} =\{\textbf{S}^{\prime 1} \; \textbf{S}^{\prime 2} \; ... \; \textbf{S}^{\prime n} \} \in \mathbb{R}^{N_{x}\times\ N_{y}\times n}$ are the fluctuations around this mean. Next the scaling of $\mathbf{S}^{\prime}$ is conducted as 
\begin{equation} \label{dn_cran_2}
\mathbf{S}_{s}^{\prime i}=\frac{\mathbf{S}^{\prime i} -\mathbf{S}_{\min }^{\prime}}{\mathbf{S}_{\max }^{\prime}-\mathbf{S}_{\min }^{\prime}},  \qquad i=1,2,...,n,
\end{equation}
where $\mathbf{S}_{\max}^{\prime}$ and  $\mathbf{S}_{\min}^{\prime}$ are the maximum and the minimum vales in $\mathbf{S}^{\prime}$, respectively, and arranged as a matrix. The obtained dataset matrix $\mathbf{S}_{s}^{\prime} = \left\{\mathbf{S}_{s}^{\prime 1} \; \mathbf{S}_{s}^{\prime 2} \; \ldots\; \mathbf{S}_{s}^{\prime n} \right\} $ is further broken up into a set of ${N}_{s}$ finite time training sequences, where each training sequence consists of $N_{t}$ snapshots. The training dataset with the above modifications has the following form: 
\begin{equation} \label{dn_cran_3}
\mathcal{S}=\left\{\mathcal{S}_{s}^{\prime 1}\; \mathcal{S}_{s}^{\prime 2}\; \ldots\; \mathcal{S}_{s}^{\prime N_{s} } \right\} \in[0,1]^{N_{x} \times N_{y} \times N_{t} \times N_{s}}
\end{equation}
where each training sample $\mathcal{S}_{s}^{\prime j}=\left[\mathbf{S}_{s, j}^{\prime 1} \; \mathbf{S}_{s, j}^{\prime 2} \ldots \mathbf{S}_{s, j}^{N_{t}}\right]$ is a matrix consisting of the scaled database.

\textbf{Optimization:} A loss function is constructed that equally weights the error in the full-state reconstruction and the evolution of the low-dimensional representations. The target of the training is to find the CRAN parameters $\theta_c = \{ \theta_{c,enc}, \theta_{c,dec}, \theta_{c,evolver}\}$ such that for any sequence $\mathcal{S}_{s}^{\prime j}=\left[\mathbf{S}_{s, j}^{\prime 1} \; \mathbf{S}_{s, j}^{\prime 2} \ldots \mathbf{S}_{s, j}^{N_{t}}\right]$ and its corresponding low-dimensional representation $ \left[\mathbf{A}_{c,j}^{1} \; \mathbf{A}_{c,j}^{2} \ldots \mathbf{A}_{c,j}^{N_{t}}\right] $ the following error between the truth and the prediction is minimized:

\begin{equation}  \label{CRAN_loss}
\theta_c 
=
\mathrm{argmin}_{\theta_{c}}
\left[
\frac{\beta}{N_{t}} \sum_{i=1}^{N_{t}} \frac{\left\|\mathbf{S}_{s,j}^{\prime i}- f_{c}^{-1}(f_{c}(\mathbf{S}_{s,j}^{\prime i}; \theta_{c,enc}); \theta_{c,dec})\right\|_{2}^{2} }{\left\|\mathbf{S}_{s,j}^{\prime i}\right\|_{2}^{2}}
+
\frac{(1-\beta)}{N_{t}-1} \sum_{i=2}^{N_{t}} \frac{\left\|\mathbf{A}^{i}_{c,j}- g_c(\mathbf{A}^{i}_{c,j}; \theta_{c,evolver})\right\|_{2}^{2}}{\left\|\mathbf{A}^{i}_{c,j}\right\|_{2}^{2}}
\right]
\end{equation}
where $\beta=0.5$. The value of $\beta$ is chosen in order to give equal weightage to the errors in the full state reconstruction and feature predictions, respectively. This loss function is iteratively minimised for all the $j = 1,2,\dots,N_{s}$ training sequences using the adaptive moment optimization \cite{kingma2014adam}. 
}

\section{Snapshot-field transfer and load recovery} \label{interface-load-recovery}
We introduce a novel procedure for the partitioned mesh-to-mesh field transfer between the full-order grid and the neural network grid.
A convolutional neural network relies on a spatially uniform input data stream for identifying patterns in datasets. This is accounted for the uniform dimension 
and operation of the adaptive CNN kernels. 
Given that most practical FSI problems are modeled in a highly unstructured and body conformal mesh, there is a need to develop a flow field data processing step to interpolate the scattered information as snapshot images before learning them in the CRAN framework. Consequently, in this section, we introduce a general data processing step that can be utilized for mapping the flow field variables from an unstructured grid of the full-order model to a uniform  grid of the DL-ROM. We achieve this via interpolation and projection of the field information in an iterative process that allows a recoverable interface force data loss.  Once this loss is observed in the training forces, we correct them by reconstructing to a higher-order CFD force. We select the interface force as the primary criteria because the boundary layer forces are highly grid sensitive. This iterative process is  cyclically performed as shown in Fig.~\ref{load_recovery}. We refer to the entire cyclic process as a snapshot-field transfer and load recovery { (snapshot-FTLR)} method.       
\begin{figure}
\centering
\includegraphics[width=0.9\textwidth]{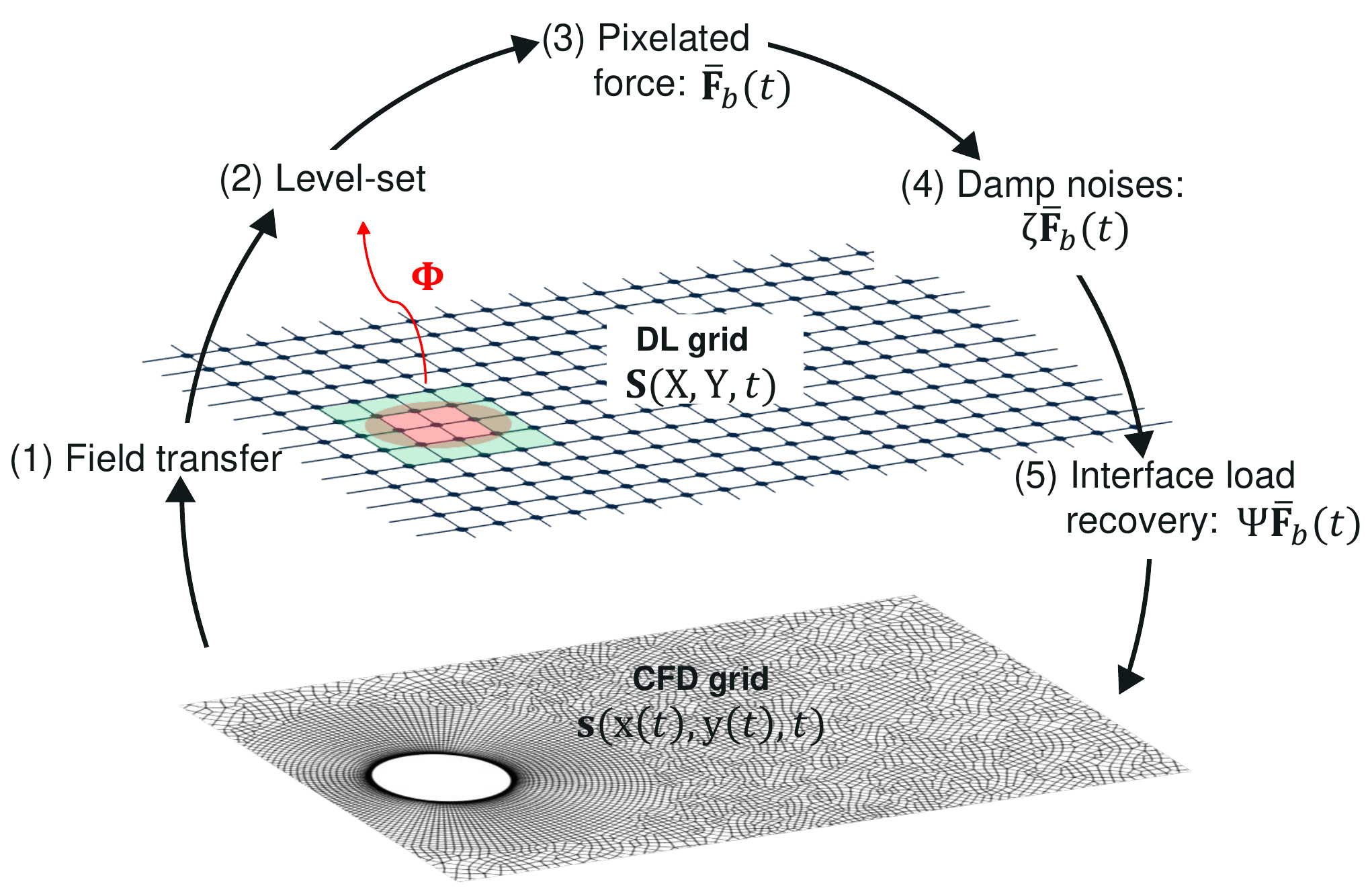}
\caption{Illustration of an iterative cycle for mesh-to-mesh field transfer and load recovery to determine the CRAN snapshot grid. See the details 
of all variables in the main text.}
\label{load_recovery}
\end{figure}

Herein, our intent is to find the best DL-ROM grid that can recover the full-order interface load correctly and capture the Lagrangian-Eulerian interface. 
While the loss of data information is observed in the training data, 
it is assumed that the loss is unchanged and can be used as correction 
during the predictions. The proposed snapshot-FTLR technique can be divided into five key steps, which are as follows:
\begin{enumerate}
\item Field transfer: The first step involves the field transfer from a highly unstructured moving point cloud ($\text{x}(t),\text{y}(t)$) to a reference uniform grid ($\text{X},\text{Y}$). The size of the grid can be chosen $N_{x} \times N_{y}$. We use Scipy's \emph{griddata} function \cite{SciPy} to interpolate the scattered CFD data and fit a surface. This function generates the interpolant using Clough-Tocher scheme \cite{alfeld1984trivariate} by triangulating the scattered  data $\textbf{s}(\text{x}(t),\text{y}(t),t)$ with Quickhull algorithm and forming a piecewise cubic Bezier interpolating polynomial on each triangle. 
The gradients of the interpolant are chosen such that the curvature of the interpolated surface is approximately minimized \cite{nielson1983method, renka1984triangle}. After achieving this minimization, the field values at the static Eulerian probes $N_{x} \times N_{y}$ are generated as $\textbf{S}(\text{X},\text{Y},t)$. 

\item Level-set: With the aid of learned interface description, a level $\Phi$ is assigned for all the cells on the uniform grid. The exact interface description is known for a stationary solid boundary, which is then used for synchronous structural motion prediction via the POD-RNN framework. The process is further demonstrated in section \ref{pod_rnn}. This step provides the identification of the solid, fluid and interface cells on the DL-ROM grid.

\item Pixelated force\footnote{This section has been presented in our previous work from Bukka \textit{et al.} in \cite{bukka2021assessment}. It is outlined here for the sake of completeness.}: The next step is the calculation of the forces exerted by the interface cells that contain the rigid body description. We refer to these forces as pixelated forces. The method of measuring the pixelated force incorporates the construction of the Cauchy stress tensor in the interface pixels.
As shown in Fig. \ref{figinterface-cells}, for an interface cell $\mathrm{k}$ at time instant $n$, the pixelated force $\overline{\textbf{f}}_{\mathrm{k}}^{n}$ for a Newtonian fluid can be written as
\begin{equation}
\begin{aligned}
     \overline{\textbf{f}}_{\mathrm{k}}^{n} =  (\stnf_{\mathrm{a};\mathrm{k}}^{n} - \stnf_{\mathrm{b};\mathrm{k}}^{n}).\mathrm{\mathbf{n}}_{\mathrm{x}} \mathrm{\Delta} \mathrm{y} + 
                  (\stnf_{\mathrm{c};\mathrm{k}}^{n} - \stnf_{\mathrm{d};\mathrm{k}}^{n}).\mathrm{\mathbf{n}}_{\mathrm{y}} \mathrm{\Delta} \mathrm{x},
\end{aligned}
\label{faceSigma}
\end{equation}
where $\stnf_{*;\mathrm{k}}^{n}$ is the Cauchy stress tensor for any point inside the cell $\mathrm{k}$. As depicted in Fig.~\ref{figinterface-cells}, we calculate this tensor at the mid-points of cell faces $\mathrm{a}-\mathrm{d}-\mathrm{b}-\mathrm{c}$ via finite difference approximation.  $\mathrm{\mathbf{n}}_{\mathrm{x}}$ and $\mathrm{\mathbf{n}}_{\mathrm{y}}$ are normals in the $\mathrm{x}$ and $\mathrm{y}$-direction, respectively, whereby $\mathrm{\Delta} \mathrm{x}$ and $\mathrm{\Delta} \mathrm{y}$ denote the cell sizes. For the present case, we only consider the pressure component while calculating $\overline{\textbf{f}}_{\mathrm{k}}^{n}$.

\begin{figure}
\centering
\includegraphics[width=0.8\textwidth]{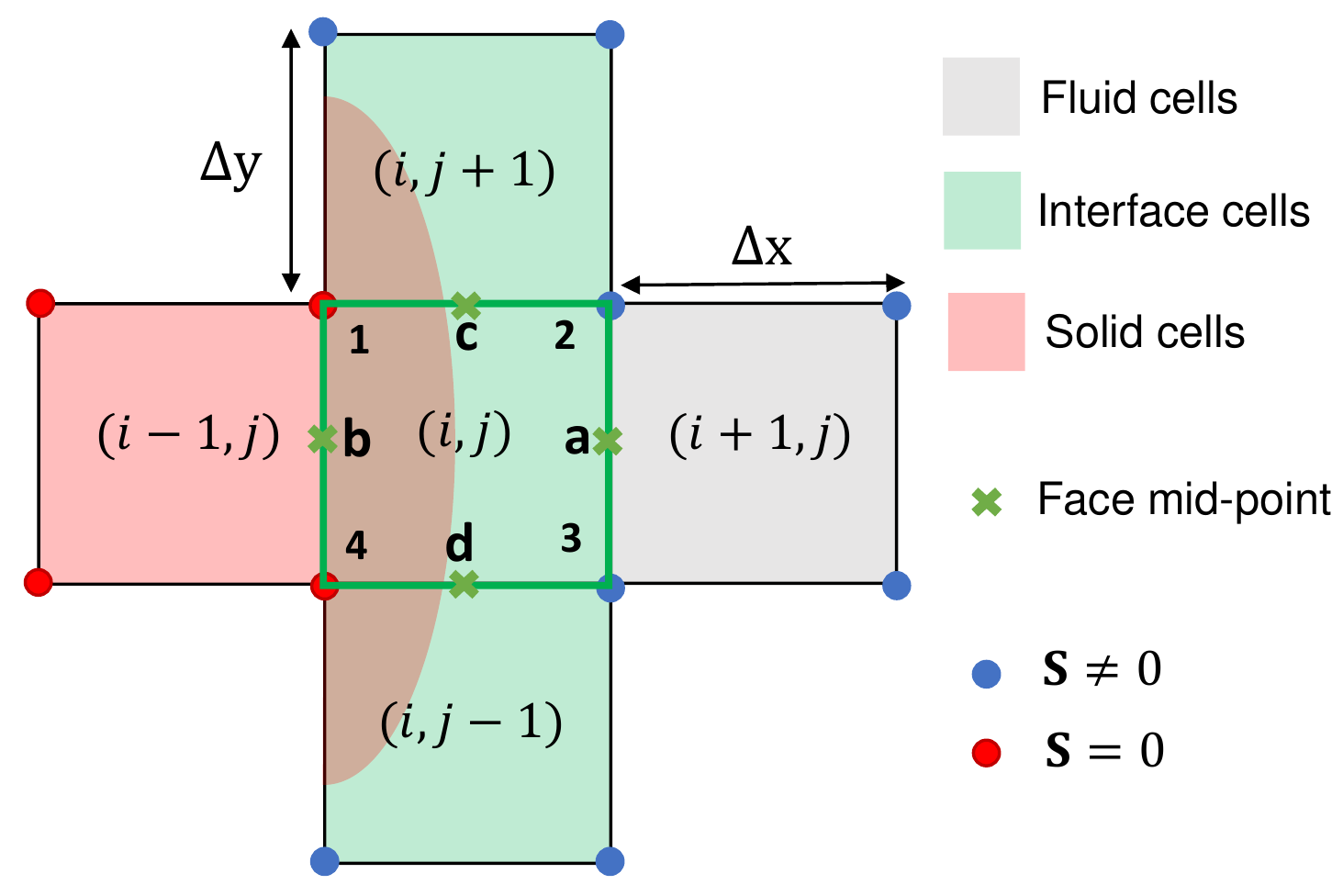}
\caption{Identification of interface cells in the DL-ROM grid and finite difference interpolation of field at the cell faces (green cross). The blue nodes contain the flow field values and the red dots are marked as zero.}
\label{figinterface-cells}
\end{figure}
To obtain the total pixelated force signal, these individual pixelated forces are summed over all the interface cells   as
\begin{equation}
\begin{aligned}
\overline{\textbf{F}}_{{b}}^{n} = \sum_{\mathrm{k}=1}^{N_{F}} \overline{\textbf{f}}_{\mathrm{k}}^{n}.
\end{aligned}
\label{totalDiscForce}
\end{equation}
where $N_{F}$ denotes the number of interface cells.
These total pixelated force signals $\overline{\textbf{F}}_{b} = \{ \overline{\textbf{F}}_{b}^{1} \; \overline{\textbf{F}}_{b}^{2} \; ... \; \overline{\textbf{F}}_{b}^{n} \}$, however, can be corrupted with missing or noisy information compared to its higher-order counterpart. This is accounted due to the loss of sub-grid information that leads to a coarser calculation of forces on the sharp interface. Thus, these output signals need to be corrected. 

\item Damp noises: To correct the pixelated signals, we first apply a moving average Gaussian filter $\mathbf{\zeta}$ to smoothen the signals. This is achieved by damping the high-frequency noises in these signals that are encountered due to a full-order interface movement on a low-resolution uniform grid. This damping smoothens the force propagation for easier reconstruction. To smooth a pixelated force signal $\overline{\textbf{F}}_{b}^{n}$ at a time step $n$, we perform the following operations:
  \begin{enumerate}
	
	\item collect the time series high-frequency noisy signals $\overline{\textbf{F}}_{b}$ of window length $2f+1$ 
	\begin{equation} \label{zeta_1}
	\overline{\textbf{F}}_{b} = \{  \overline{\textbf{F}}_{b}^{n-f}\; ...\; \overline{\textbf{F}}_{b}^{n-1}\; \overline{\textbf{F}}_{b}^{n} \;\overline{\textbf{F}}_{b}^{n+1} \;... \;\overline{\textbf{F}}_{b}^{n+f}  \}, 
	\end{equation}
	\item select Gaussian-like weights of the specified window length with mean weight $w^{n}$ for $\overline{\textbf{F}}_{b}^{n}$
	\begin{equation}
	\begin{split}
	w = \{ w^{n-f}\; ...\; w^{n-1}\; w^{n} \;w^{n+1} \; ... \; w^{n+f}  \},    
	\qquad \text{with} \sum_{i=n-f}^{n+f} \; w^i = 1 
	\end{split}
	\end{equation}
	\item apply the weighted average filter to damp the noise at $\overline{\textbf{F}}_{b}^{n}$
	\begin{equation} \label{zeta_3}
	\overline{\textbf{F}}_{b}^{n} := \; \sum_{i=n-f}^{n+f} \; w^i \overline{\textbf{F}}_{b}^{i} \; = \zeta \overline{\textbf{F}}_{b}^{n}.     
	\end{equation}
	
  \end{enumerate}
\item Interface force recovery: The smooth pixelated force propagation $\zeta \overline{\textbf{F}}_{b}$ can still lack the mean and derivative effects compared to its full-order counterpart $\overline{\textbf{F}}_{\ifsnt}$. This is where we introduce the functional mapping $\Psi$ as described in Algorithm \ref{lab1}. If this mapping can recover the bulk forces correctly, then we can utilize the selected grid. If not, then refine the grid and repeat the above steps from (1)-(5), until we find the best snapshot grid that recovers these bulk quantities with acceptable accuracy. 
\end{enumerate}

\begin{remark}
The mesh-to-mesh field transfer and load recovery for the snapshot DL-ROM grid search are expressed in a similar procedure as Algorithm \ref{lab2}. Once selected, the chosen uniform grid is utilized for learning the flow variables in the CRAN framework with the POD-RNN operating on the moving interface. Algorithm \ref{lab3} essentially employs the aforementioned interface load recovery to extract the higher-order forces from the synchronized field and interface predictions. 
\end{remark}


\newpage
{\small Algorithm \labeltext{1}{lab1}: Functional reconstruction mapping $\Psi$ for a higher-order force recovery on the DL-ROM grid.}
\noindent
\begin{mdframed}[
        linecolor=red,linewidth=1pt,%
        frametitlerule=true,%
        frametitlerulewidth=1pt, innertopmargin=\topskip,
        frametitle={Algorithm 1},
        outerlinewidth=0.75pt
    ]      
      \textbf{Input}: De-noised pixelated force signals $\overline{\textbf{F}}_{b}$, full-order FSI forces $\overline{\textbf{F}}_{\ifsnt}$  \\
  \textbf{Output}: Reconstructed pixelated force signal $\Psi \overline{\textbf{F}}_{b}$  \\ \\
  To reconstruct smooth pixelated force signals $\overline{\textbf{F}}_{b}$: 
  \begin{enumerate}
	\item Collect $n$ time steps of denoised pixelated force and full-order FSI forces: \\
	$  \overline{\textbf{F}}_{b} = \{ \overline{\textbf{F}}_{b}^{1} \; \overline{\textbf{F}}_{b}^{2} \; ...\; \overline{\textbf{F}}_{b}^{n-1} \;\overline{\textbf{F}}_{b}^{n} \}$; 
	\\
	$  \overline{\textbf{F}}_{\ifsnt} = \{ \overline{\textbf{F}}_{\ifsnt}^{1} \; \overline{\textbf{F}}_{\ifsnt}^{2} \; ... \; \overline{\textbf{F}}_{\ifsnt}^{n-1} \;\overline{\textbf{F}}_{\ifsnt}^{n} \}$;
	\item Get the mean and fluctuating force components:\\ 
	$  \overline{\textbf{F}}_{b}^{'} = \overline{\textbf{F}}_{b} - \mathrm{mean}(\overline{\textbf{F}}_{b}) $;
	\\ 
	$  \overline{\textbf{F}}_{\ifsnt}^{'} = \overline{\textbf{F}}_{\ifsnt} - \mathrm{mean}(\overline{\textbf{F}}_{\ifsnt}) $;
	\item Define the time-dependent derivative error using $\overline{E}_{c}$: \\
	 $\overline{E}_{c} = (\overline{\textbf{F}}_{\ifsnt}^{'} - \overline{\textbf{F}}_{b}^{'}) ./(\overline{\textbf{F}}_{b}^{'})$ with $\overline{\textbf{F}}_{b}^{'} \neq 0$; 
	 \item Reconstruct the smooth pixelated force signals with mean and derivative corrections: \\ 
		$  \overline{\textbf{F}}_{b} := \overline{\textbf{F}}_{b}^{'} + \mathrm{mean}(\overline{\textbf{F}}_{\ifsnt}) + \mathrm{mean}(\overline{E}_c) \overline{\textbf{F}}_{b}^{'} = \Psi \overline{\textbf{F}}_{b}$; 
	
  \end{enumerate}
\end{mdframed}

\noindent
{\small Algorithm \labeltext{2}{lab2}: Iterative snapshot DL-ROM grid search via interface force recovery.}
\noindent
\begin{mdframed}[
        linecolor=red,linewidth=1pt,%
        frametitlerule=true,%
        frametitlerulewidth=1pt, innertopmargin=\topskip,
        frametitle={Algorithm 2},
        outerlinewidth=0.75pt
    ]     
  
  \textbf{Input}: Full-order field data $ {\textbf{s}} = \{ {\textbf{s}}^{1} \; {\textbf{s}}^{2} \; ... \; {\textbf{s}}^{n} \}$ and FSI forces  $ {\overline{\textbf{F}}}_{\ifsnt} =  \{ {\overline{\textbf{F}}}_{\ifsnt}^{1} \; {\overline{\textbf{F}}}_{\ifsnt}^{2} \; ... \; {\overline{\textbf{F}}}_{\ifsnt}^{n} \}$ \\
  \textbf{Output}: Grid selection $N_{x} \times N_{y}$  \\ \\
  Initialise a uniform grid ($N_{x} \times N_{y}$) of cell size $\Delta \mathrm{x}$ = $\Delta \mathrm{y}$: \\
  \textbf{while}:
  \begin{enumerate}
	\item Project point cloud CFD data $\textbf{s}(\text{x}(t),\text{y}(t)) \in \mathbb{R}^{m \times n} $ on snapshot grid  ${\textbf{S}}(\text{X},\text{Y})$: \\
	${\textbf{S}} \leftarrow \textbf{s}$ \; \; s.t. ${\textbf{S}} \in \mathbb{R}^{N_{x} \times N_{y} \times n}  ;$
  	\item Apply learned level-set function ${\Phi} \in \mathbb{R}^{N_{x} \times N_{y} \times n}$ on ${\textbf{S}}(\text{X},\text{Y})$ element wise: \\
  	  ${\textbf{S}} := \Phi * {\textbf{S}};$
  	\item Calculate pixelated force signals $\overline{\textbf{F}}_{b}$ from interface cells $\ifsnt$ in ${\textbf{S}}$ using Eqs. (\ref{faceSigma}) - (\ref{totalDiscForce});
  	\item Damp the high-frequency noises in pixelated signals $\overline{\textbf{F}}_{b}$ via moving average Gaussian filter $\zeta$: \\
  	 $\overline{\textbf{F}}_{b} := \zeta \overline{\textbf{F}}_{b} \;$ using Eqs. (\ref{zeta_1})-(\ref{zeta_3}); 
  	\item Reconstruct the smooth pixelated force $\overline{\textbf{F}}_{b}$ to full-order by functional mapping $\Psi$: \\
  $\overline{\textbf{F}}_{b} :\approx \Psi \overline{\textbf{F}}_{b} \;$ using Algorithm \ref{lab1};
    \item \textbf{if} $\;\;$ ($  ( \| \overline{\textbf{F}}_{b} - \overline{\textbf{F}}_{\ifsnt} \|  / \| \overline{\textbf{F}}_{\ifsnt}  \| )  \leq \epsilon$ ): \\
     \textbf{break}; \\ 
     \textbf{else}: \\ 
     Refine the grid $\;\;$ $N_{x}:= 2 N_{x}, \; N_{y}:= 2 N_{y};$
  \end{enumerate}
\end{mdframed}
%
\newpage
\noindent
{\small Algorithm \labeltext{3}{lab3}: Extracting FSI forces via predicted fields on the DL-ROM grid and the ALE displacement.}
\noindent
\begin{mdframed}[
        linecolor=red,linewidth=1pt,%
        frametitlerule=true,%
        frametitlerulewidth=1pt, innertopmargin=\topskip,
        frametitle={Algorithm 3},
        outerlinewidth=0.75pt
    ]      
  
      \textbf{Input}: Field predictions from CRAN $ \hat{\textbf{S}} = \{ \hat{\textbf{S}}^{n+1} \; \hat{\textbf{S}}^{n+2} \; ... \; \hat{\textbf{S}}^{n+p} \}$, displacement predictions from POD-RNN $ \hat{\textbf{y}} = \{ \hat{\textbf{y}}^{n+1} \; \hat{\textbf{y}}^{n+2} \; ... \; \hat{\textbf{y}}^{n+p} \}$, calculated interface mapping $\Psi$ from training data \\
  \textbf{Output}: Full-order force predictions:  $ \hat{\overline{\textbf{F}}}_{\ifsnt} =  \{ \hat{\overline{\textbf{F}}}_{\ifsnt}^{n+1} \; \hat{\overline{\textbf{F}}}_{\ifsnt}^{n+2} \; ... \; \hat{\overline{\textbf{F}}}_{\ifsnt}^{n+p} \}$  \\ \\
  To extract full-order FSI forces: 
  \begin{enumerate}
	\item Apply predicted level-set function $\hat{\Phi} \in \mathbb{R}^{N_{x} \times N_{y} \times p}$ from POD-RNN on $\hat{\textbf{S}} \in \mathbb{R}^{N_{x} \times N_{y} \times p}$ element wise: \\
  	  $\hat{\textbf{S}} := \Phi * \hat{\textbf{S}};$
  	\item Calculate pixelated force signals $\hat{\overline{\textbf{F}}}_{b}$ from interface cells $\ifsnt$ in $\hat{\textbf{S}}$ using Eqs. (\ref{faceSigma}) - (\ref{totalDiscForce}); 
  	\item Damp high-frequency noises in extracted pixelated signals $\hat{\overline{\textbf{F}}}_{b}$ via moving average gaussian filter $\zeta$: \\
  	 $\hat{\overline{\textbf{F}}}_{b} := \zeta \hat{\overline{\textbf{F}}}_{b} \;$ using Eqs. (\ref{zeta_1})-(\ref{zeta_3});
  	\item Reconstruct the extracted smooth pixelated force $\hat{\overline{\textbf{F}}}_{b}$ to full-order by known functional mapping $\Psi$: \\
  	 $\hat{\overline{\textbf{F}}}_{\ifsnt} \approx \Psi \hat{\overline{\textbf{F}}}_{b} \;$ 
  \end{enumerate}
\end{mdframed}

\section{Application: Vortex-induced vibration (VIV)} \label{VIV_application}
This section demonstrates our hybrid partitioned DL-ROM methodology for a benchmark fluid-structure interaction problem.
While the proposed technique can be generalized to any fluid-structure
system involving the interaction dynamics of flexible structures with an unsteady wake-vortex system, we consider an unsteady elastically-mounted cylinder problem in an external flow. The wake flow behind the bluff body generates switching vortex loads on the solid, which causes its vibration. This structural vibration, in turn, affects the near-wake flow dynamics resulting in a two-way synchronized oscillating system. For this phenomenon of vortex-induced vibration, the flow physics is highly sensitive and directly related to the interaction dynamics between the downstream vortex patterns and the cylinder's motion. Of particular interest is to learn and predict such synchronized wake-body interaction using data-driven computing. We consider the parameter sets corresponding to a limit cycle response as well as non-stationary vibration response.

\begin{figure} 
\centering
\subfloat[]{\includegraphics[width =0.7\textwidth]{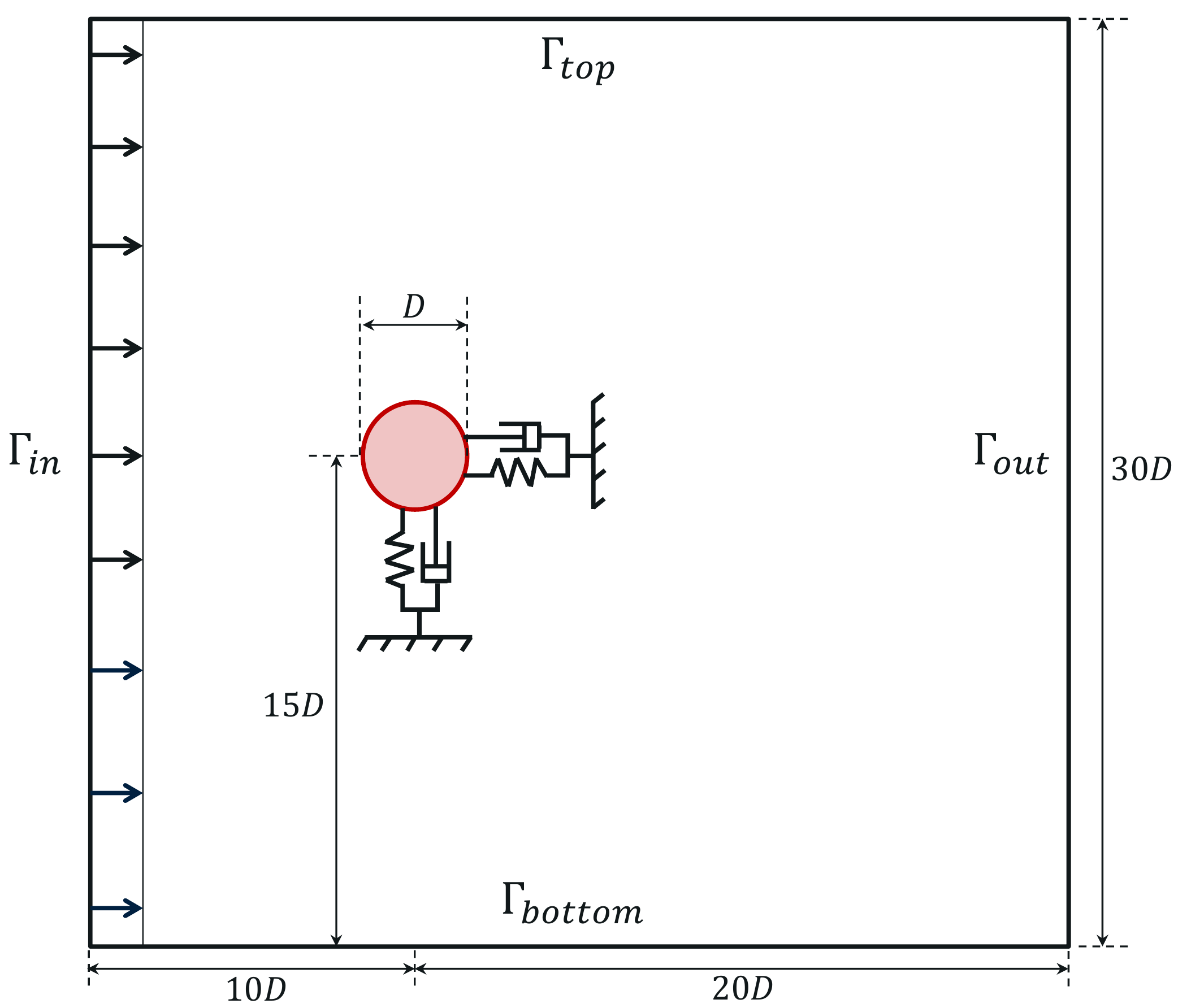}} \\
\subfloat[]{\includegraphics[width =0.6\textwidth]{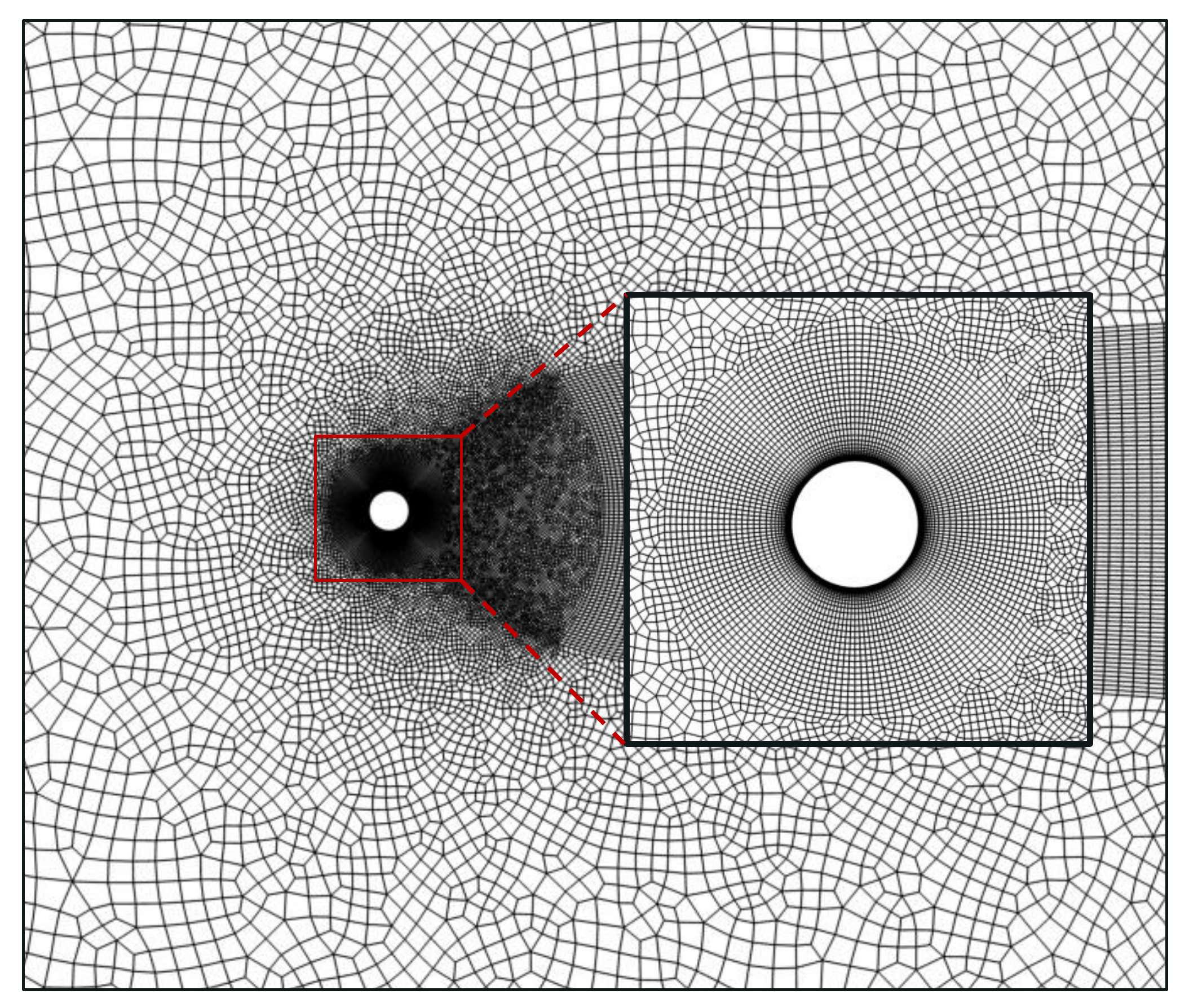}}
\caption{Freely vibrating circular cylinder in a uniform flow: (a) Schematic of an elastically-mounted cylinder undergoing VIV, (b) representative full-order domain and the near cylinder unstructured mesh.}
\label{setup_single}
\end{figure}

The freely vibrating structural system is installed in a 2D computational domain. The center of the cylinder is at sufficient distances from the boundaries to capture the downstream wake dynamics (see Fig.~\ref{setup_single} (a)).
A no-slip and traction continuity condition is enforced on the cylinder surface (Eqs. (\ref{eq4})-(\ref{eq5})), while a uniform velocity profile $U_{\infty}$ is maintained at the inlet $\Gamma_{in}$. The traction-free boundary is implemented on the outlet $\Gamma_{out}$ and slip condition is specified at the top $\Gamma_{top}$ and bottom $\Gamma_{bottom}$. We discretize the fluid domain via unstructured finite element mesh. The final mesh, which is obtained after following the standard rules of mesh convergence, is presented in Fig.~\ref{setup_single} (b). It contains a total of $25,916$ quadrilateral elements with $26,114$ nodes. The full-order simulation is carried out via finite element incompressible Navier-Stokes solver in the ALE framework. The output details are tabulated in Table~\ref{tab:validation} for validation. By constructing the Cauchy stress tensor on the fluid-solid surface, the fluid force along the interface is computed. The drag and lift force coefficients are calculated as
\begin{equation}
\begin{aligned}
    \mathrm{C}_{\mathrm{D}} = \frac{1}{\frac{1}{2}\rho^{f}U_{\infty}^{2}D} {\int_{\ifsnt} (\stf.\mathrm{\mathbf{n}}).\mathrm{\mathbf{n}}_{\mathrm{x}} \mathrm{d}\Gamma },\\
    \mathrm{C}_{\mathrm{L}} = \frac{1}{\frac{1}{2}\rho^{f}U_{\infty}^{2}D} {\int_{\ifsnt} (\bm{\stf}.\mathrm{\mathbf{n}}).\mathrm{\mathbf{n}}_{\mathrm{y}} \mathrm{d}\Gamma },
\end{aligned}
\label{force_eqns}
\end{equation}
where $\mathrm{C}_{\mathrm{D}}$ and $\mathrm{C}_{\mathrm{L}}$ denote the drag and lift force coefficients, respectively, and
$\Gamma^\mathrm{fs}$ is the fluid-solid interface. $\mathbf{n}_{\mathrm{x}}$ and $\mathbf{n}_{\mathrm{y}}$ are the Cartesian components of the unit normal vector $\mathbf{n}$.

\begin{table}[]
\small
\centering
\caption{An elastically-mounted circular cylinder undergoing vortex-induced vibration: Comparison of the full-order forces and the displacements in the present study and the benchmark data. $\mathrm{C}_{\mathrm{D}}$ and $\mathrm{C}_{\mathrm{L}}$ represent the drag and lift force coefficients, respectively. $\mbox{A}_{\mbox{x}}$ and $\mbox{A}_{\mbox{y}}$ represent the x and y displacements of the cylinder, respectively. }
\begin{tabular}{P{1.3cm}P{1.1cm}P{1.1cm}P{1.5cm}P{1.5cm}P{1.5cm}P{1.5cm}}
\toprule
\toprule
Mesh & $N_{cyl}^{*}$ & $N_{fluid}^{**}$ & $\mathrm{mean}(\mathrm{C}_{\mathrm{D}})$ & $(\mathrm{C}_{\mathrm{L}})_{\mathrm{rms}}$ & $(\mbox{A}_{\mbox{x}})_{\mathrm{rms}} $ & $\left(\mbox{A}_{\mbox{y}}\right)_{\mathrm{m a x}}$ \\\\
\bottomrule
\bottomrule
Present & 124 & 25,916 & 2.0292 &  0.1124 & 0.0055 &  0.5452\\
\hline
Ref. \cite{li2016vortex} & 168 & 25,988 & 2.0645 & 0.0901 & 0.0088 & 0.5548 \\
\hline       
\end{tabular} \\ 
$^{*}$ $N_{cyl}$ are the number of cylinder surface elements\\
$^{**}$ $N_{fluid}$ are the number of Eulerian fluid elements\\
\label{tab:validation}
\end{table}

\subsection{Limit cycle VIV response} \label{lco}
We start by applying our proposed methodology for the VIV at frequency lock-in. A rigid cylinder is mounted in the transverse and streamwise directions with linear and homogeneous springs and no damping. This results in alike natural frequencies $f_{nx} = f_{ny}$ in both Cartesian directions. Computation is carried at a fixed cylinder mass ratio $m^{*}= \frac{4 m}{\pi \rho^{f}  D^{2}}=10$, reduced velocity $U_{r}= \frac{U_{\infty}}{f_{n} D}=\frac{U_{\infty}}{\frac{1}{2 \pi} \sqrt{\frac{k}{m}} D}=5$,  and Reynolds number $Re=\frac{\rho^{\mathrm{f}} U_{\infty} D }{\mu^{\mathrm{f}}}=200$. $m$ is the solid mass, $f_{n}$ is the natural frequency of the cylinder and $U_{\infty}$ is the uniform inlet $\mathrm{x}$-velocity. The remaining other constants imply their usual meaning. At these non-dimensional parameters, a high and sensitive limit cycle oscillations are obtained \cite{li2016vortex}.

The full-order simulation is carried out for $tU_{\infty}/ D = 250$ with a time step of $0.025\;tU_{\infty}/ D$. From these full-order data, $n_{tr}=5000$ snapshots (100-225 $tU_{\infty}/D$) are used for training and $n_{ts}=1000$ (225-250 $tU_{\infty}/D$) are kept for testing. Following the data generation, we apply our multi-level framework and the snapshot-FTLR method for predicting the flow fields, the ALE displacements, and the pressure force coefficients.

\subsubsection{Structural displacements via POD-RNN driver}\label{podrnn_stat}

Time instances of the ALE $\mathrm{x}$-displacements or $\mathrm{y}$-displacements $\textbf{Y} = \left\lbrace\textbf{y}^{1}\;\textbf{y}^{2}\;\dots\;\textbf{y}^{N}\right\rbrace \in \mathbb{R}^{m\times N}$  are decomposed using the POD projection. Here, $m=26,114$ and $N=6000$. The fluctuation matrix  $\textbf{Y}^{'} \in \mathbb{R}^{m\times N}$ and the offline database of the mean field $\bar{\textbf{y}} \in \mathbb{R}^{m}$ and the POD modes $\boldsymbol{\nu} \in \mathbb{R}^{m\times N}$ are obtained.
Eigenvalues $\boldsymbol{\Lambda}^{N\times N}$ of the covariance matrix $\textbf{Y}^{'\mathrm{T}}\textbf{Y}^{'}  \in \mathbb{R}^{N\times N}$ are extracted as their magnitude measures the energy of the respective POD modes. Cumulative and percentage of the modal energies are plotted in Figs.~\ref{ce_te_stat_cyl_1} and \ref{ce_te_stat_cyl_2} with respect to the number of modes.   
\begin{figure}
\centering
\subfloat[]{\includegraphics[width = 0.47\textwidth]{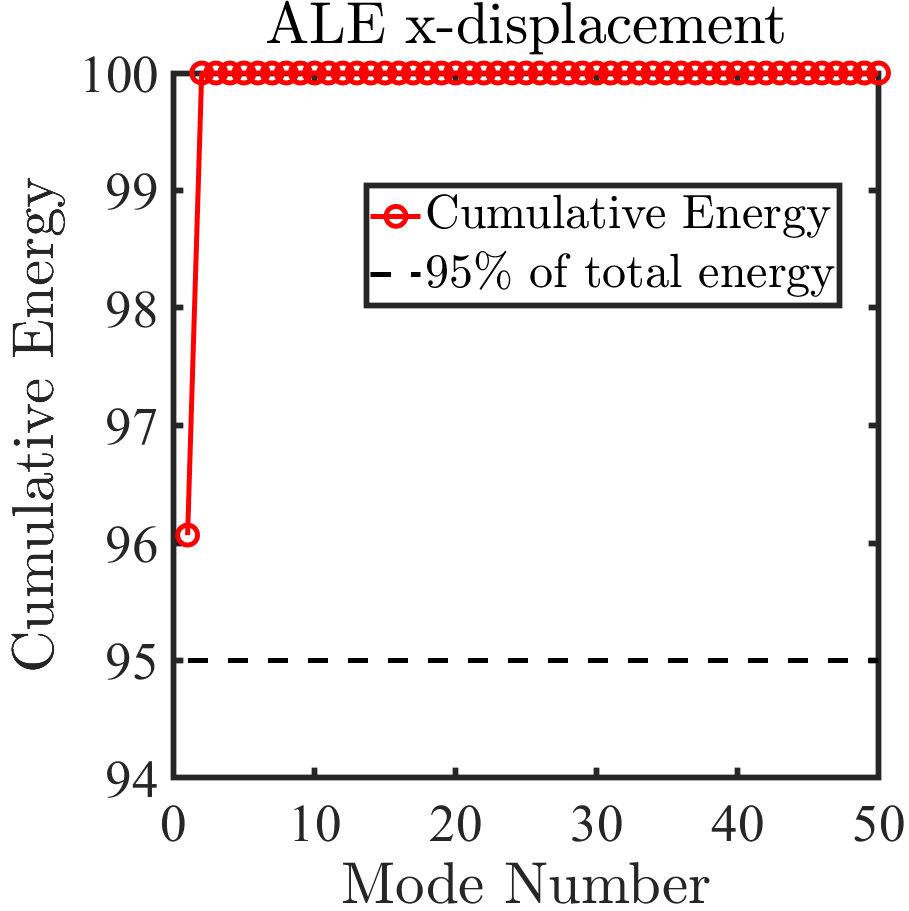}}
\hspace{0.02\textwidth}
\subfloat[]{\includegraphics[width = 0.485\textwidth]{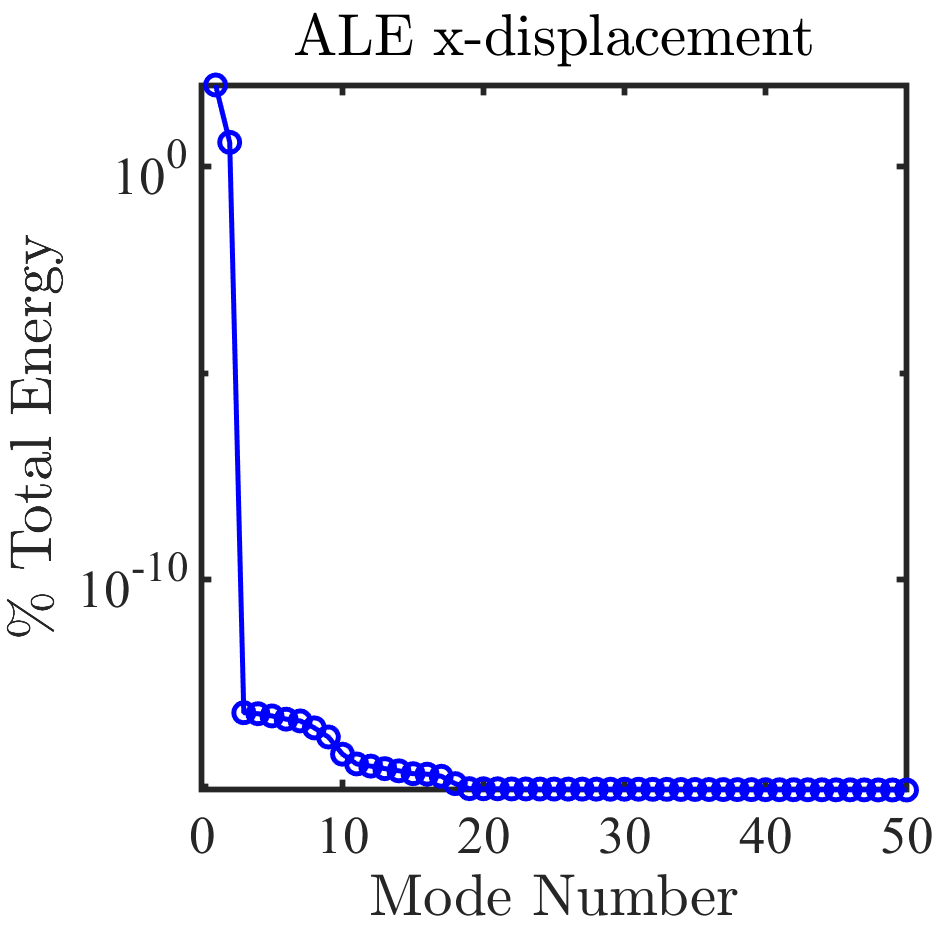}} 
\caption{Freely vibrating circular cylinder in a uniform flow ($m^{*}=10,U_r=5, Re=200$): (a) Cumulative and (b) percentage of the modal energies for the ALE x-displacements $\textbf{Y}_{\textbf{x}}$.}
\label{ce_te_stat_cyl_1}
\end{figure}
\begin{figure}
\centering
\subfloat[]{\includegraphics[width = 0.47\textwidth]{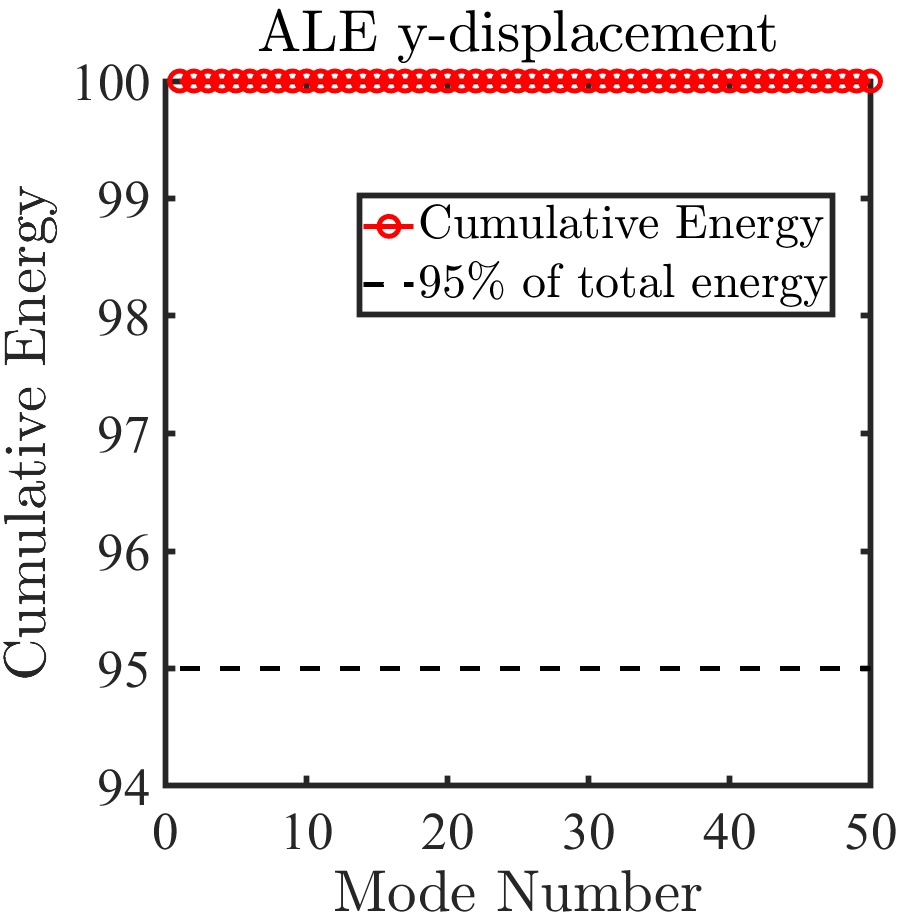}}
\hspace{0.02\textwidth}
\subfloat[]{\includegraphics[width = 0.49\textwidth]{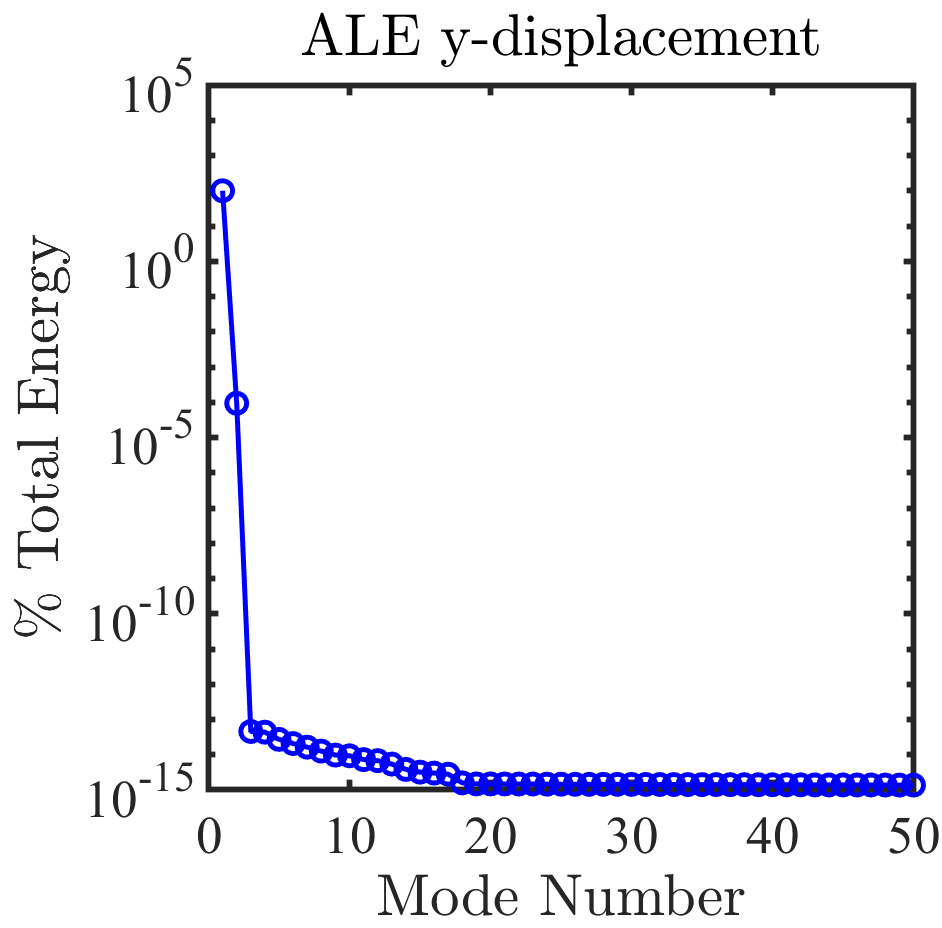}} 
\caption{Freely vibrating circular cylinder in a uniform flow ($m^{*}=10,U_r=5, Re=200$): (a) Cumulative and (b) percentage of the modal energies for the ALE y-displacements $\textbf{Y}_{\textbf{y}}$.}
\label{ce_te_stat_cyl_2}
\end{figure}
It can be observed that the system energy of $>99.99\%$ is concentrated in the first $1-2$ POD modes. This observation implies that the ALE field can potentially be reduced with just $k=2$ POD modes instead of $m=26,114$ fluids nodes. One reason for this observation is that the majority of the variance about $\bar{\textbf{y}}$ is concentrated at the moving interface and propagates linearly to the far-field boundaries. 
Interestingly, with these two modes, the full-order mesh can be reconstructed within a maximum absolute error $ < 1.0 \times 10^{-15}$.
Fig.~\ref{podrnnvivth} shows the dynamical evolution of the dominant POD modes $\boldsymbol{\nu} \in \mathbb{R}^{m\times k}$ using $\mathbf{A}_{\nu} = \boldsymbol{\nu}^{T} \textbf{Y}^{'} \in \mathbb{R}^{k \times N}$. The time coefficients are nearly periodic with a limit cycle oscillation as inherent in the high-dimensional information.
\begin{figure}
\centering
\subfloat[]{\includegraphics[width =0.9\textwidth]{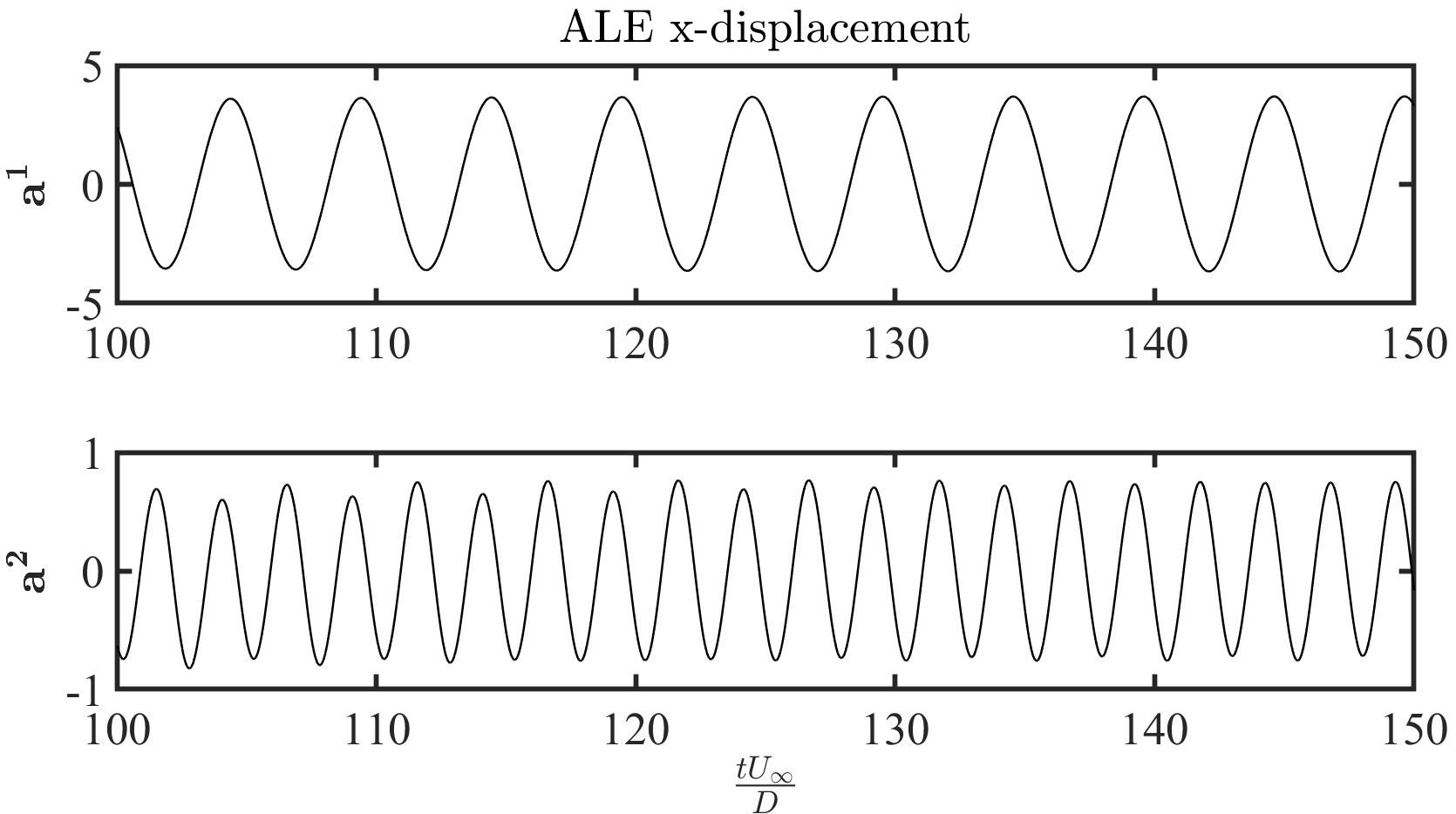}}  \\
\subfloat[]{\includegraphics[width =0.9\textwidth]{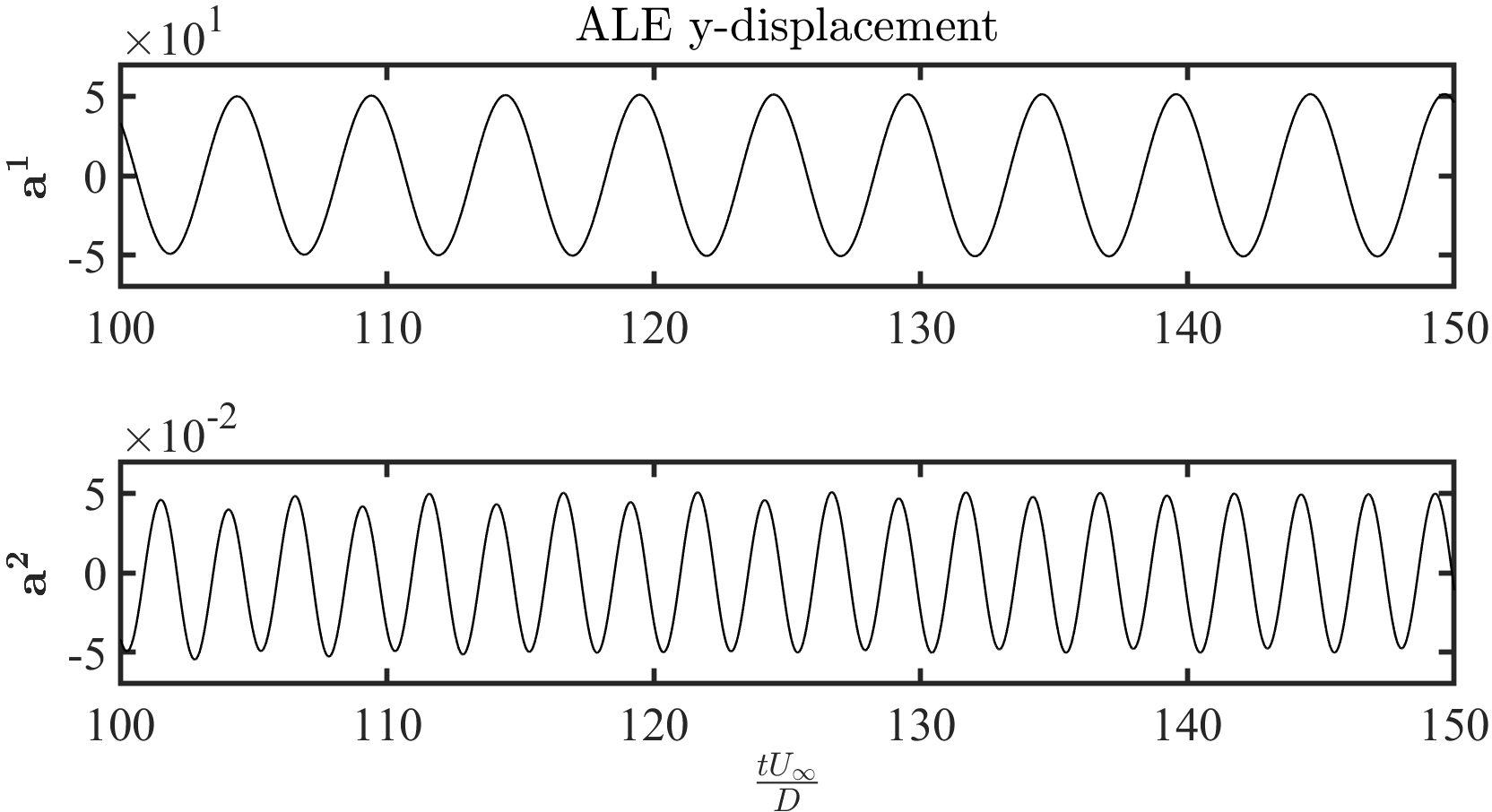}}
\caption{Freely vibrating circular cylinder in a uniform flow ($m^{*}=10,U_r=5, Re=200$): Time history of the $k=2$ modal coefficients shown from $100$ till $150\;tU_{\infty}/D$ for the (a) ALE x-displacements $\textbf{Y}_{\textbf{x}}$ and the (b) ALE y-displacements $\textbf{Y}_{\textbf{y}}$.}
\label{podrnnvivth}
\end{figure}

The $N$ temporal coefficients are divided into the training and the testing part. We train a closed-loop LSTM  in a time-delayed fashion for learning these time coefficients. This training implies that the predicted output is used as an input for the next time step. The input to the closed-loop recurrent neural network is a vector of dimension $k$ so that all $k$ modes are learned and predicted in one time instant. The nearly periodic trend in the modes allows an easier hyperparameter construction of the LSTM-RNN, albeit different phases and amplitudes of the modes exist. Further details about the network can be 
found in Table~\ref{tab:lstm_network}. The training took around 5 minutes on a single graphics processing unit, and the prediction of the time coefficients $\hat{\textbf{A}}_{\nu}$ is depicted in Fig.~\ref{podrnn_th_predict_stat} for testing. We keep the multi-step prediction cycle to $p=100$ while testing the output. This implies that one test time step is used to predict the next 100 time coefficients until a ground input is fed. It is worth mentioning that an encoder-decoder LSTM architecture would have been preferred to extract the temporal coefficients in the case of chaotic dynamics. In the present case of periodic oscillations, a simple closed-loop LSTM-RNN is sufficient for the point cloud tracking.     

\begin{table}
\small
\centering
\caption{Freely vibrating circular cylinder in a uniform flow ($m^{*}=10,U_r=5, Re=200$): Network and hyperparameters details of the closed-loop RNN.}
\begin{tabular}{P{1.1cm}P{0.8cm}P{1.1cm}P{1cm}P{1.3cm}P{1.1cm}P{1.1cm}P{1.4cm}P{1.5cm}}
\toprule
\toprule
ALE  & $k$ & Cell & $N_{h}^{*}$ & Optimizer & $\alpha^{*}$ & Epochs & Learning rate decay & Decay step \\
\bottomrule
\bottomrule
x or y & 2 & LSTM & 256 & Adam$^{*}$ & 0.001 &  1600 & 0.2 & 125 epochs\\
\hline
\end{tabular}\\
$N_h\;:$ LSTM hidden cell dimension\\
$\alpha\;:$ Initial learning rate\\
Adam$^{*}$: Adaptive moment optimization \cite{kingma2014adam} 
\label{tab:lstm_network}
\end{table}

\begin{figure}
\centering
\subfloat[]{\includegraphics[width = 0.90\textwidth]{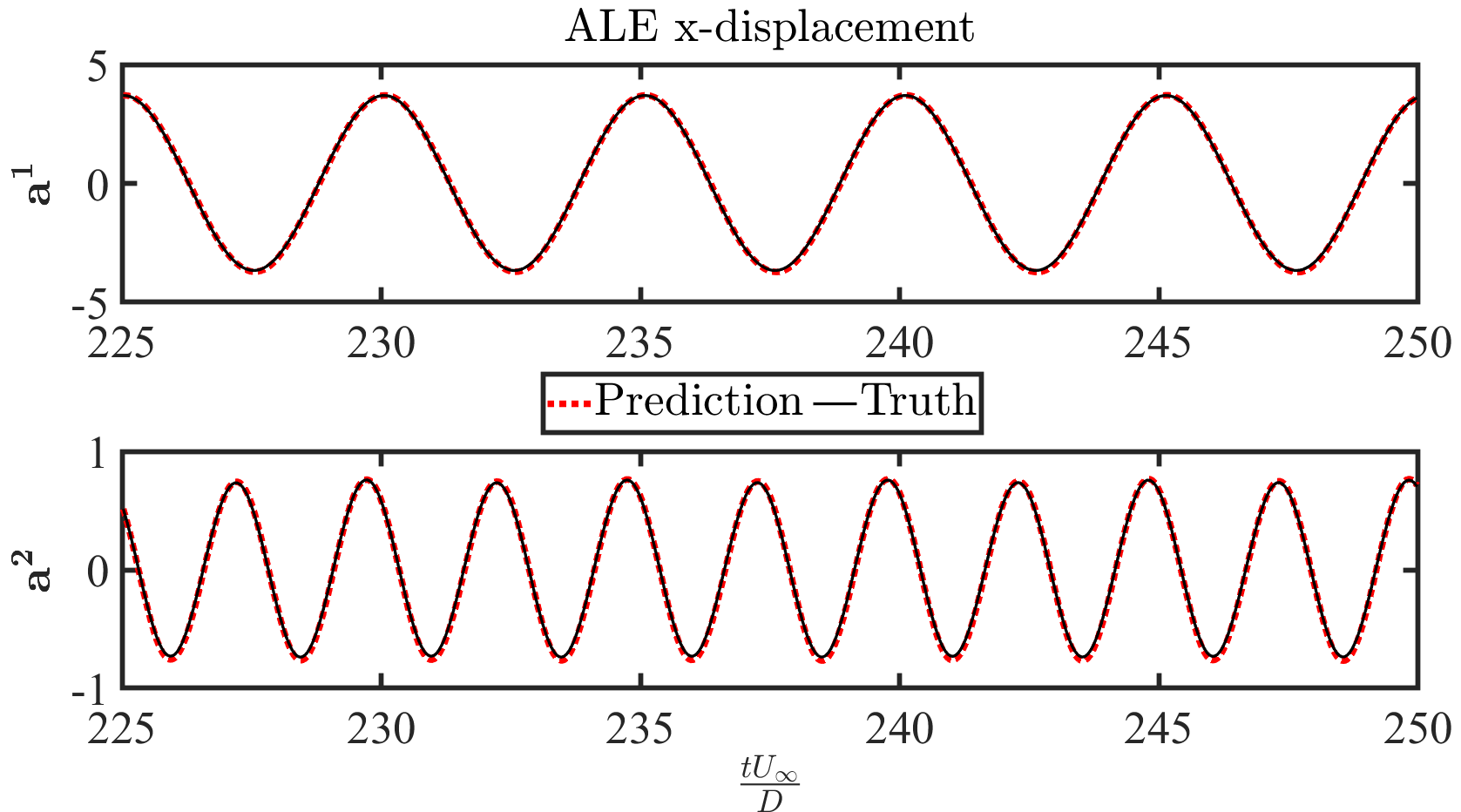}}\\
\subfloat[]{\includegraphics[width = 0.90\textwidth]{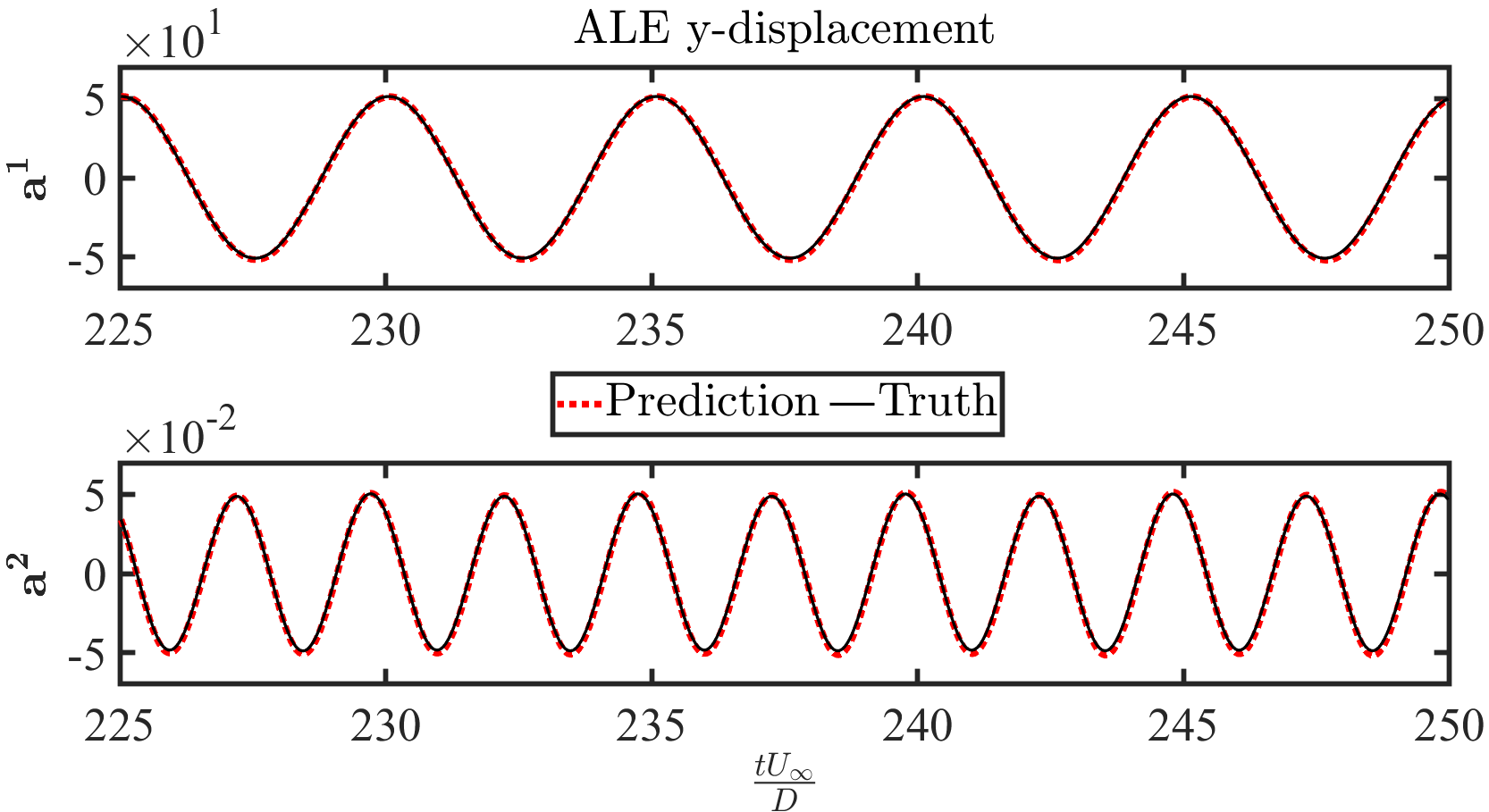}}
\caption{Freely vibrating circular cylinder in a uniform flow ($m^{*}=10,U_r=5, Re=200$): Prediction of the modal coefficients on the test data for the (a) ALE x-displacements $\textbf{Y}_{\textbf{x}}$ and the (b) ALE y-displacements $\textbf{Y}_{\textbf{y}}$.}
\label{podrnn_th_predict_stat}
\end{figure}

The predicted modal coefficients at any time step $i$, $\hat{\textbf{A}}_{\nu}^{i} \in \mathbb{R}^{k}$, can simply be reconstructed back to the point cloud $\hat{\textbf{y}}^{i} \in \mathbb{R}^{m}$ using the mean field $\bar{\textbf{y}} \in \mathbb{R}^{m}$ and the $k$ spatial POD modes $\boldsymbol{\nu} \in \mathbb{R}^{m\times k}$. Fig. \ref{disp_pred} compares the predicted and true values of the VIV motion in the $\mathrm{x}$ and $\mathrm{y}$-directions of the cylinder. The results indicate that the POD-RNN can accurately predict the position of the moving FSI interface for a limit cycle oscillation. These results are inferred in a closed-loop fashion by predicting $p=100$ steps from one demonstrator at a time. The incorporation of the ground data demonstrators in the predictions can circumvent the compounding effect of errors and thereby boost the long term predictive abilities of the neural networks \cite{venkatraman2015improving}. This description completes the moving point cloud prediction for the chosen parameter set.

\begin{figure}
\centering
\subfloat[]{\includegraphics[width = 0.49\textwidth]{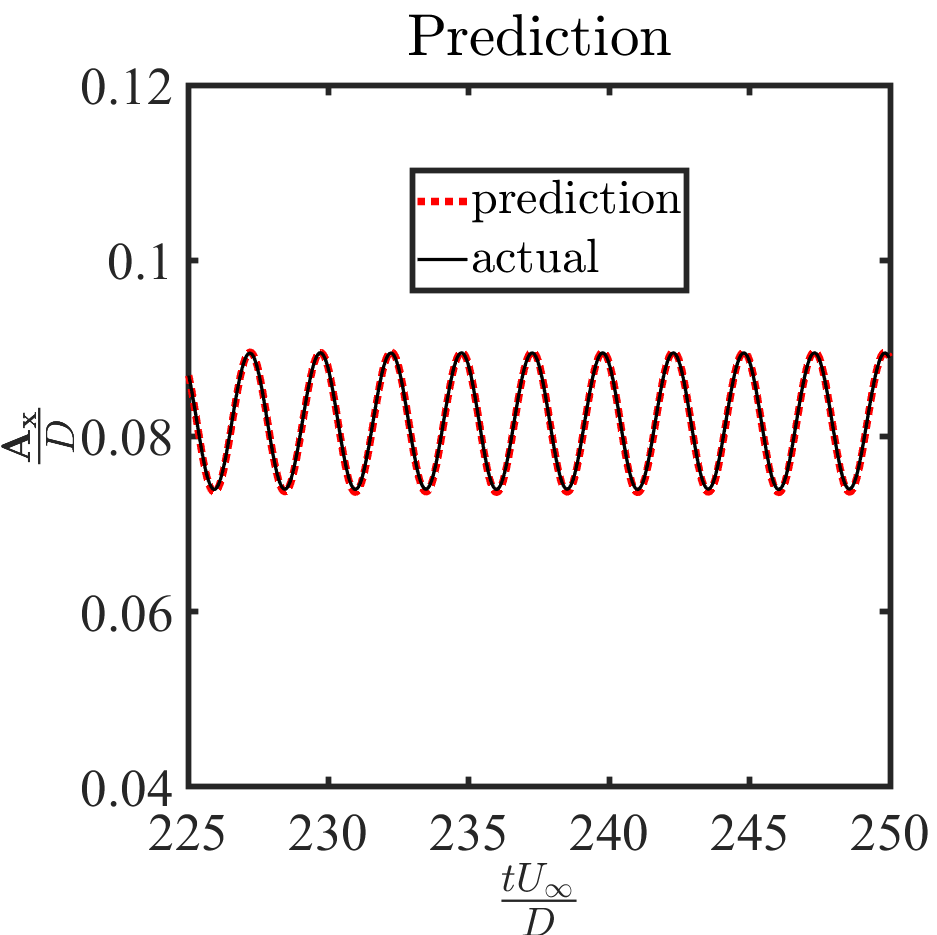}} 
\subfloat[]{\includegraphics[width = 0.47\textwidth]{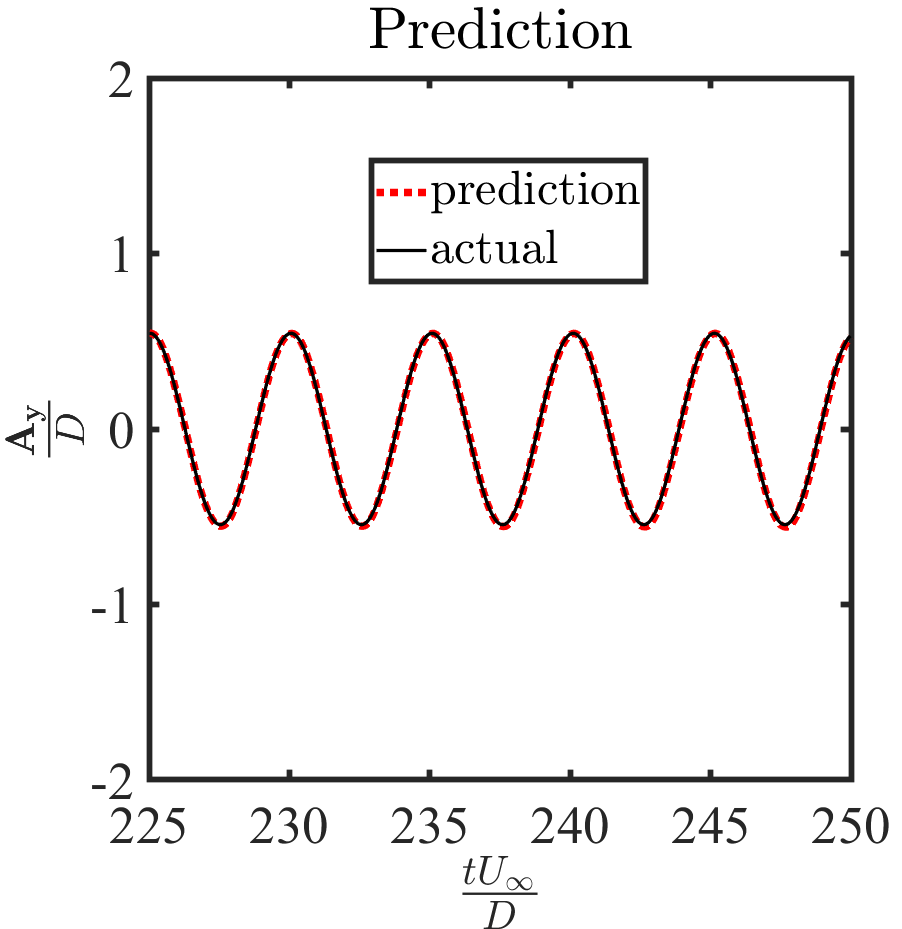}}
\caption{Freely vibrating circular cylinder in a uniform flow ($m^{*}=10,U_r=5, Re=200$): Predicted and actual (a) x-position and (b) y-position of the interface normalised by the diameter of cylinder $D$. The root mean squared error between the true and predicted is $2.17\times 10^{-4}$ and $1.08\times 10^{-2}$ for $\mbox{A}_{\mbox{x}} / D$ and $\mbox{A}_{\mbox{y}} /D$, respectively.}
\label{disp_pred}
\end{figure}

\subsubsection{Snapshot-FTLR }\label{int_load_results}
As mentioned earlier, the POD-RNN driver tracks the moving interface and extracts the $\Phi$ level-set feature. However, the best snapshot DL-ROM grid ($N_{x} \times N_{y}$) must be selected for the field predictions in the CRAN framework. The snapshot-FTLR estimates the point cloud field data on a uniform grid where a higher-order load recovery is possible at the practical level. 
We project the point cloud flow field data, for instance the pressure fields, ${\textbf{s}} = \{ {\textbf{s}}^{1} \; {\textbf{s}}^{2} \;  ... \;  {\textbf{s}}^{N} \} \in \mathbb{R}^{m \times N}$ as spatially uniform snapshots ${\textbf{S}} = \{ {\textbf{S}}^{1} \;  {\textbf{S}}^{2} \;  ... \;  {\textbf{S}}^{N} \} \in \mathbb{R}^{N_{x} \times N_{y} \times N}$. This uniformity is achieved via the SciPy's $griddata$ function \cite{SciPy} by mapping the $m$ dimensional unstructured data on a 2D uniform grid. Here, $N_{x}=N_{y}=32,64,128,256,512,1024$. Note that the DL-ROM grid for this case is of the size $8D \times 8D$ with $ \approx 5D$ length for the downstream cylinder wake. 
To assess the sensitivity of grid distribution, we compare the accuracy of surface fit provided by the grid data using nearest, linear and cubic interpolations. This is achieved by sampling the field's maximum and the minimum values with pixel numbers on the fitted surface process. As seen from Fig.~\ref{pressure_convergence}, for any pressure time step $t U_{\infty} / D = 125$, the nearest levels-off to the true $p_{max}$ and  $p_{min}$ during grid refinement. This reason is that this method assigns the value of the nearest neighbor in the distributed triangulation-based information. 

\begin{remark}
The linear and cubic interpolation techniques involve triangulations  with $C^0$ and $C^1$ continuity. The linear method linearly converges to the true values on grid refinement. {\color{black} The cubic interpolation approach, however, converges well on a coarser grid $N_{x}=N_{y}=64,128$, but deviates on subsequent grid refinement $N_{x}=N_{y}=256,512,1024$ with small relative errors.} 
\end{remark}

\begin{figure}
\centering
\subfloat[]{\includegraphics[width = 0.482\textwidth]{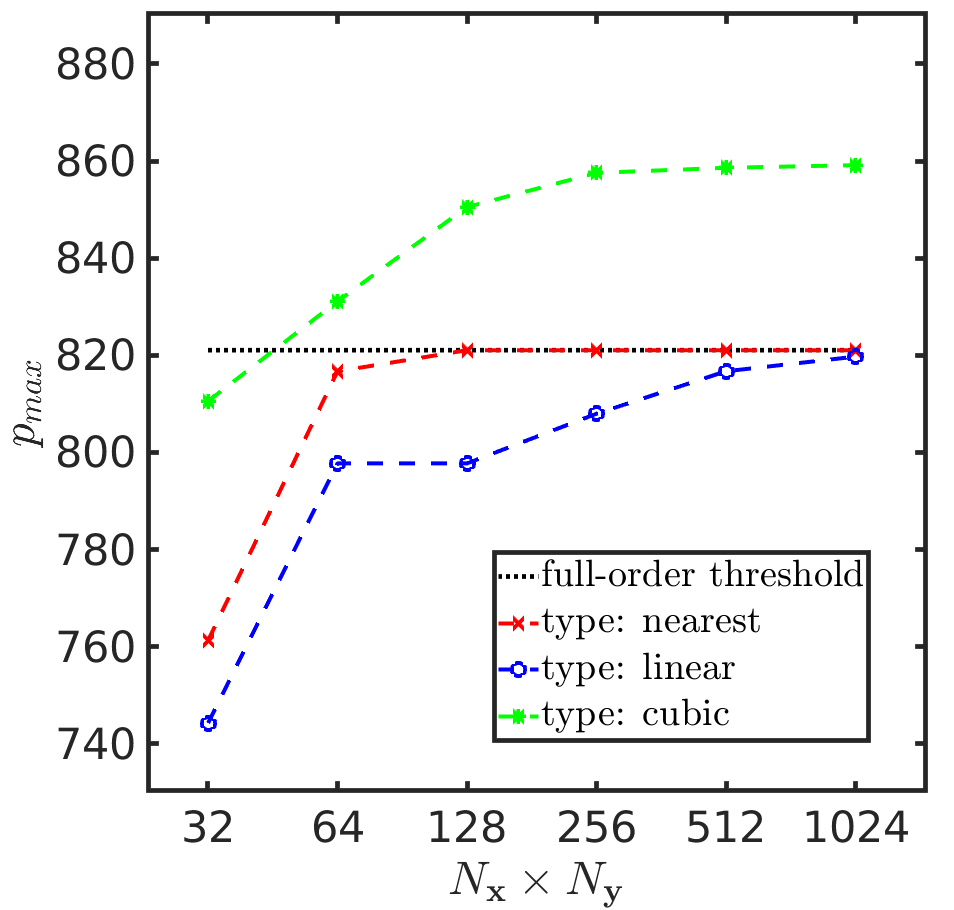}}
\hspace{0.01\textwidth}
\subfloat[]{\includegraphics[width = 0.48\textwidth]{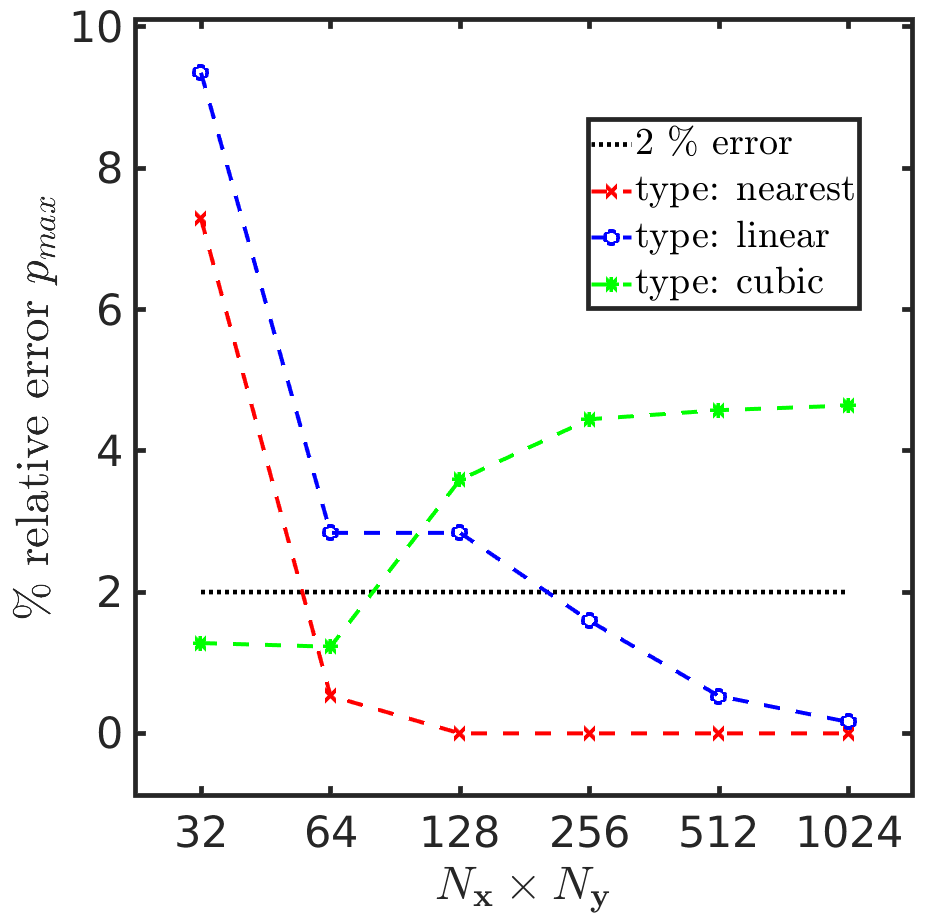}} \\
\subfloat[]{\includegraphics[width = 0.498\textwidth]{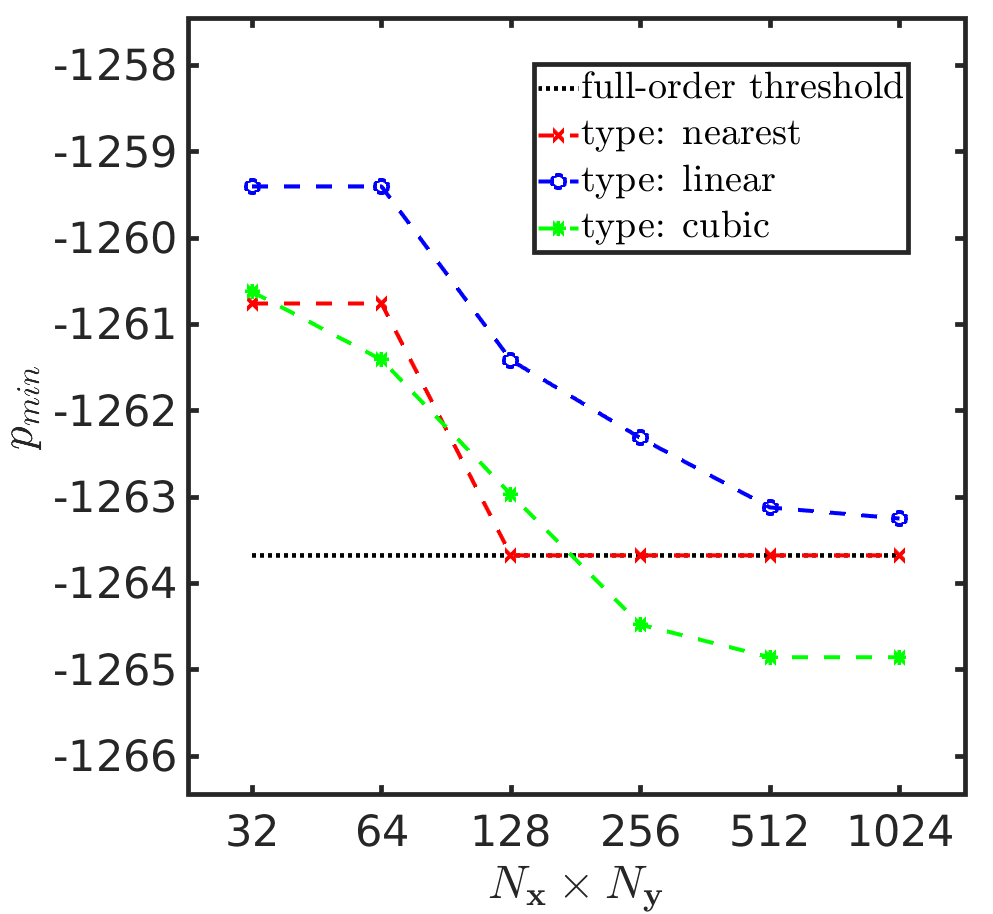}}
\hspace{0.01\textwidth}
\subfloat[]{\includegraphics[width = 0.48\textwidth]{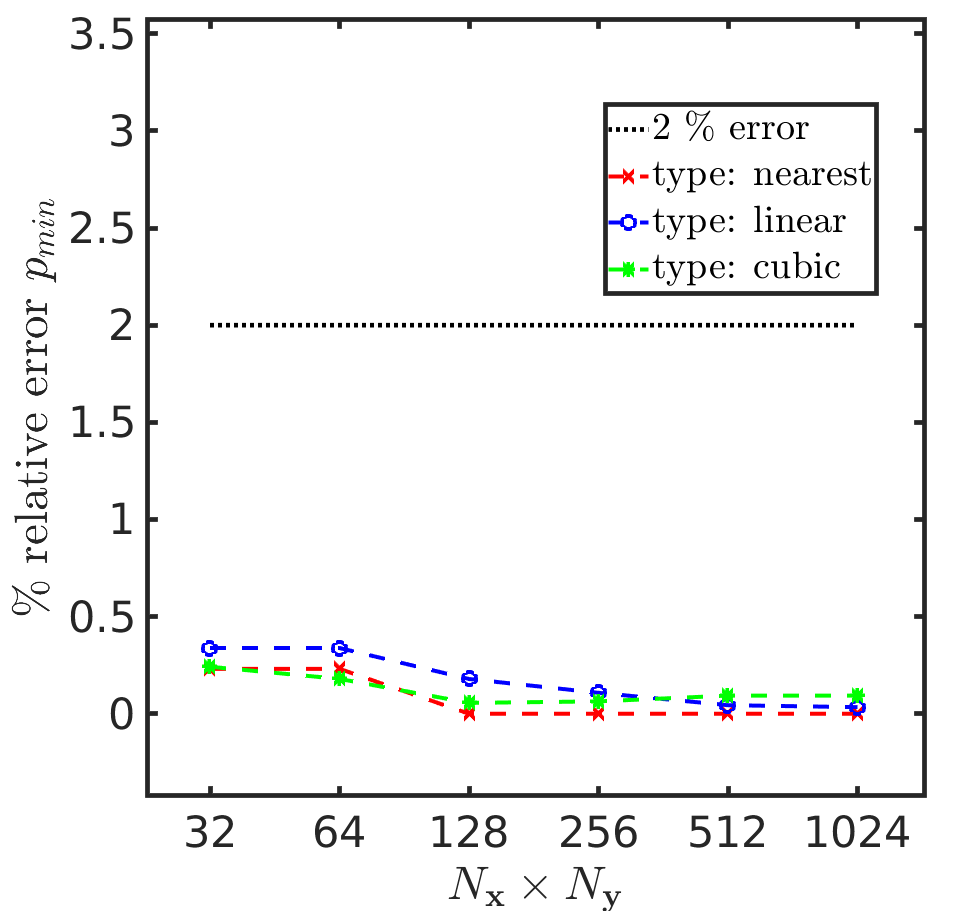}} \\
\caption{Freely vibrating circular cylinder in a uniform flow ($m^{*}=10,U_r=5, Re=200$): Convergence of the pressure field with the number of pixels for different interpolation schemes at the time step $t U_{\infty} / D = 125$. (a)-(b) Maximum pressure values and the respective errors, (c)-(d) minimum pressure values and the respective errors.}
\label{pressure_convergence}
\end{figure}
Fig.~\ref{pressure_behavior} compares the interpolation methods for the pressure field with respect to the full-order for $t U_{\infty} / D = 125$ on the $512 \times 512$ DL pixelated grid. {\color{black} It can be observed that the nearest contains oscillations compared to the full-order description due to a discontinuous assignment of the field at the selected probes. A closer look in Fig.~\ref{pressure_behavior} (d) depicts some local fitting deviations compared to the full-order for the cubic due to a higher-order interpolation. Conversely, the linear and the nearest look much closer to the full-order results.} Hence, we rely on the linear technique for a coarse-grain field transfer in the remainder of our analysis. 
\begin{figure}
\centering
\subfloat[]{\includegraphics[width = 0.49\textwidth]{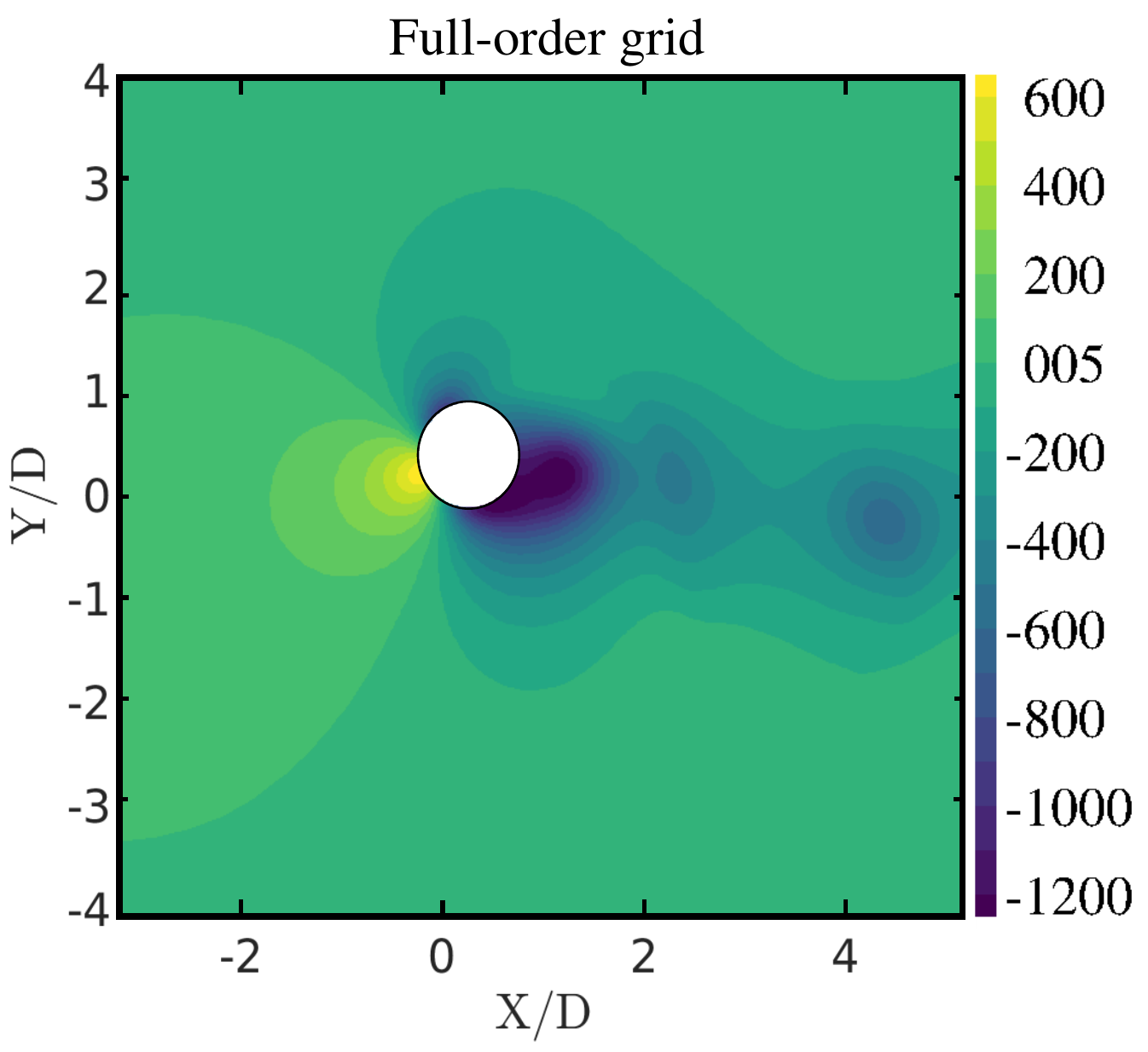}}
\hspace{0.01\textwidth} 
\subfloat[]{\includegraphics[width = 0.49\textwidth]{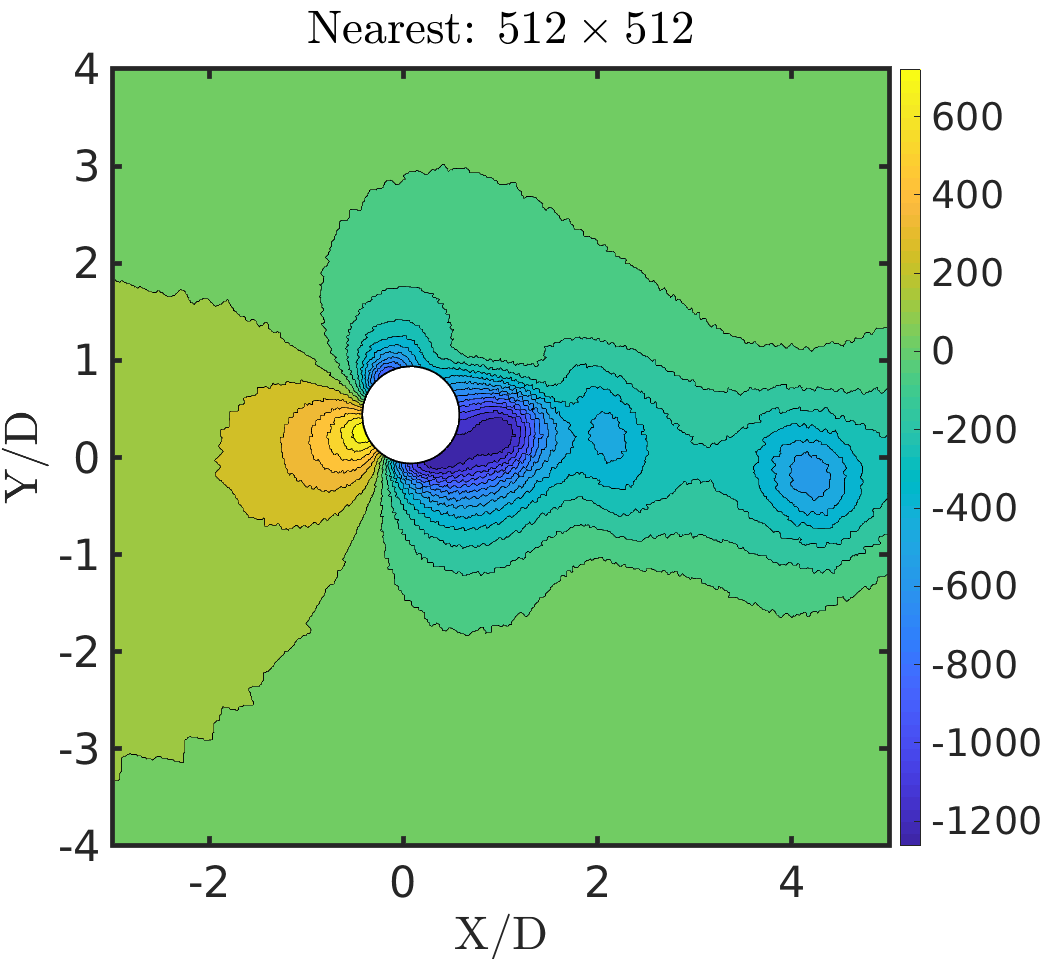}}\\
\subfloat[]{\includegraphics[width = 0.49\textwidth]{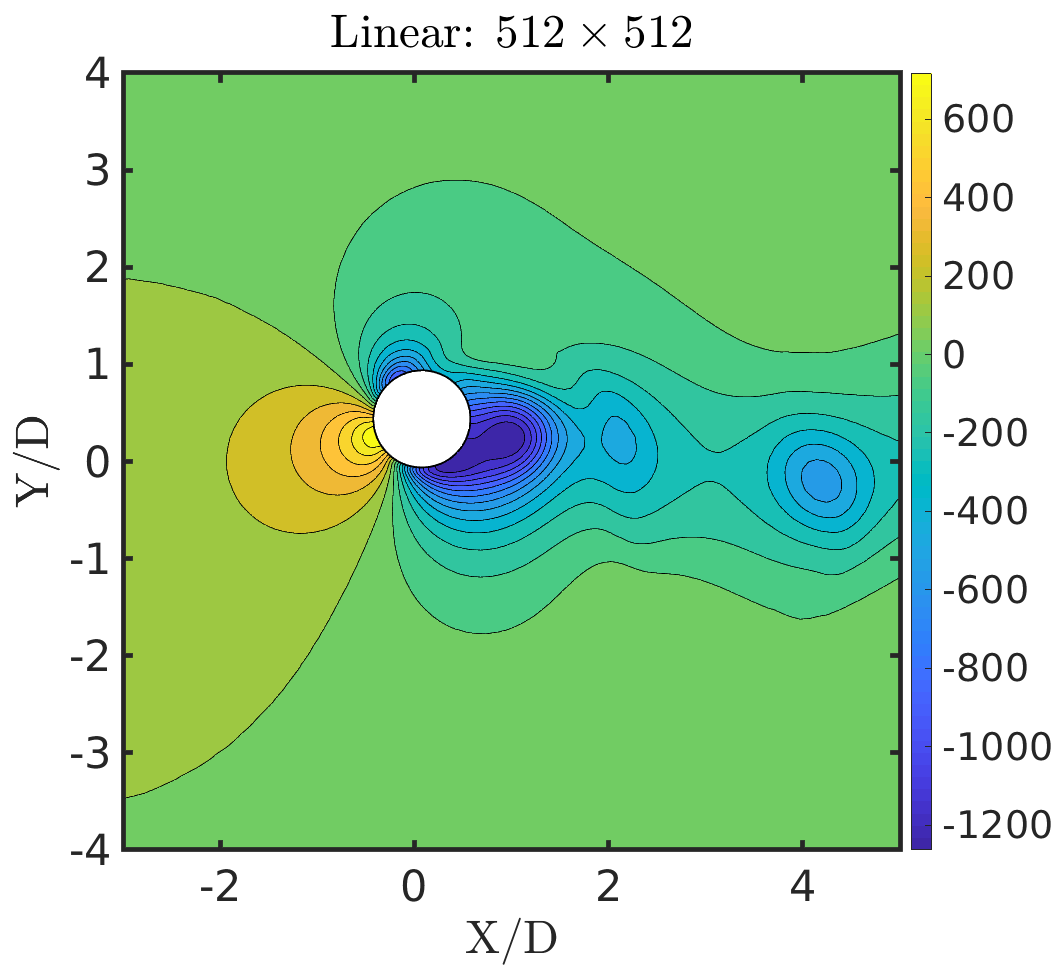}}
\hspace{0.01\textwidth}
\subfloat[]{\includegraphics[width = 0.49\textwidth]{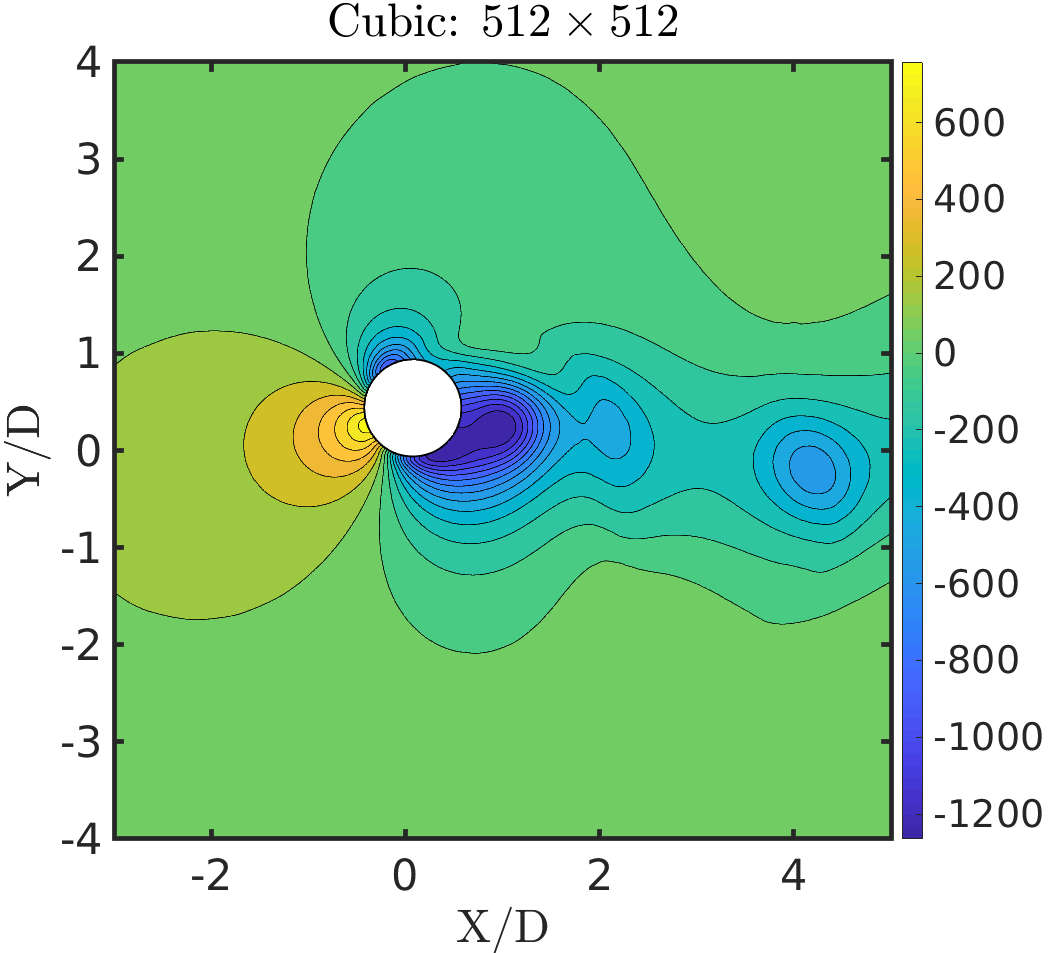}}
\caption{Freely vibrating circular cylinder in a uniform flow ($m^{*}=10,U_r=5, Re=200$): Qualitative behavior of the different interpolation methods for the NS pressure field in the ALE reference at the time step $t U_{\infty} / D = 125$: (b) Nearest neighbor, (c) linear and (d) piece-wise cubic type values on the snapshot $512 \times 512$ uniform grid with respect to (a) full-order CFD grid.}
\label{pressure_behavior}
\end{figure}
%
Fig.~\ref{forces_behavior_pixels_1} demonstrates the grid dependence of the normalized pressure pixelated forces $\overline{\textbf{F}}_{b} / 0.5 \rho^{f}U_{\infty}^{2}D $ vs the full-order forces $\overline{\textbf{F}}_{\ifsnt} / 0.5 \rho^{f}U_{\infty}^{2}D$. 
We observe that the interface coarsening leads to the pixelated force propagation to contain some noises, especially in the direction where the solid displacements are large. These noisy signals become dominant on reducing the time step and increasing the cell size. The reasons are attributed to the loss of the interface information due to lower fidelity and a linear way of the pixelated force calculation. Interestingly, these noises are somewhat invariant to a sharp fluid-solid interface propagation on a uniform grid. These force signals are devoid of any high-frequency noises for the static boundaries as discussed in our previous work
\cite{bukka2021assessment}. 
\begin{figure}
\centering
\subfloat[]{\includegraphics[width = 0.48\textwidth]{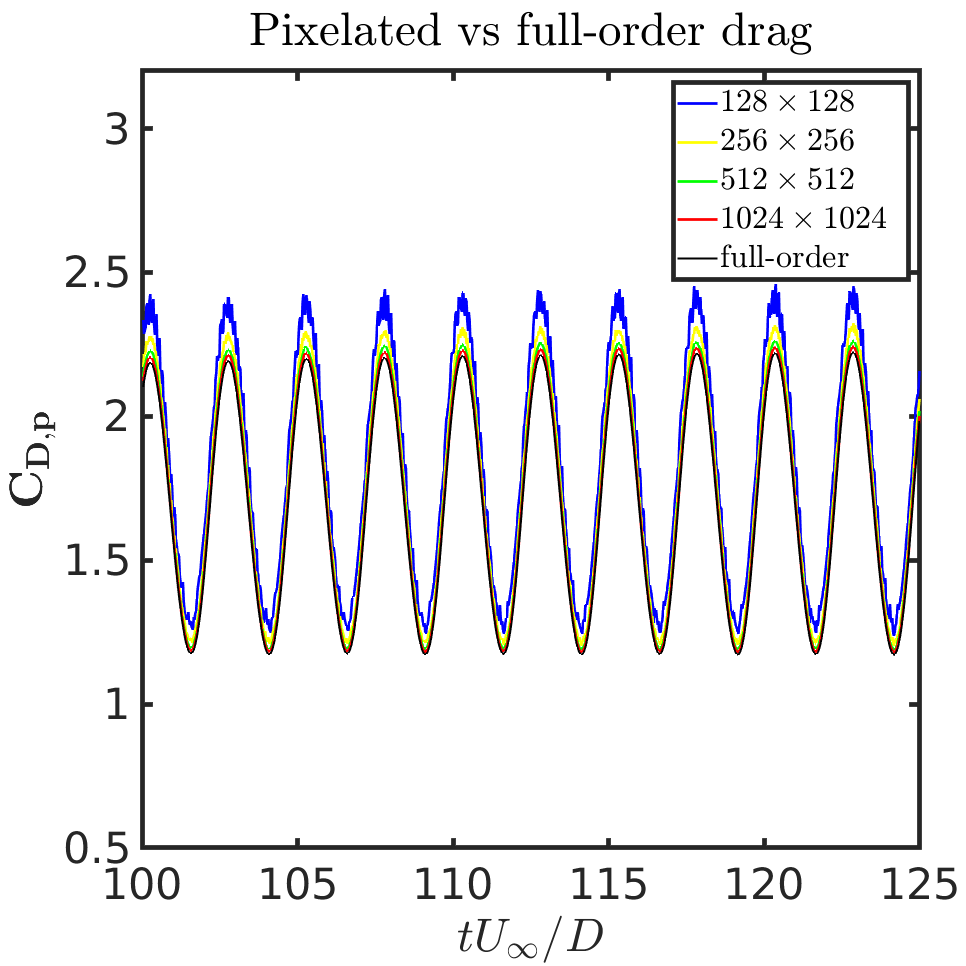}} 
\hspace{0.02\textwidth}
\subfloat[]{\includegraphics[width = 0.482\textwidth]{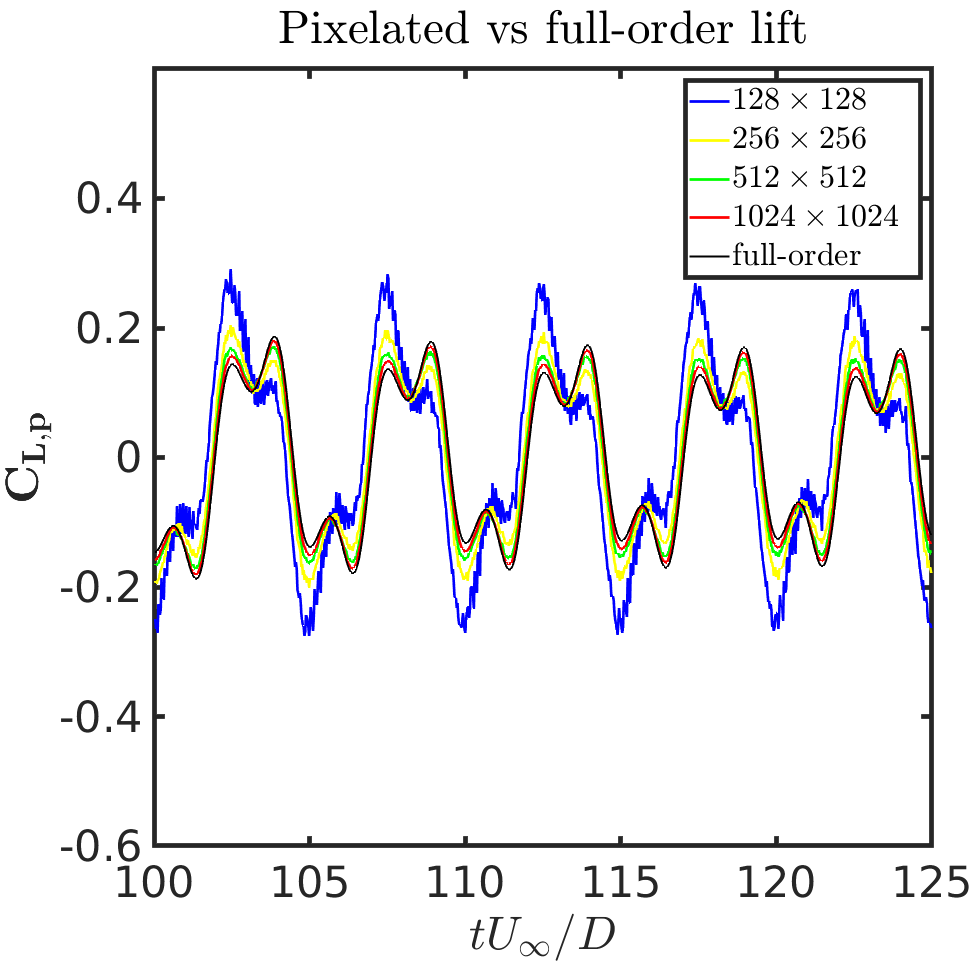}}
\caption{Freely vibrating circular cylinder in a uniform flow ($m^{*}=10,U_r=5, Re=200$): Total pixelated force propagation (from 100-125 $t U_{\infty}/D$) on the snapshot DL-ROM grids vs the full-order for (a) the pressure drag coefficient $\mathrm{C}_{\mathrm{D},\mathrm{p}}$, and (b) the pressure lift coefficient $\mathrm{C}_{\mathrm{L}, \mathrm{p}}$ on the training field.}
\label{forces_behavior_pixels_1}
\end{figure}
%

To correct these forces, while still maintaining a lower interface grid resolution, we first apply the Gaussian filtering $\zeta$  on the pixelated force propagation to damp the high-frequency noises. Fig.~\ref{forces_behavior_pixels} (a)-(b) depict the smooth trend $\zeta \overline{\textbf{F}}_{b}$ in the normalized forces on the various DL-ROM grids for the drag and lift coefficients via a Gaussian filter length of $20$. It is worth noting that the pixelation process results in the mean and derivative variation in the bulk quantities from the full-order counterpart. These errors reduce with super-resolution. However, we are interested in refining the uniform grid to an extent where the bulk quantities can be linearly reconstructed to the full-order within the mean and derivative correction using Algorithm \ref{lab1}. 

The pixelated force and the reconstruction process $\Psi$ on the different resolutions of the DL-ROM grid is summarized in Table \ref{tab:gridvals}. Columns 3 and 4 tabulate the mean and the root mean square of the total pixelated drag and lift coefficients, respectively, for various grid sizes. Column 5 in Table \ref{tab:gridvals} similarly denote the mean of time-dependent derivative correction $\overline{E}_{c}$ observed from the Algorithm \ref{lab1}. Columns 6 and 7 depict the reconstruction error $\epsilon$ in the corrected forces $ \Psi \overline{\textbf{F}}_{b}$ and the full-order $\overline{\textbf{F}}_{\ifsnt}$ as calculated from step 6 of Algorithm \ref{lab2}. The reconstruction accuracy is nearly $99.8 \% $ and $96.5 \%$ for the drag and the lift, respectively, onward grid $512 \times 512$. This process facilitates an optimal uniform grid to carry the neural prediction on while still recovering the bulk quantities within reasonable accuracy.

\begin{remark}
It is worth mentioning that the propagation of the pixelated force depends on the problem at hand; however, the present reconstruction offers a general data recovery methodology. Fig.~\ref{forces_behavior_pixels} (c)-(d)  depict the force correction by observing the $\Psi$ mapping and correcting on the training forces. We select the $512 \times 512$ as the snapshot  DL-ROM grid for the flow field predictions. It accounts for a reasonable force recovery while avoiding the necessity of super-resolution at the interface and  bypassing the unstructured mesh complexity of the moving point cloud.  
\end{remark}

\begin{table}[]
\small
\centering
\caption{Freely vibrating circular cylinder in a uniform flow ($m^{*}=10,U_r=5, Re=200$): Pixelated forces and the reconstruction accuracy on the various snapshot DL-ROM grids with respect to the full-order grid.}
\begin{tabular}{P{1.1cm}P{1.5cm}P{1.5cm}P{1.5cm}P{1.2cm}P{1.5cm}P{1.5cm}}
\toprule
\toprule
Grid & Cell-size & $\mathrm{mean}(\zeta \mathrm{C}_{\mathrm{D}, \mathrm{p}})$ & $(\zeta\mathrm{C}_{\mathrm{L}, \mathrm{p}})_{\mathrm{rms}}$ & $\mathrm{mean}(\overline{E}_{c})$ & $ \epsilon ( \mathrm{C}_{\mathrm{D}, \mathrm{p}} ) $ & $\epsilon ( \mathrm{C}_{\mathrm{L}, \mathrm{p}} )$ \\\\
\bottomrule
\bottomrule
128 & 0.0625 & 1.8295 & 0.1491 &  0.8534 & 7.6 $\times 10^{-3}$ & 3.2 $\times 10^{-1}$\\
\hline
256  & 0.0313 & 1.7470 & 0.1212 & 0.9124 & 3.3 $\times 10^{-3}$ & 1.5 $\times 10^{-1}$\\
\hline
512  & 0.0157 & 1.7130 & 0.1180 & 0.9307 & 2.9 $\times 10^{-3}$ & 4.7 $\times 10^{-2}$\\
\hline
1024 & 0.0078 & 1.6979 & 0.1168 & 0.9545 & 2.8 $\times 10^{-3}$ & 2.8 $\times 10^{-2}$\\
\hline
FOM & 2.2 $\times 10^{-5}$ & 1.6834 & 0.1174 & -- & -- & --\\
\hline
\end{tabular} 
\label{tab:gridvals}
\end{table}

\begin{figure}
\centering
\subfloat[]{\includegraphics[width = 0.48\textwidth]{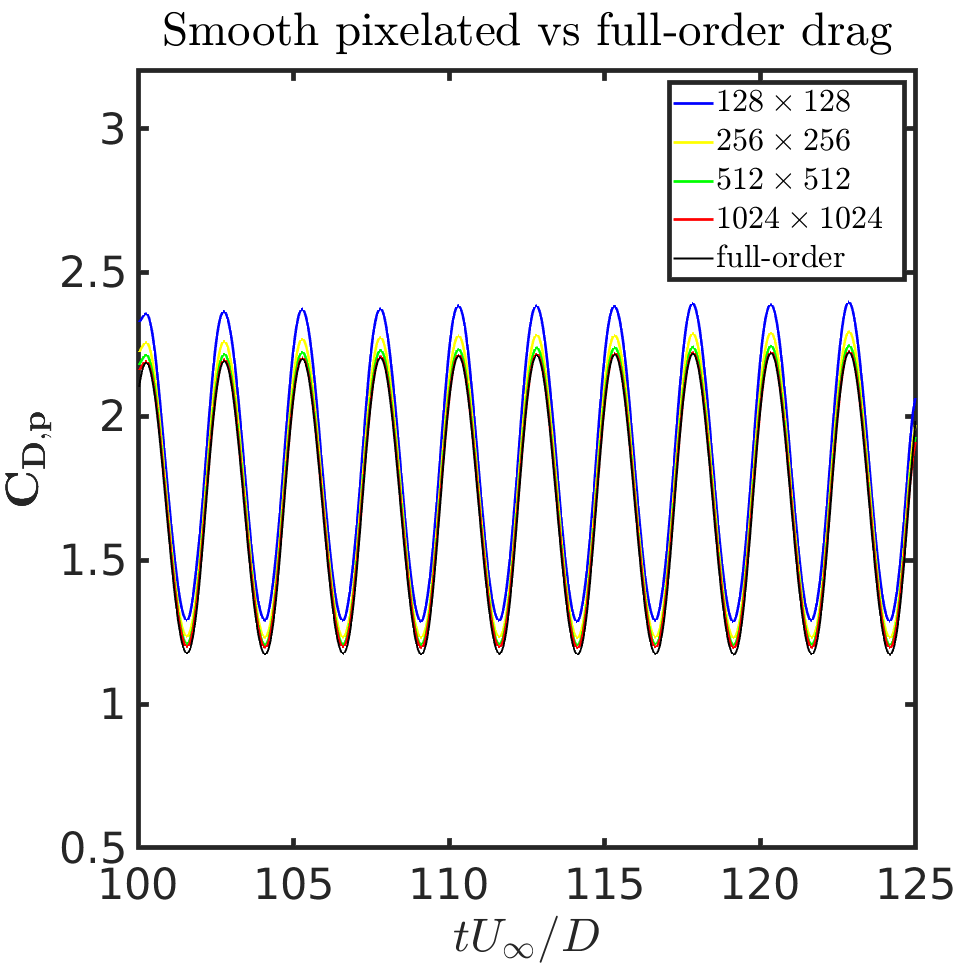}} 
\hspace{0.02\textwidth}
\subfloat[]{\includegraphics[width = 0.481\textwidth]{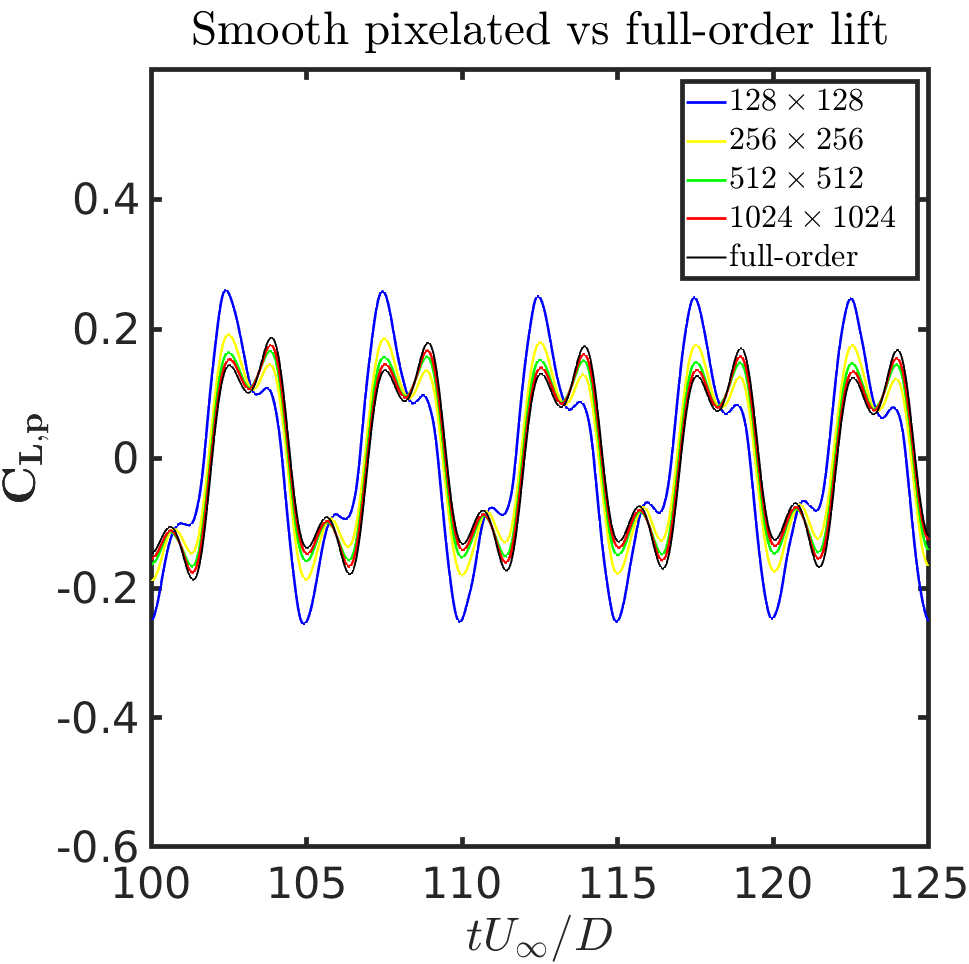}} \\ 
\subfloat[]{\includegraphics[width = 0.48\textwidth]{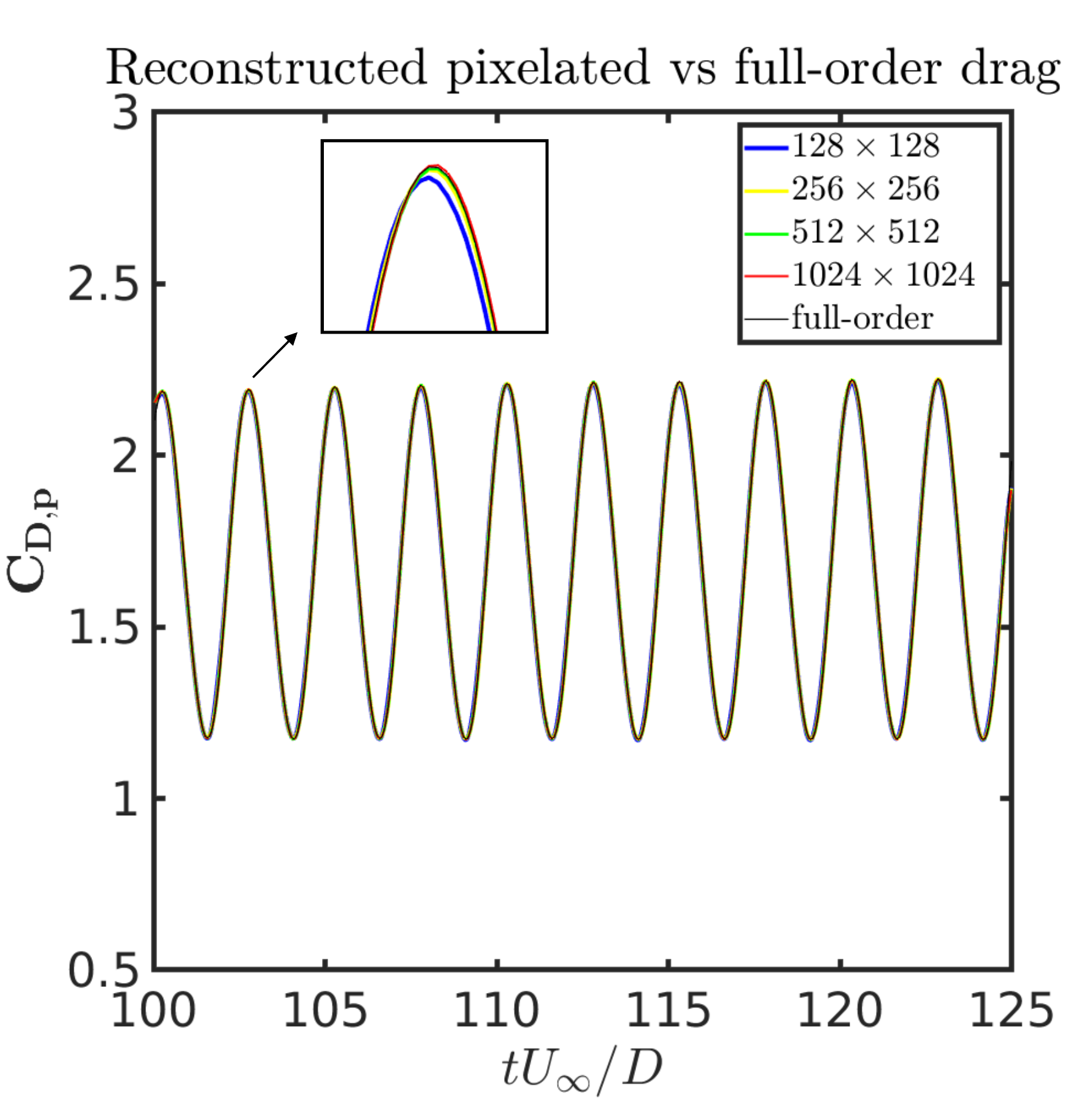}} 
\hspace{0.02\textwidth}
\subfloat[]{\includegraphics[width = 0.489\textwidth]{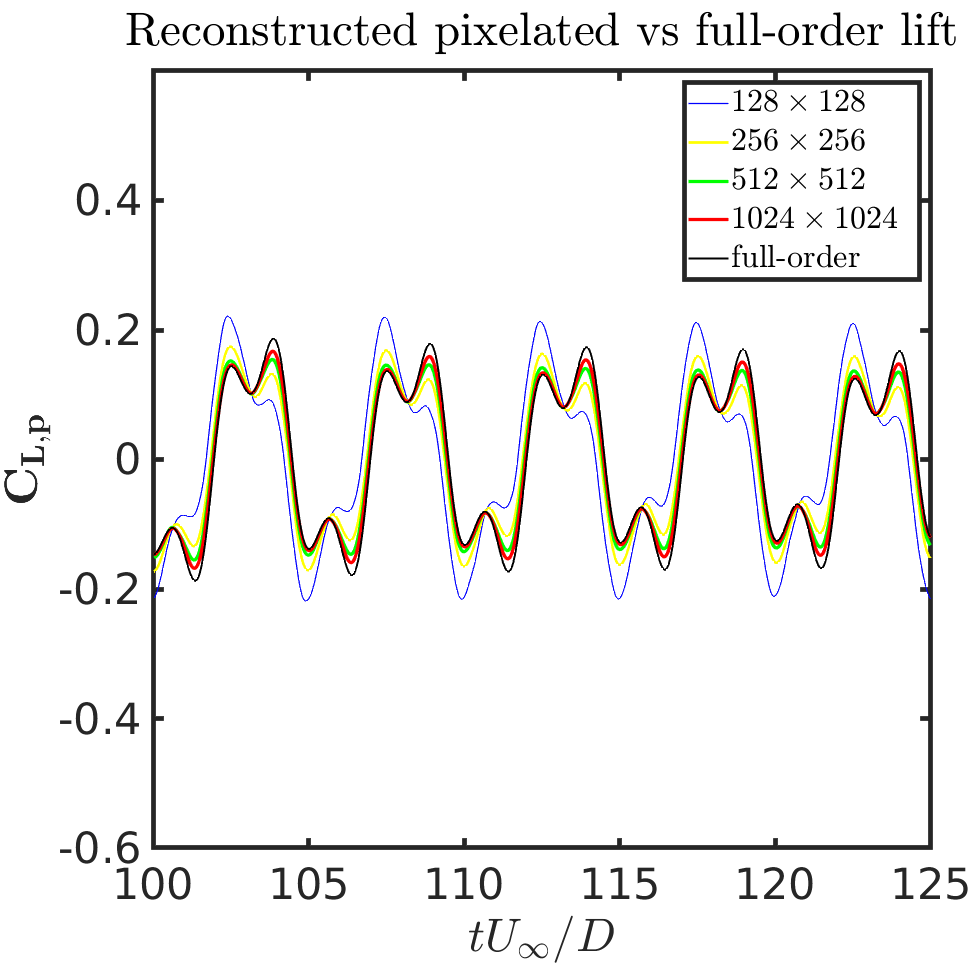}} 
\caption{Freely vibrating circular cylinder in a uniform flow ($m^{*}=10,U_r=5, Re=200$): Interface load behavior vs time (from 100-125 $t U_{\infty}/D$) on the snapshot DL-ROM grids. (a)-(b) denote the smoothen force propagation and (c)-(d) depicts recovered interface force information for drag and lift coefficients.}
\label{forces_behavior_pixels}
\end{figure}

\subsubsection{Near-wake dynamics using the CRAN driver}\label{cran_results}
With the chosen grid $N_{x}=N_{y}=512$, the end-to-end nonlinear learning based on the CRAN is used for the flow field predictions. Algorithm \ref{lab3} can be employed to recover the bulk quantities from the field data. 
As described in section \ref{int_load_results}, a point cloud flow dataset  $\textbf{s} = \left\lbrace\textbf{s}^{1}\; \textbf{s}^{2}\; \dots \; \textbf{s}^{N}\right\rbrace \in \mathbb{R}^{m\times N}$ from the full-order solver is transferred to the uniform grid using the snapshot-FTLR method. The generated field information $\textbf{S} = \left\lbrace\textbf{S}^{1}\; \textbf{S}^{2}\; \dots\; \textbf{S}^{N}\right\rbrace \in \mathbb{R}^{N_{x} \times N_{y} \times N}$ are decomposed into $n_{tr}=5000$ training data (100-225 $tU_{\infty}/D$) and $n_{ts}=1000$ testing data (225-250 $tU_{\infty}/D$).
The training time steps $n_{tr}$ are further sparsed for every $0.05\;tU_{\infty}/ D$, thereby reducing the total number of trainable steps to $n_{tr} /2$. This is essentially carried out to speed-up the CRAN I/O data processing, while still maintaining the same time horizon of training. {\color{black} As detailed in section \ref{cnn_rnn} from Eqs. (\ref{dn_cran_1})-(\ref{dn_cran_3}), the standard principles of data normalization and batch-wise arrangement is adopted to generate the scaled featured input 
$
\mathcal{S}=\left\{\mathcal{S}_{s}^{\prime 1} \; \mathcal{S}_{s}^{\prime 2}  \; \ldots \; \mathcal{S}_{s}^{\prime N_{s} } \right\} \in[0,1]^{N_{x} \times N_{y} \times N_{t} \times N_{s}},
$
where $N_s$ = 100 and $N_t = 25$. While training, a mini-batch of size $n_{s}=1$ is maintained. For every training iteration, a training sample $\mathcal{S}_{s}^{\prime j}=\left[\mathbf{S}_{s, j}^{\prime 1} \; \mathbf{S}_{s, j}^{\prime 2} \; \ldots \; \mathbf{S}_{s, j}^{N_{t}}\right]$ is randomly shuffled from the scaled featured input to update and refresh the CRAN neural parameters.}

The encoding and decoding space of the CRAN is summarized in Table \ref{tab:cranlevels} in a way that the spatial downsampling and prolongation are achieved gradually. We keep the low-dimensional encoder state $\textbf{A}_c$ and the LSTM-RNN evolver cells of the sizes: $h=N_h=32,64,128$ (see Fig. 3). These modes are obtained from a nonlinear space of the autoencoder network and are found to be the most sensitive hyperparameter in this architecture. Since these projections are self-supervised, we experiment with the different sizes of the modes based on the convergence during training. We first instantiate the training of all the three CRAN models for $N_{train} = 500,000$ iterations starting from the pressure fields. We select the $N_h=128$ based on a faster decay in the overall loss (Eq. (\ref{CRAN_loss})) compared with the $N_h = 32,64$ evolver states. $N_h=128$ pressure model is further optimized until the total iterations reach to $N_{train} = 10^{6}$ with the overall loss $\approx 4.93 \times 10^{-5}$. Once the pressure training is completed, we transfer the learning to the x-velocity field and optimize the CRAN network subsequently for $N_{train} = 750,000$ iterations.

\begin{table}[]
\small
\centering
\caption{Freely vibrating circular cylinder in a uniform flowr ($m^{*}=10,U_r=5, Re=200$): Deep CRAN layer details: Convolutional kernel sizes, numbers and stride. Layers 6,7,8 represent the fully connected feed-forward encoding, while 9,10,11 depict the similar decoding. The low-dimensional state $\textbf{A}_c$ is evolved between the layers 8 and 9 with the LSTM-RNN. }
\begin{tabular}{P{1.0cm}P{1.5cm}P{1.0cm}P{1.0cm}|P{1.0cm}P{1.5cm}P{1.0cm}P{1.0cm}}
\toprule
\toprule
\multicolumn{4}{c}{Conv2D encoder} &  \multicolumn{4}{c}{DeConv2D decoder}\\
\midrule
Layer & kernel-size & kernels & stride & Layer & kernel-size & kernels & stride \\
\bottomrule
\bottomrule
1 & $10 \times 10$ & 2 & 4     & 12 & $5 \times 5$ & 16 & 2\\
\hline
2 & $10 \times 10$ & 4 & 4     & 13 & $5 \times 5$ & 8 & 2\\
\hline
3 & $5 \times 5$ & 8 & 2       & 14 & $5 \times 5$ & 4 & 2\\
\hline
4 & $5 \times 5$ & 16 & 2      & 15 & $10 \times 10$ & 2 & 4\\
\hline
5 & $5 \times 5$ & 32 & 2      & 16 & $10 \times 10$ & 1 & 4\\
\hline
\end{tabular} 
\label{tab:cranlevels}
\end{table}


\begin{remark}
Each CRAN model is trained on a single graphics processing unit Quadro GP100/PCIe/SSE2 with Intel Xeon(R) Gold 6136 central processing unit @ 3.00GHz × 24 processors for nearly 2.7 days. A long training time is considered so that the CRAN-based framework self-supervises with the maximum optimization.  The training process is stopped when the loss reaches $\approx 1.0\times 10^{-5}$. The test time steps depict the predictive performance of this network in terms of accuracy check and remarkable speed-ups compared with the FOM counterpart. 
\end{remark}

Herein, we are interested to demonstrate $N_{h}=128$ trained CRAN model to infer the flow field with a multi-step predictive cycle of $p=N_{t}=25$. Figs.~\ref{cran_viv_pred_p} and \ref{cran_viv_pred_velx} show the comparison
of the predicted and the true values of the pressure and $\mathrm{x}$-velocity fields, respectively, at time steps $9300$ ($232.5\;tU_{\infty}/D$), $9600$ ($240\;tU_{\infty}/D$) and $9960$ ($249\;tU_{\infty}/D$) on the test data. {\color{black} The normalized reconstruction error $E^i$, for any time step say $i$, is constructed by taking the absolute value of the difference between the true $\mathbf{S}^{i}$ and predicted field $\hat{\mathbf{S}}^{i}$ and normalizing the difference with the $L_2$ norm of the truth using the equation 
\begin{equation}
  E^{i}=\frac{|\mathbf{S}^{i}-\hat{\mathbf{S}}^{i}|}{\left\|\mathbf{S}^{i}\right\|_{2}}. 
\end{equation}
}
It is evident from these predictions that significant portions of the reconstruction errors are concentrated in the nonlinear flow separation region of the vibrating cylinder. These errors are in the order of $10^{-3}$ for the pressure and $10^{-4}$ for the x-velocity. The field predictions are accurately predicted in the expected cylinder motion, albeit with a minor deviation in the reconstruction. {\color{black}We observe that for the CRAN framework, with one input step, the coupled FSI prediction is accurate for the next 25 time steps in the future, after which the predictions can diverge. Thus, the feedback demonstrators combat the compounding errors and enforce the CRAN predictive trajectory devoid of divergence over the test data.}


The synchronous point cloud structural and coarse-grain flow field predictions are employed to calculate the integrated pressure loads $\mathrm{C}_{\mathrm{D},\mathrm{p}}$ and $\mathrm{C}_{\mathrm{L},\mathrm{p}}$ via Algorithm \ref{lab3}.
The results are shown in Fig.~\ref{forces_pred_res_correction} (in the red line). { \color{black} The interface load algorithm extracts the full-order pressure loads with consistent phases from the predicted fields on the uniform grid, largely accounting for an accurate flow separation and the near-wake predictions. Interestingly, we observe some variations in the amplitude of the force signals from a direct coarse-grain integration over the moving interface even though other quantities are predicted well. The reason for this behavior is the existence of high-frequency noises in the pixelated forces from the uniform grid, even in the training data. Since the internal training process of the CRAN does not damp these force noises directly, the predicted fields can also infer these noises in time. We note that these spurious noises are the largest when the interface displacement is high and as a result, large residuals are seen in the force plots. The development of these residuals in the CNN-based predictions can also be responsible for a finite amount of predictive ability of the CRAN architecture and the requirement of ground data to combat this problem.

\begin{remark}
One potential way to correct these residuals is to dampen the noises using a denoising recurrent neural network. These networks have been shown to perform well to denoise signals in a variety of applications such as electrocardiographic signal (ECG) \cite{antczak2018deep} and automatic speech recognition (ASR) \cite{maas2012recurrent}. For demonstration, we damp the noises in the coarse-grain force signals using a denoising LSTM-RNN. We achieve this by learning the mapping of the coarse-grained force signals to the full-order force using a standard LSTM type recurrent network on finite predicted signals. The filtered signals in Fig.~\ref{forces_pred_res_correction} (shown in the blue line) demonstrate good precision for the drag and reasonable in the lift compared to the full-order. 
Since denoising RNNs are beyond the scope of this manuscript, we refer the reader to the works in \cite{han2021hybrid} and \cite{mozer2018state} for more details.  
\end{remark}

\begin{figure}
\centering
\subfloat[]{\includegraphics[width = 0.48\textwidth]{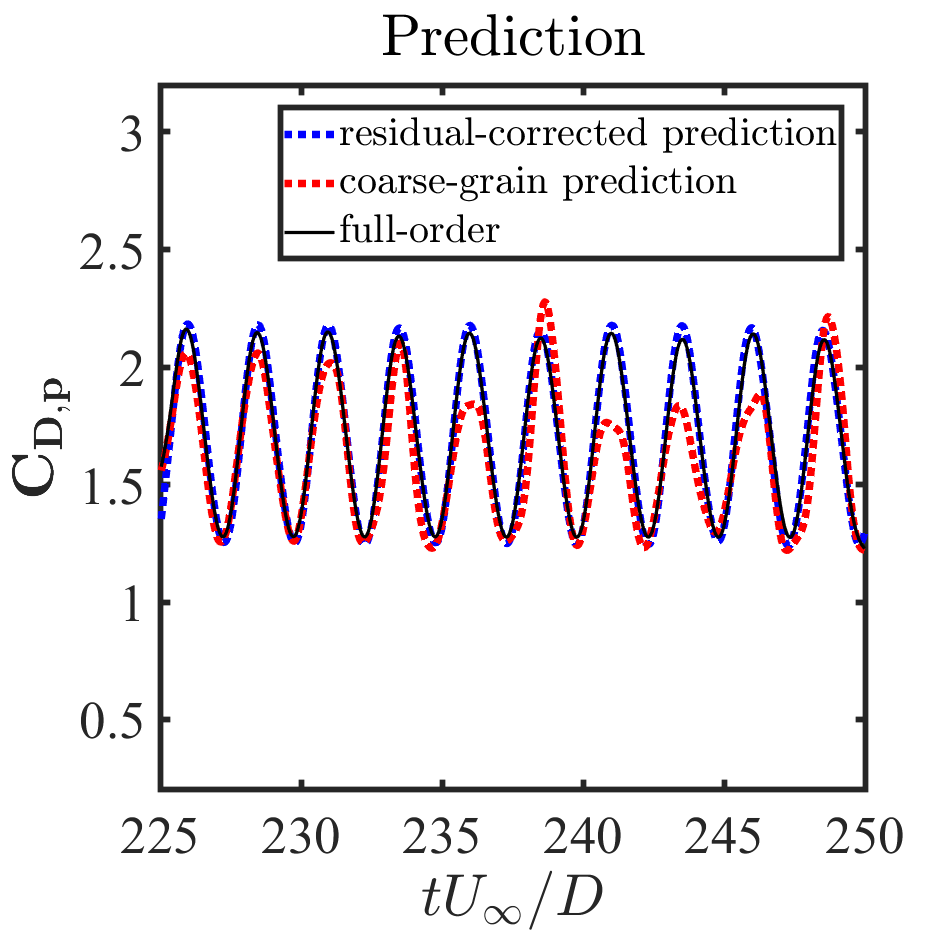}} 
\hspace{0.02\textwidth}
\subfloat[]{\includegraphics[width = 0.48\textwidth]{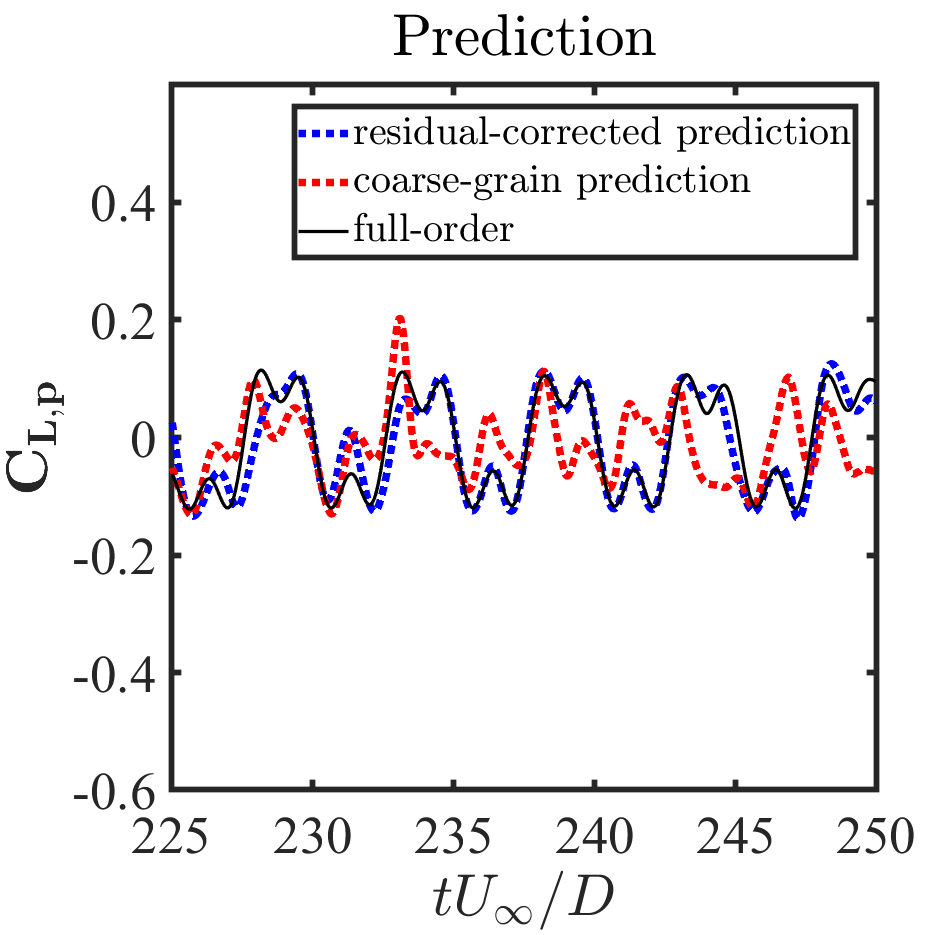}}
\caption{Freely vibrating circular cylinder in a uniform flow ($m^{*}=10,U_r=5, Re=200$): Interface force prediction on the test time steps from the predictive framework (from 225-250 $t U_{\infty}$: (a) Pressure drag coefficient $\mathrm{C}_{\mathrm{D},\mathrm{p}}$, and (b) pressure lift coefficient $\mathrm{C}_{\mathrm{L}, \mathrm{p}}$. The red lines depict the coarse-grain forces obtained from Algorithm \ref{lab3} and blue lines depict the corrected forces using a denoising LSTM-RNN. }
\label{forces_pred_res_correction}
\end{figure}
}

\begin{figure}
\begin{widepage}
\subfloat[]
{\includegraphics[width = 0.33\textwidth]{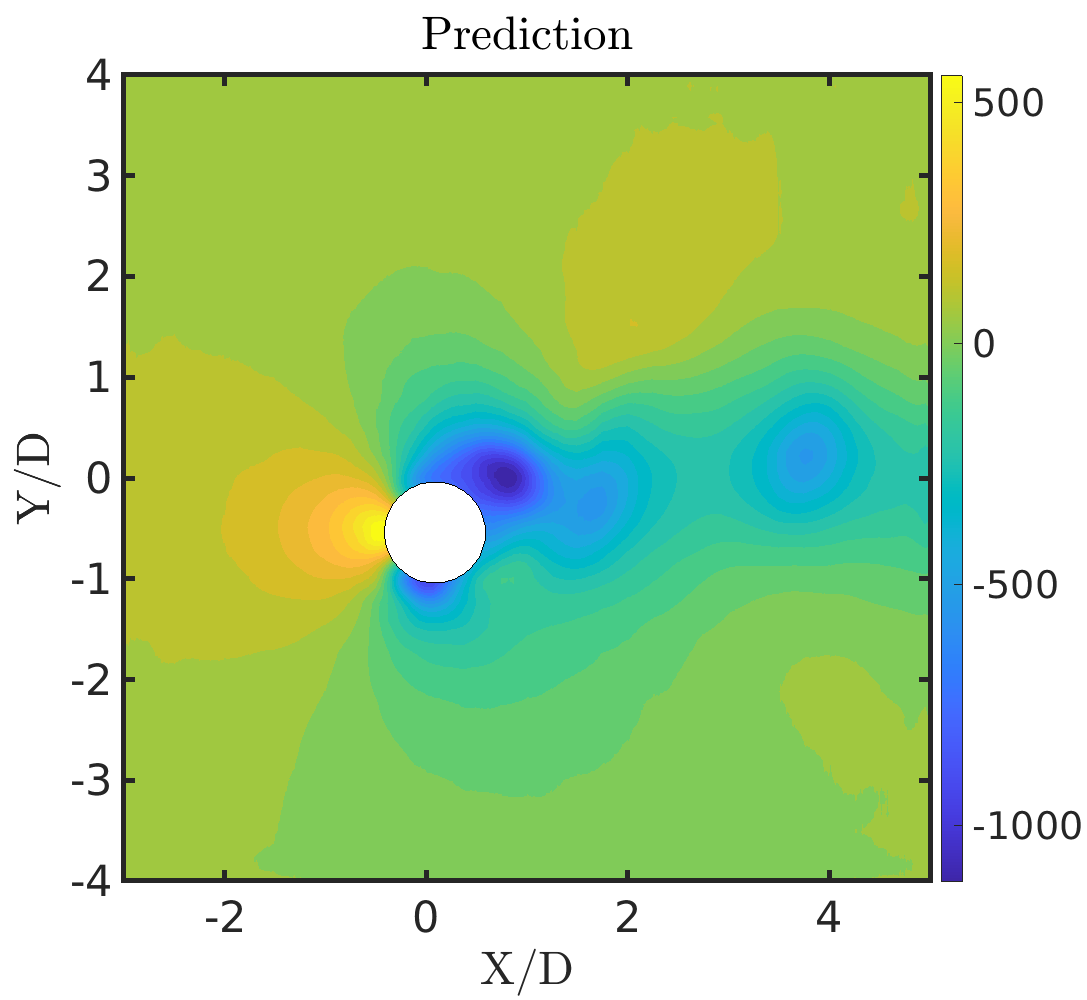}
\hspace{0.01\textwidth}
\includegraphics[width = 0.33\textwidth]{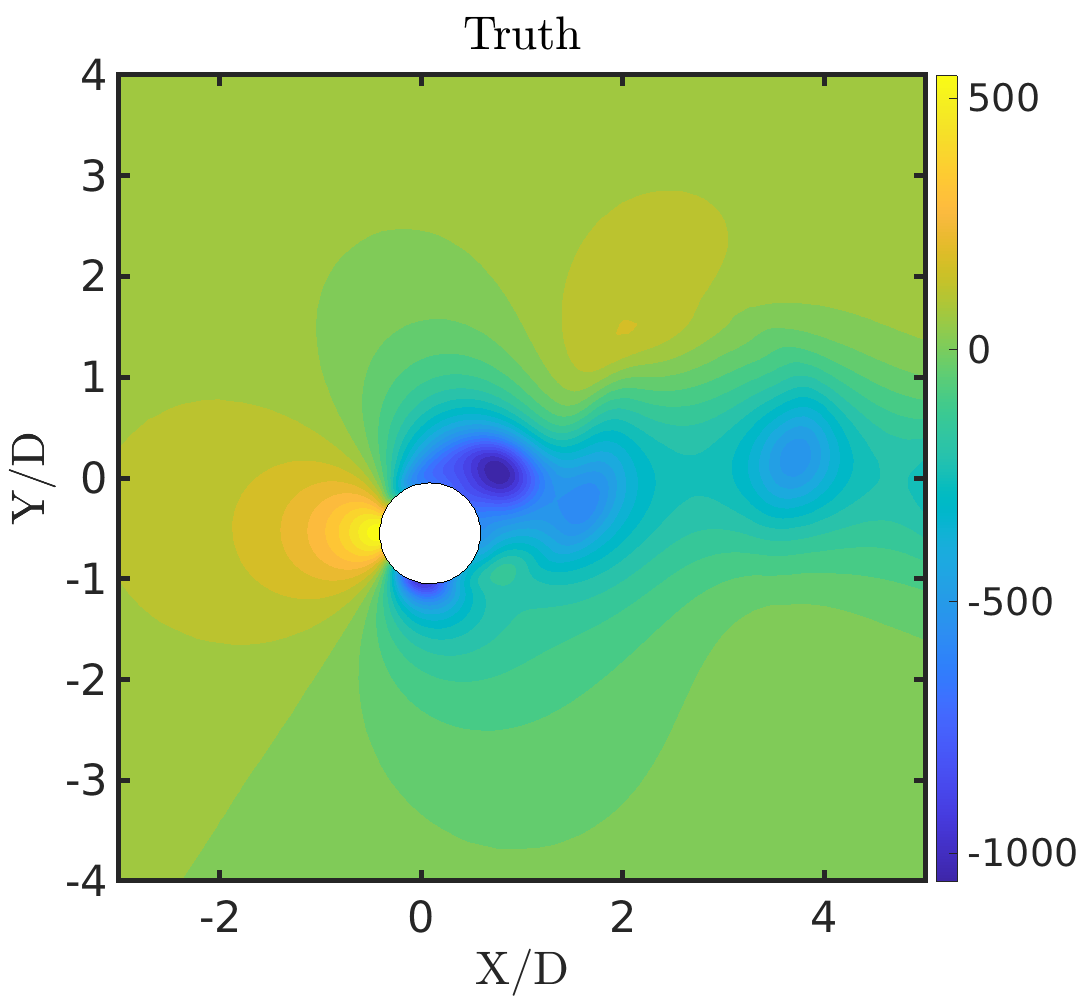}
\hspace{0.01\textwidth}
\includegraphics[width = 0.325\textwidth]{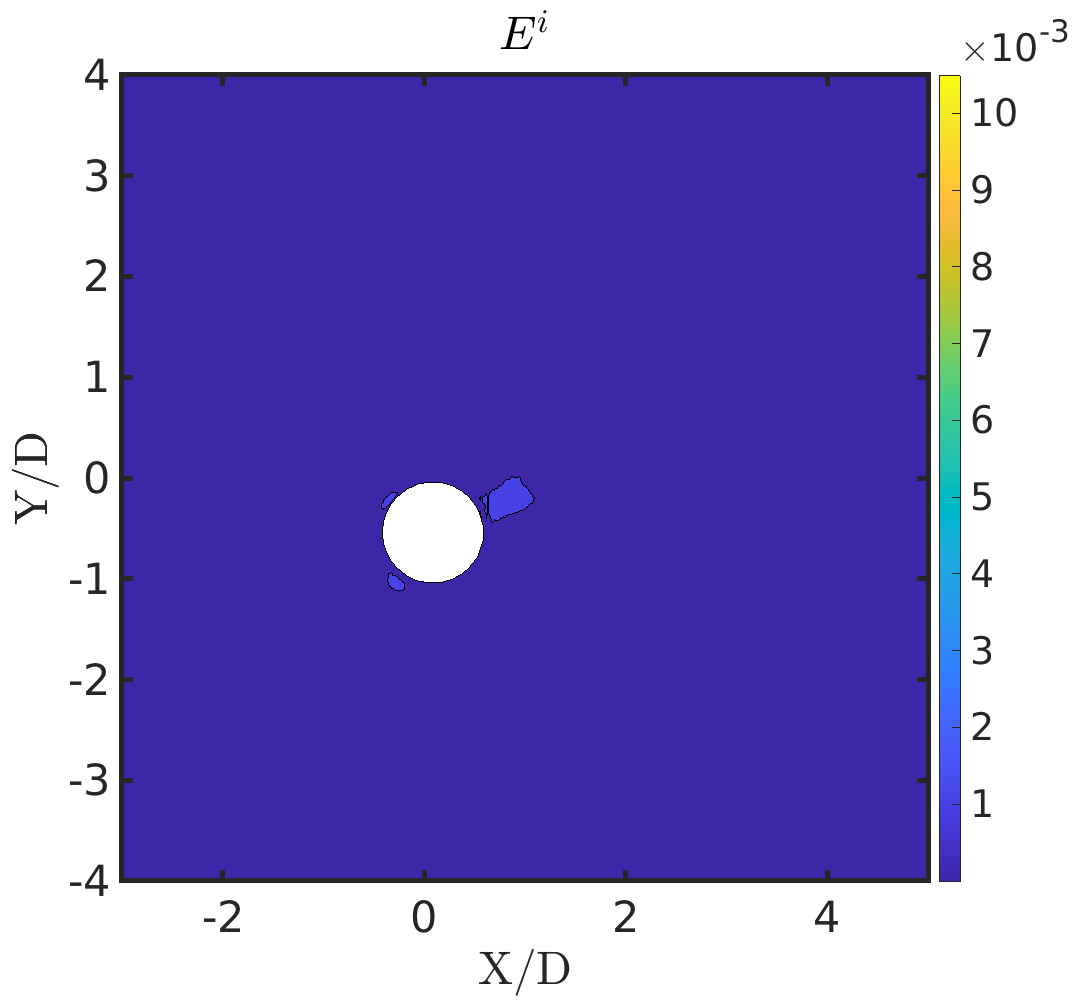}}
\\
\vspace{0.01\textwidth}
\subfloat[]
{\includegraphics[width = 0.33\textwidth]{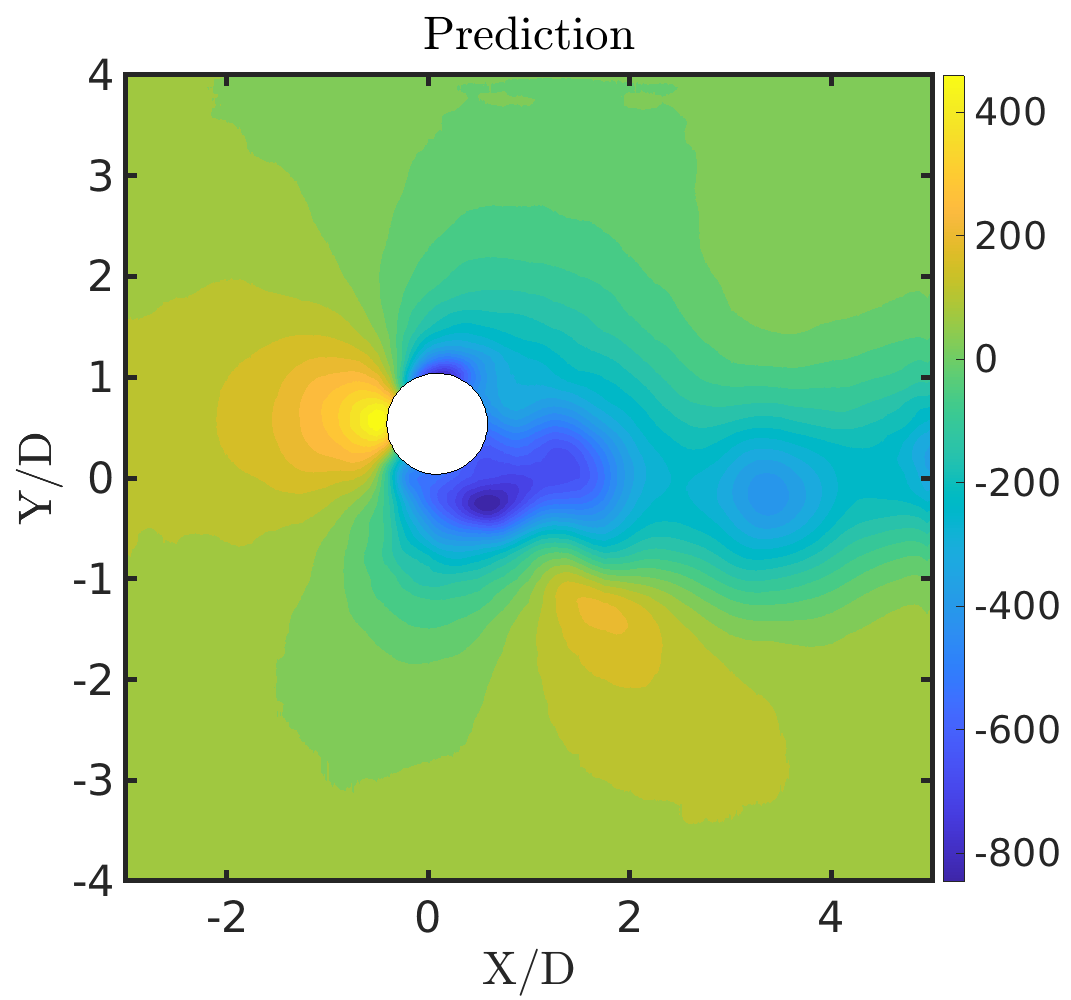}
\hspace{0.01\textwidth}
\includegraphics[width = 0.33\textwidth]{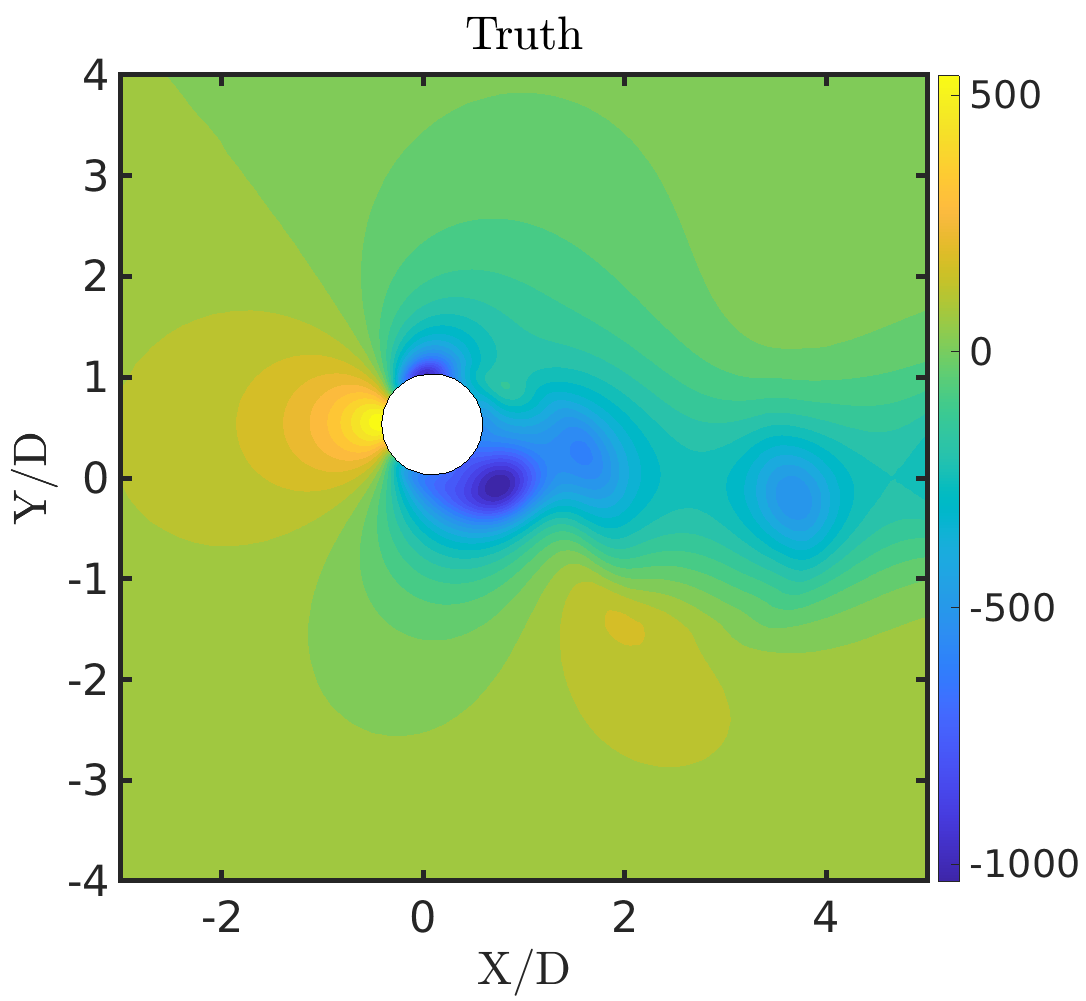}
\hspace{0.01\textwidth}
\includegraphics[width = 0.325\textwidth]{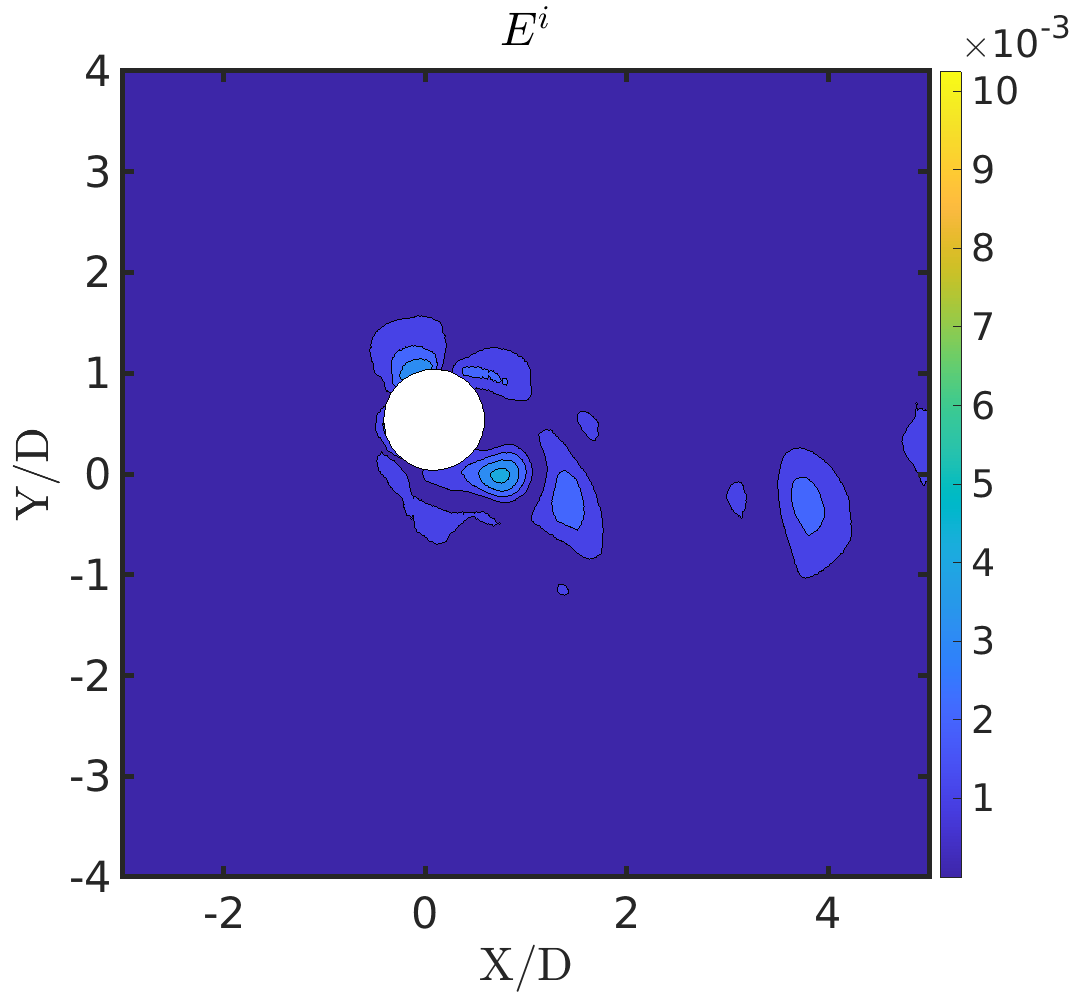}}
\\
\vspace{0.01\textwidth}
\subfloat[]
{\includegraphics[width = 0.33\textwidth]{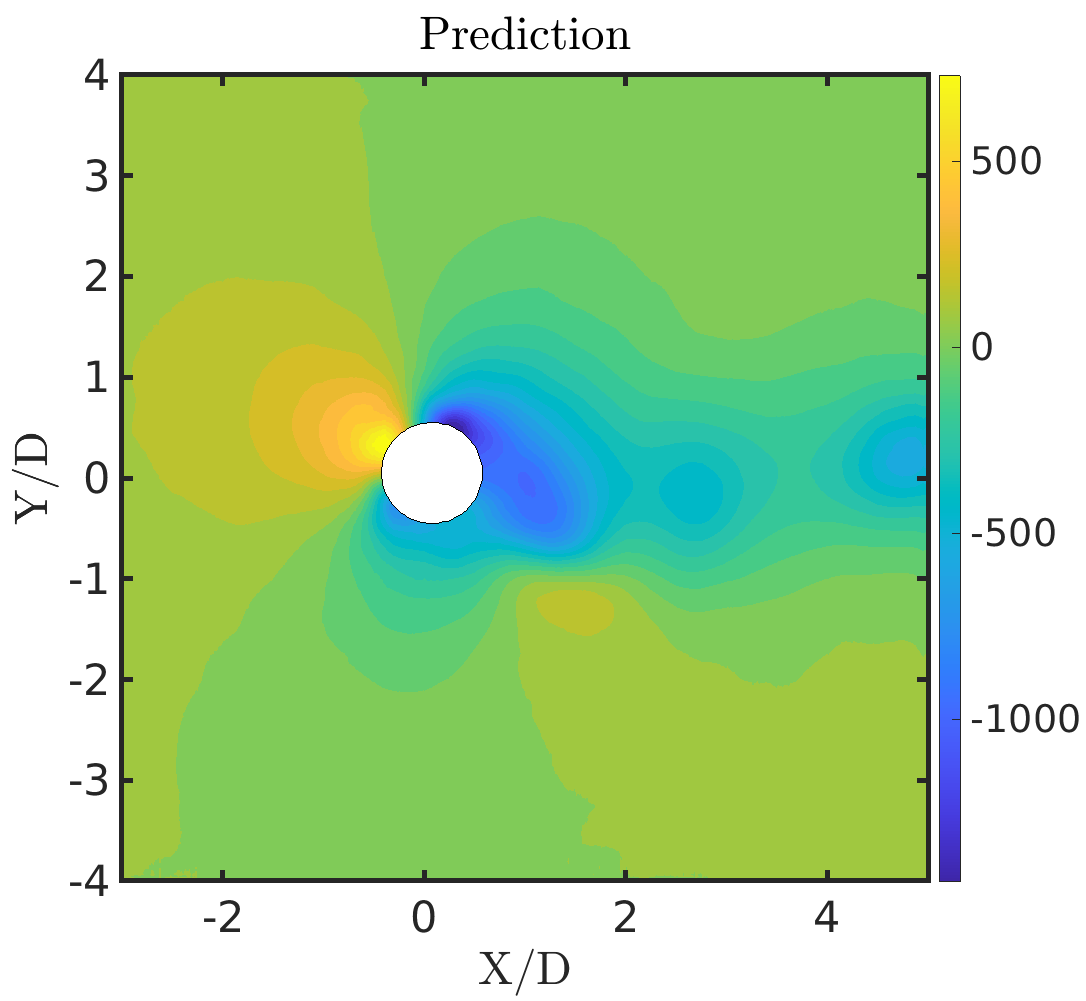}
\hspace{0.01\textwidth}
\includegraphics[width = 0.33\textwidth]{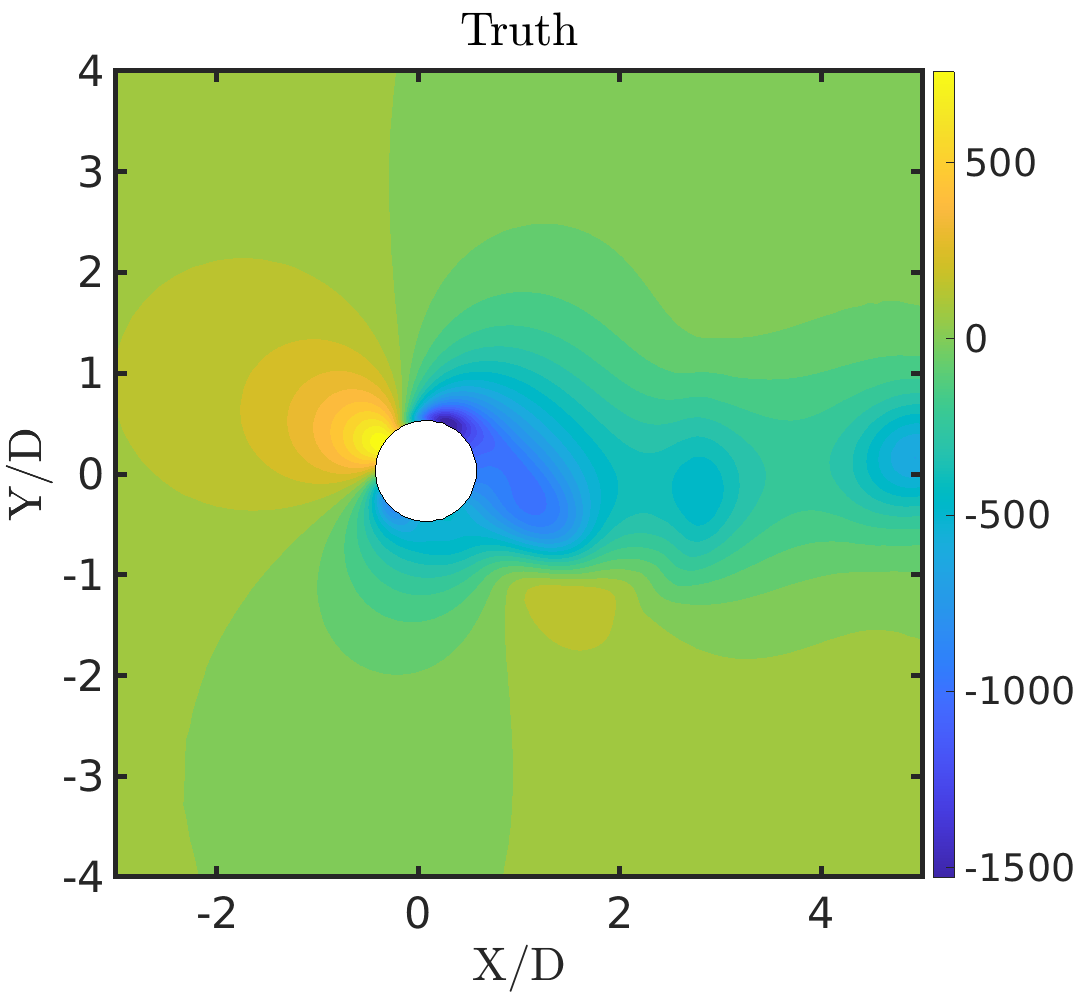}
\hspace{0.01\textwidth}
\includegraphics[width = 0.325\textwidth]{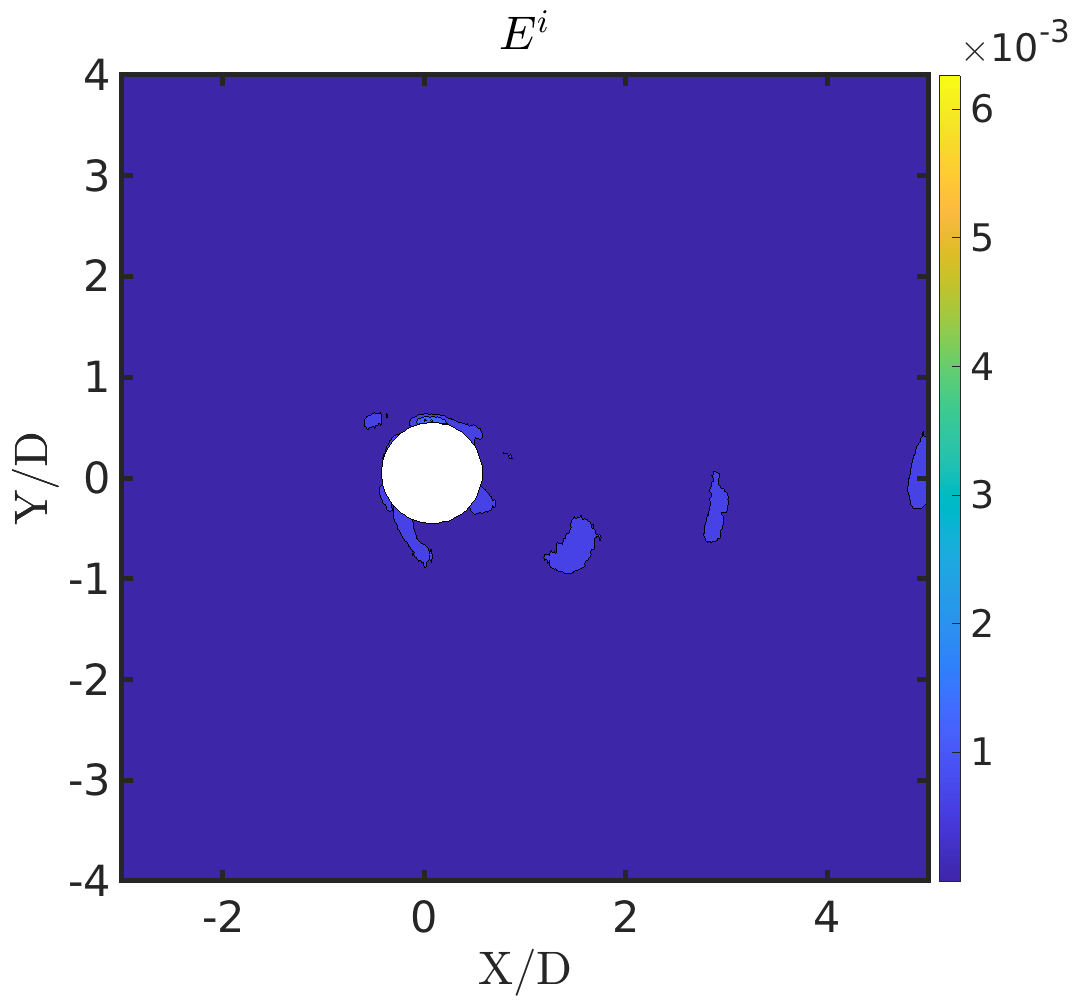}}
\end{widepage}
\caption{Freely vibrating circular cylinder in a uniform flow ($m^{*}=10,U_r=5, Re=200$): Comparison of the predicted and the true fields along with the normalized reconstruction error $E^{i}$ at $\;tU_{\infty}/D =$ (a)  232.5, (b) 240 (c)  249 for the pressure field ($P$). }
\label{cran_viv_pred_p}
\end{figure}

\begin{figure}
\begin{widepage}
\subfloat[]
{\includegraphics[width = 0.33\textwidth]{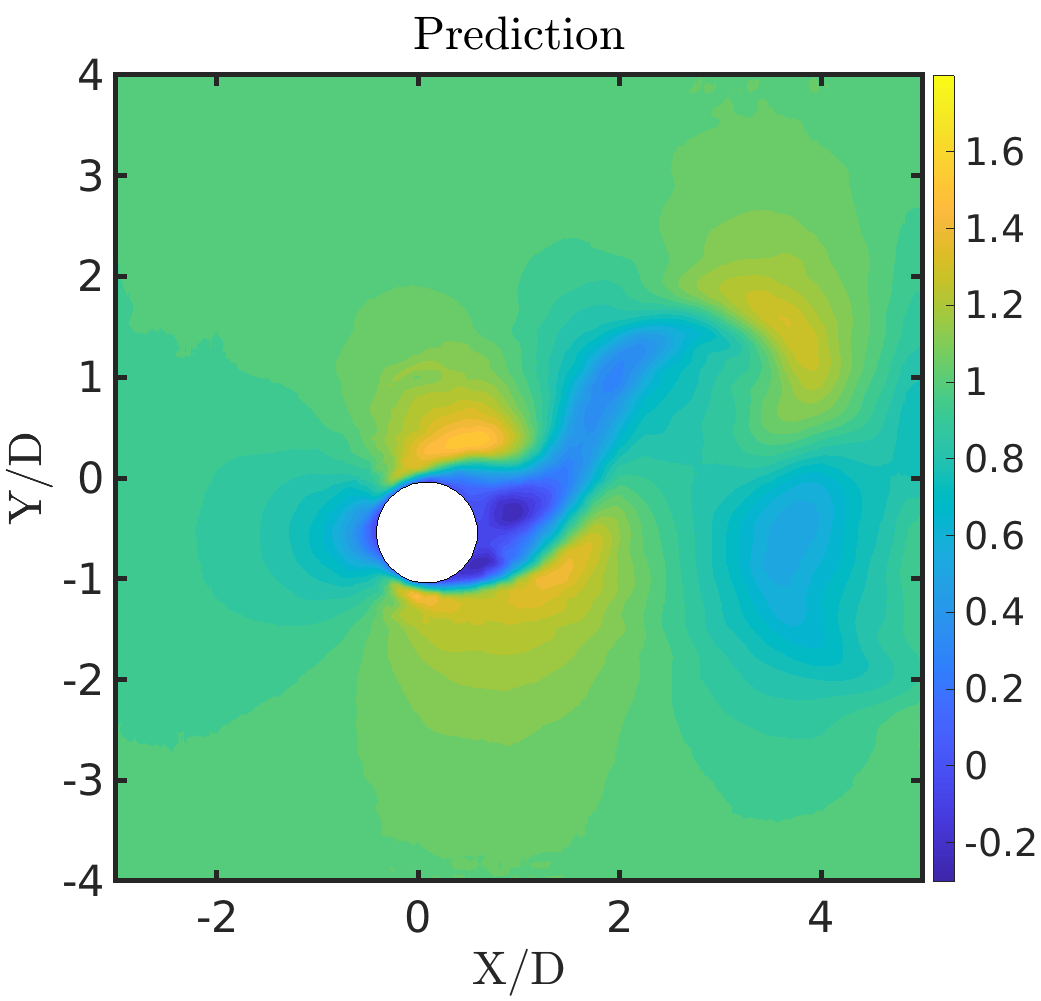}
\hspace{0.01\textwidth}
\includegraphics[width = 0.33\textwidth]{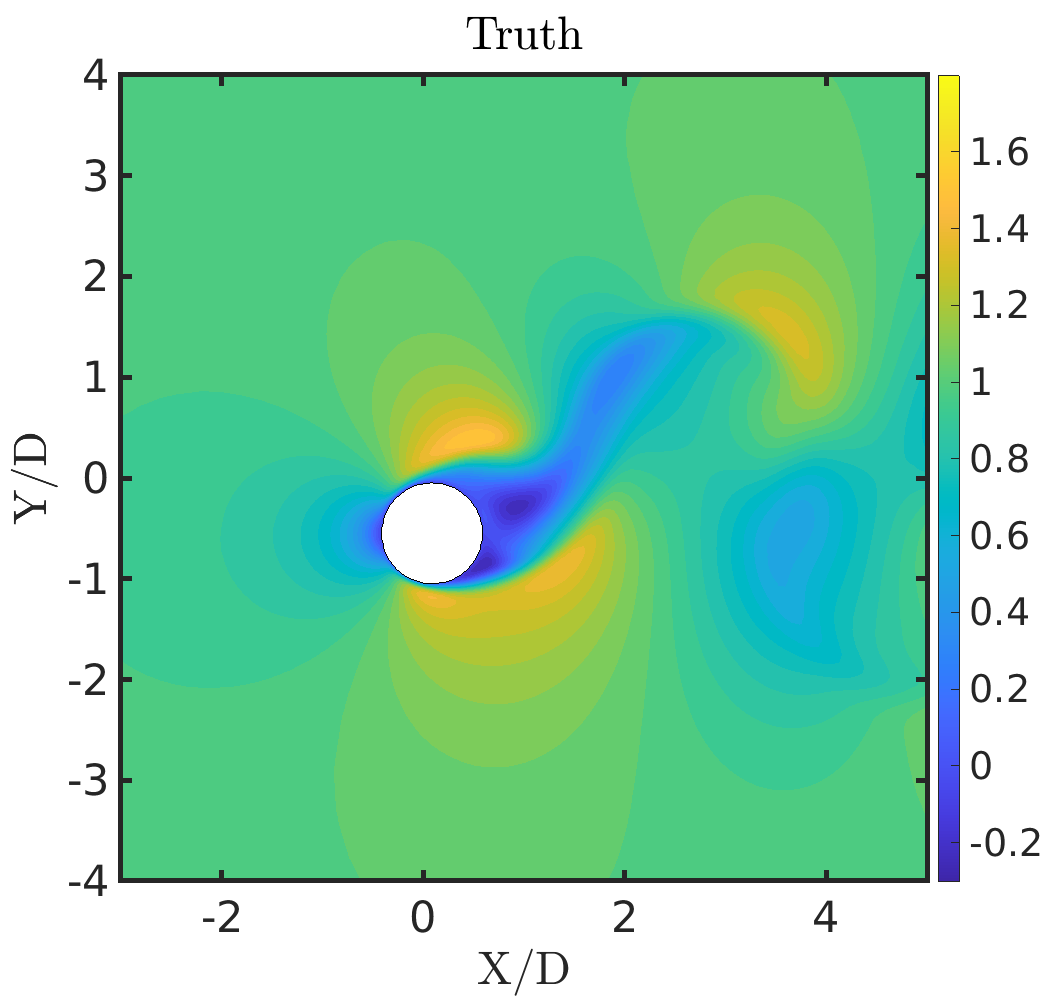}
\hspace{0.01\textwidth}
\includegraphics[width = 0.335\textwidth]{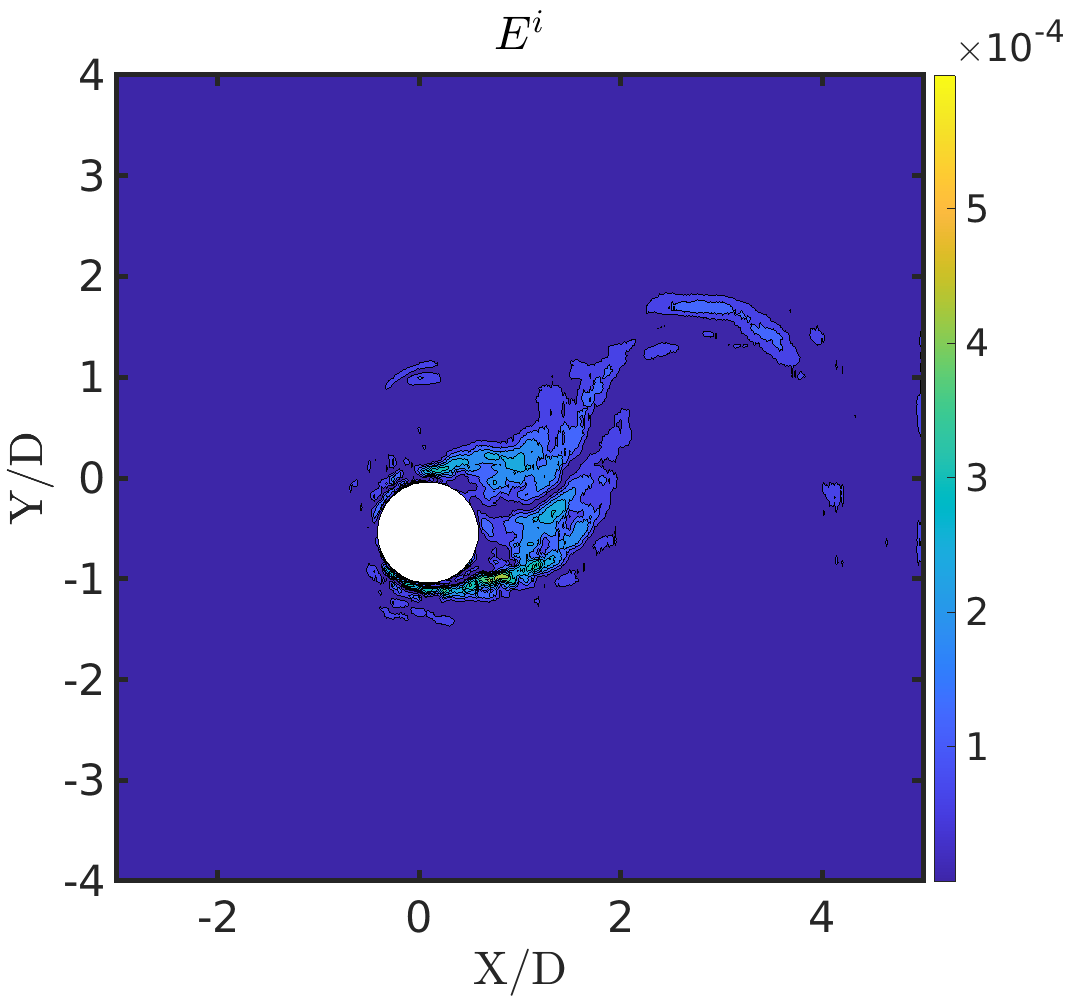}}
\\
\vspace{0.01\textwidth}
\subfloat[]
{\includegraphics[width = 0.33\textwidth]{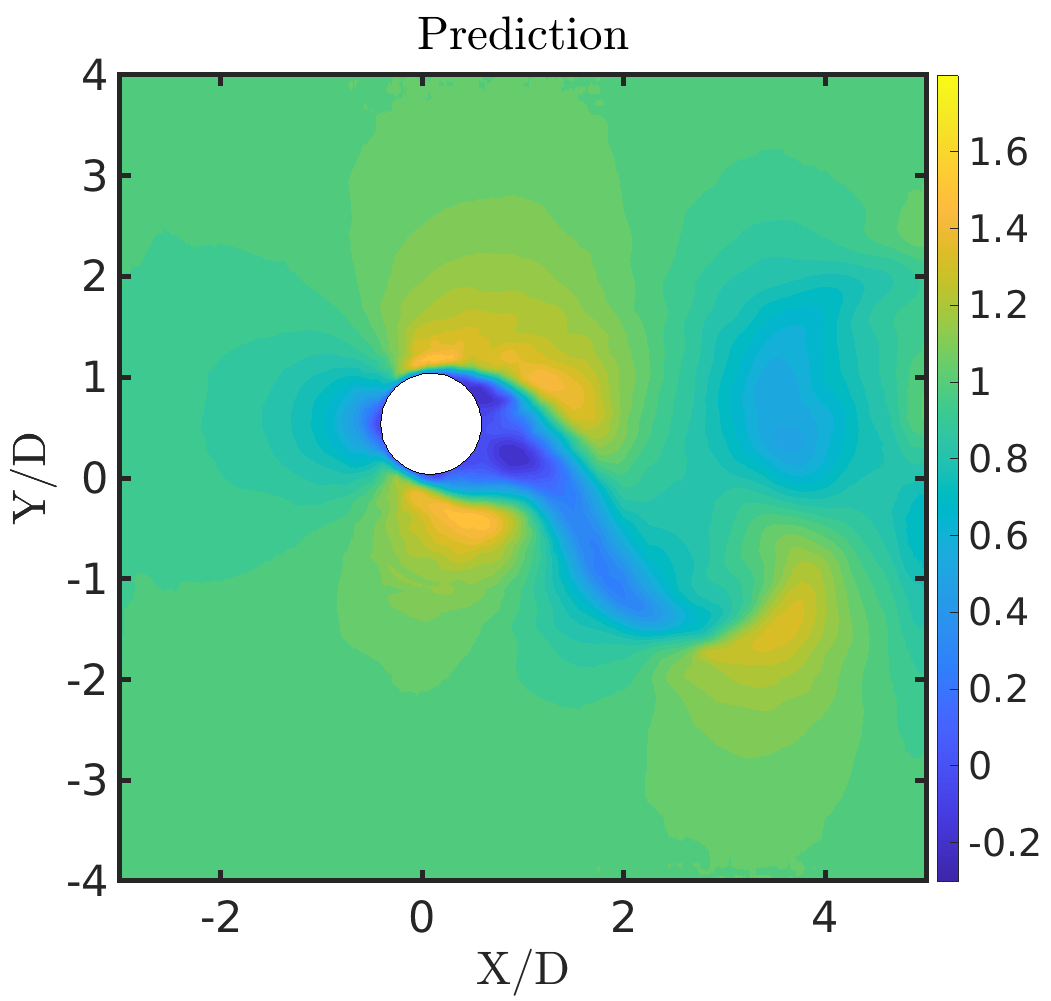}
\hspace{0.01\textwidth}
\includegraphics[width = 0.33\textwidth]{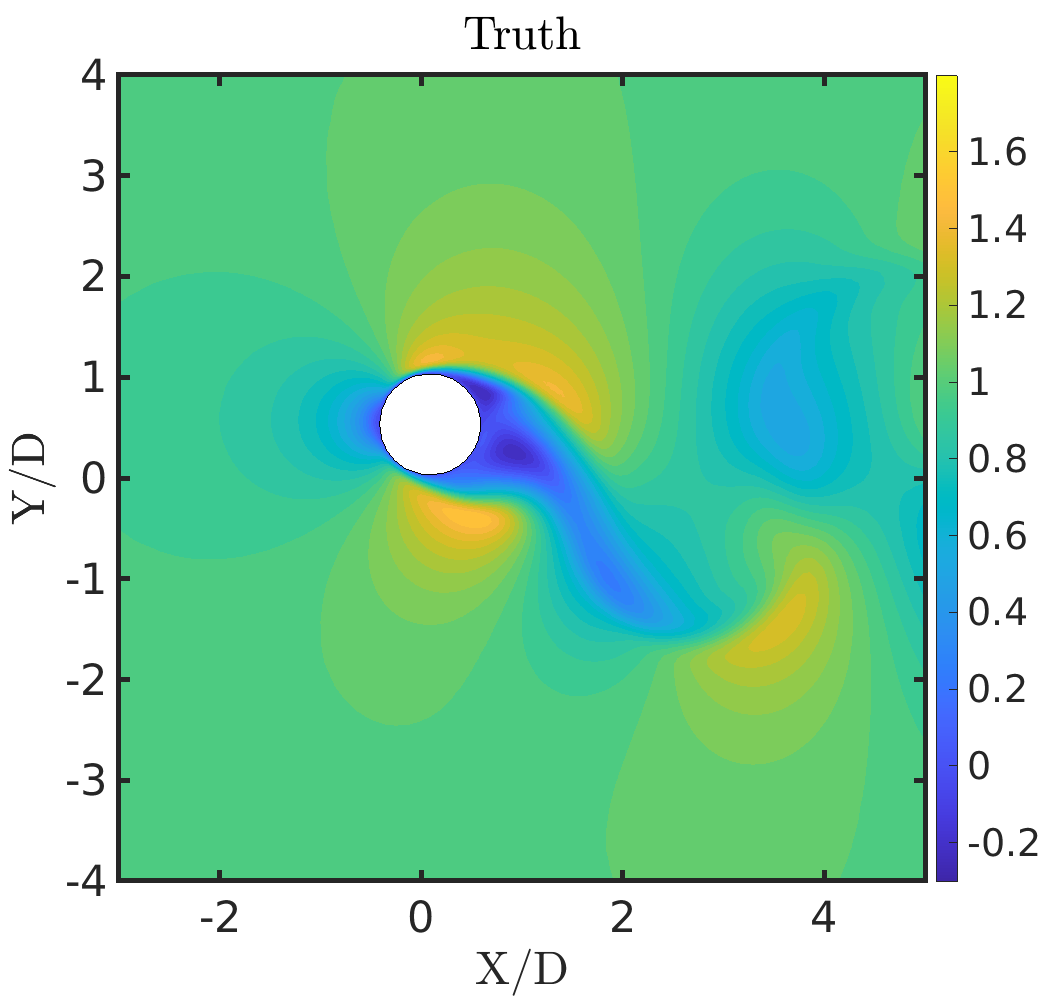}
\hspace{0.01\textwidth}
\includegraphics[width = 0.335\textwidth]{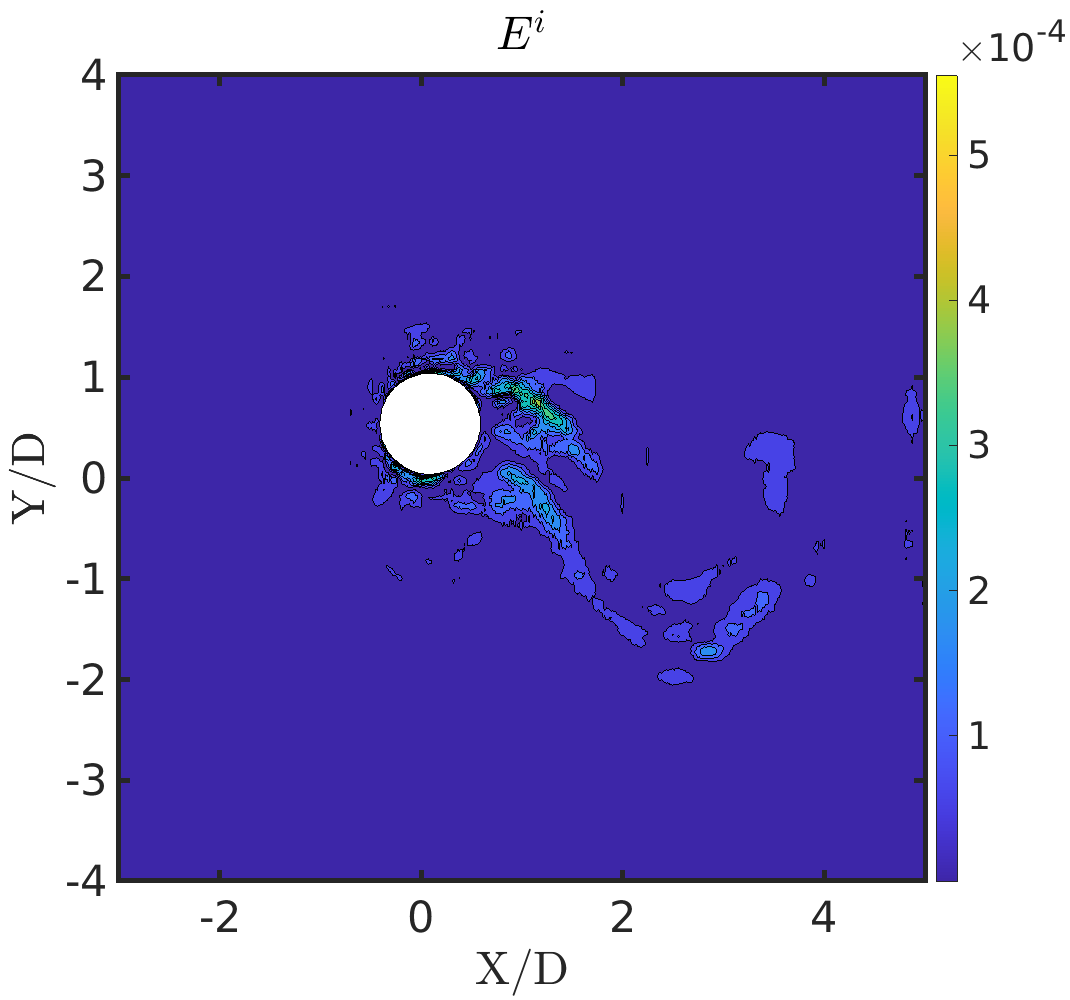}}
\\
\vspace{0.01\textwidth}
\subfloat[]
{\includegraphics[width = 0.33\textwidth]{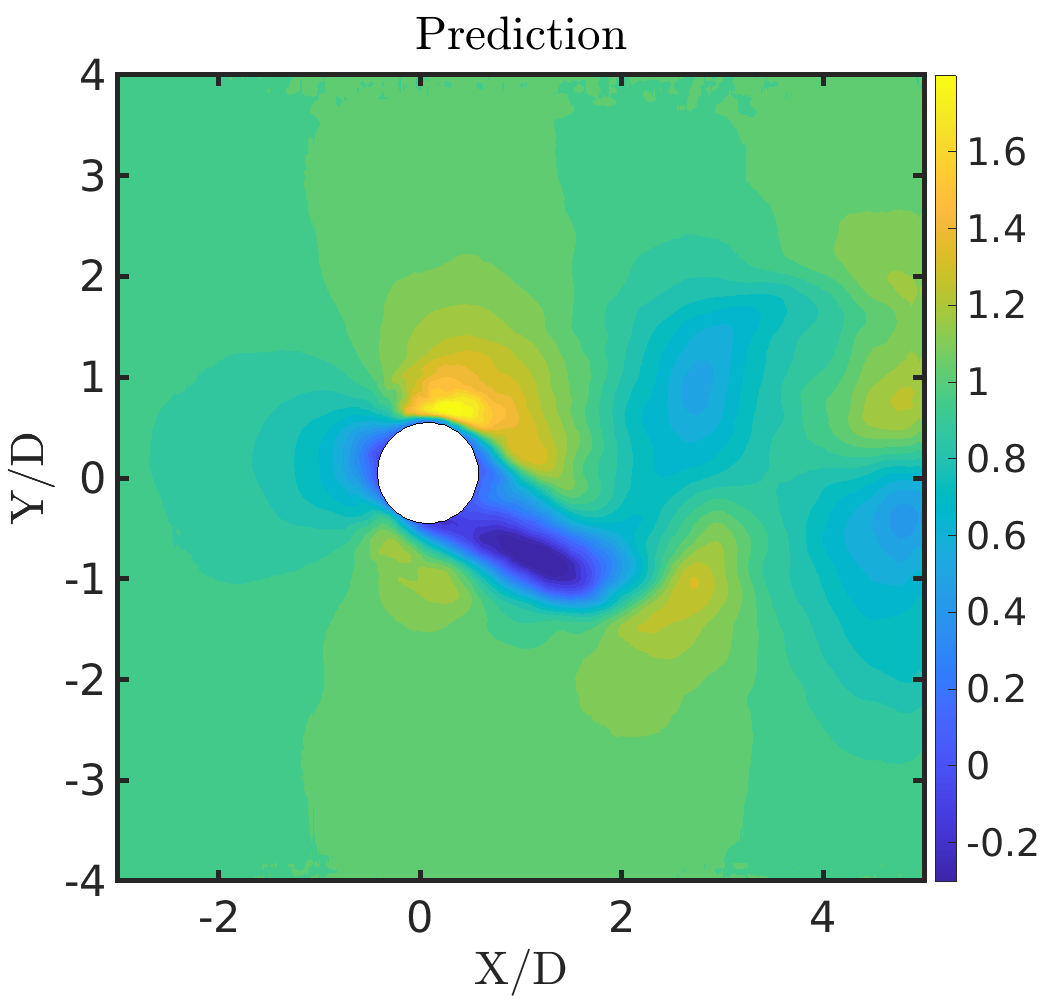}
\hspace{0.01\textwidth}
\includegraphics[width = 0.33\textwidth]{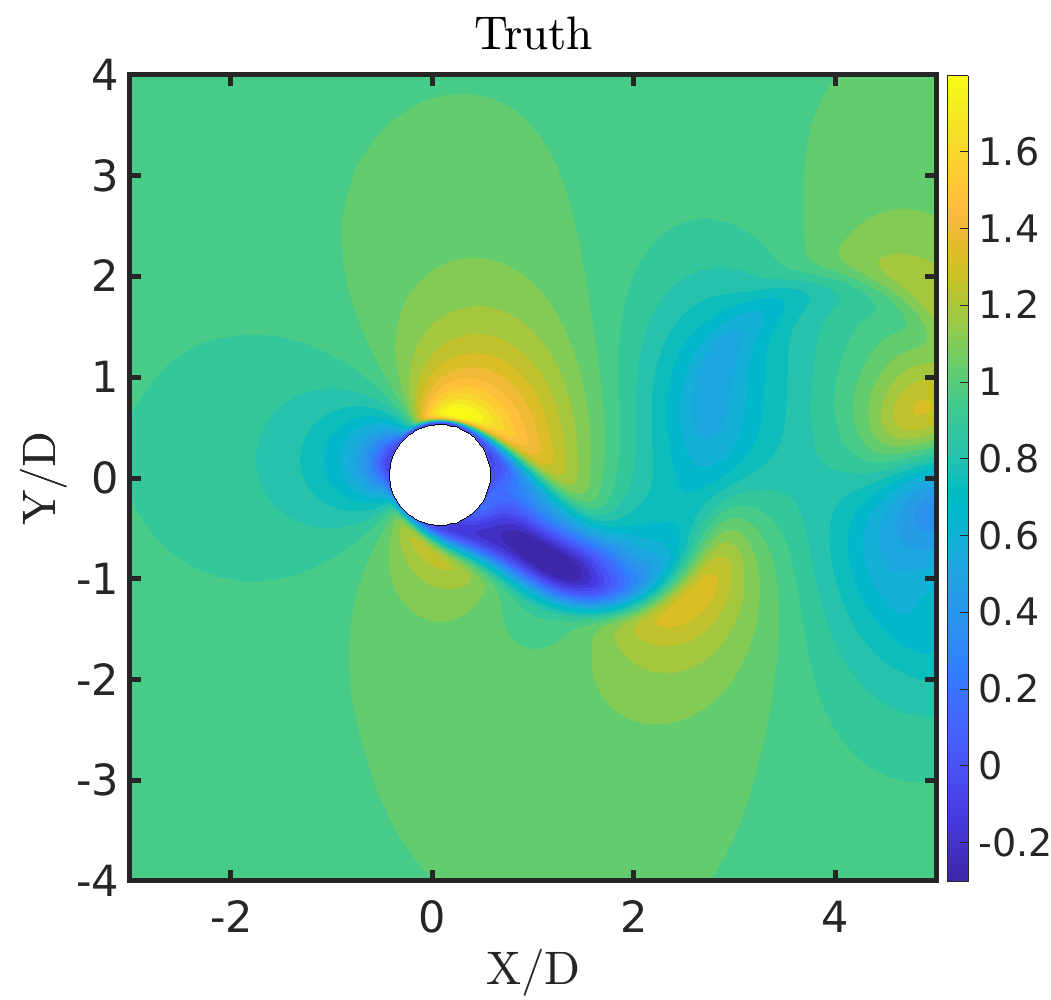}
\hspace{0.01\textwidth}
\includegraphics[width = 0.335\textwidth]{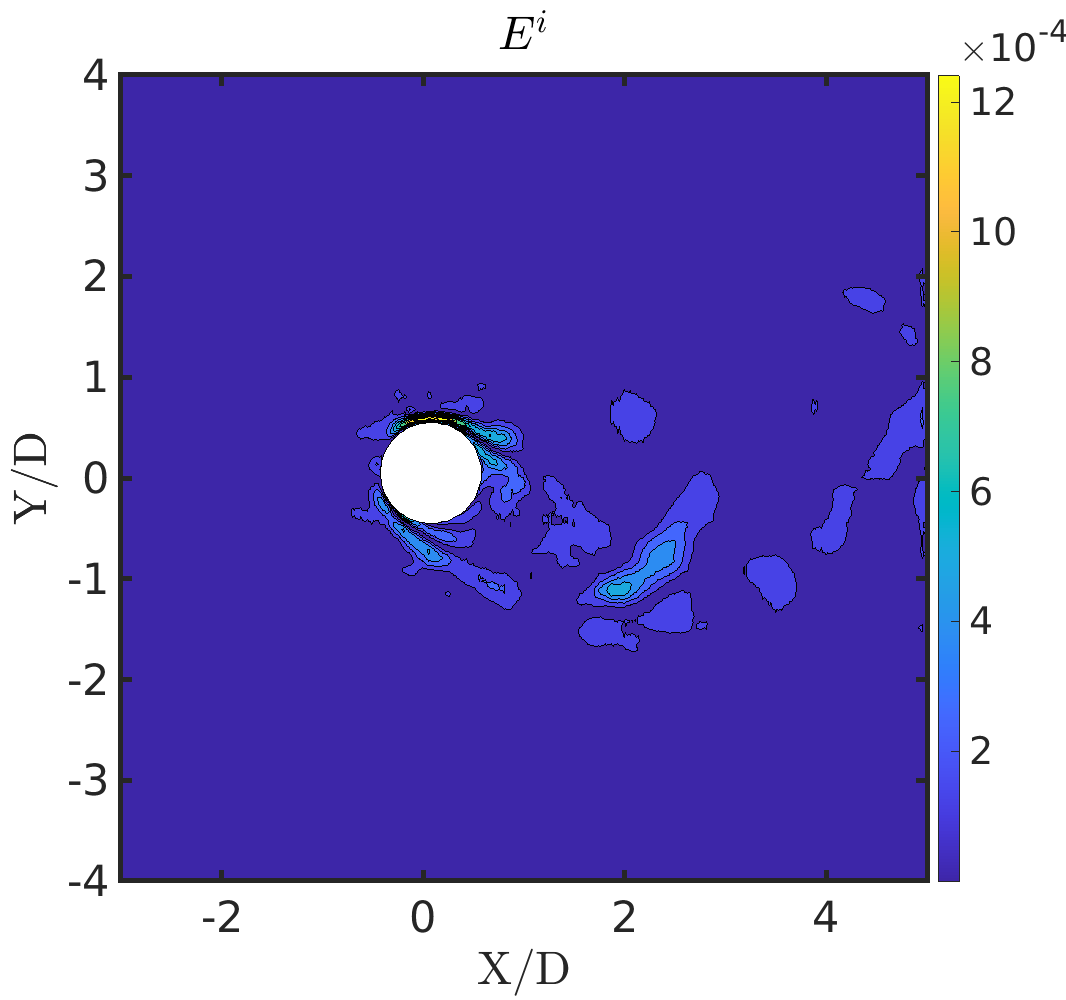}}
\end{widepage}
\caption{Freely vibrating circular cylinder in a uniform flow ($m^{*}=10,U_r=5, Re=200$): Comparison of the predicted and the true fields along with the normalized reconstruction error $E^{i}$ at $\;tU_{\infty}/D =$ (a)  232.5, (b) 240 (c)  249 for the x-velocity field ($U$). }
\label{cran_viv_pred_velx}
\end{figure}

{ \color{black}
\subsection{Non-stationary VIV response} \label{nlco}
In section \ref{lco}, we discussed the application of our hybrid partitioned DL methodology for a VIV response at a mass ratio of $m^{*}=10$, reduced velocity of $U_r=5$ and Reynolds number of $Re=200$. Although a high and sensitive VIV response is obtained at these parameters, the full-order VIV motion is a simple single frequency limit cycle oscillation. Hence, to further strengthen the application of our methodology, we now apply the partitioned DL-ROM framework for a $U_r=7$ while keeping the same Reynolds number and the mass ratio. For the considered training and testing dataset, the full-order VIV motion consists of small magnitudes of non-stationary amplitude responses with no limit cycle oscillation at these new parameters. 

The full-order time series dataset is similarly generated until $tU_{\infty}/ D=250$. A total of $2000$ snapshots are generated at every $0.125\;tU_{\infty}/D$ for the pressure field and the ALE displacements. We keep the same time duration of the training and testing range but with a reduced number of time steps: $n_{tr}=1000$ snapshots (100-225 $tU_{\infty}/D$) for training and $n_{ts}=200$ (225-250 $tU_{\infty}/D$) for testing. At these set of training and testing range, we note that there is no saturation amplitude in the dataset, which makes it a good test case to replicate the complexities involved in the VIV motion that are not necessarily single frequency limit cycle response. 


As discussed in section \ref{podrnn_stat}, time instances of the ALE $\mathrm{x}$-displacements or $\mathrm{y}$-displacements $\textbf{Y} = \left\lbrace\textbf{y}^{1}\; \textbf{y}^{2}\; \dots\; \textbf{y}^{N}\right\rbrace \in \mathbb{R}^{m\times N}$ for this full-order dataset are decomposed using the POD projection in order to reduce the point cloud information. Here, $m=26,114$ and $N=1200$. Cumulative modal energies for the ALE $\mathrm{x}$-displacements and $\mathrm{y}$-displacements are plotted in Fig.~\ref{pod_aleU7} (a) and (c), respectively, with respect to the number of modes. For the chosen parameters, it is interesting to observe that the POD spectrum for the ALE displacements are still fast decaying (similar to the limit cycle VIV oscillation in Figs.~\ref{ce_te_stat_cyl_1} and \ref{ce_te_stat_cyl_2}) and the majority of the system energy of $ > 99.9 \%$ can be described by the first 2 POD modes. This helps in reducing the ALE field with $k=2$ POD modes instead of $m=26,114$ fluid nodes for this test case. Notably, with these two modes, the full-order mesh displacement can be reconstructed within a maximum absolute error $ < 1.0 \times 10^{-14}$.

\begin{figure}
\begin{widepage}
\centering
\subfloat[]{\includegraphics[width = 0.34\textwidth]{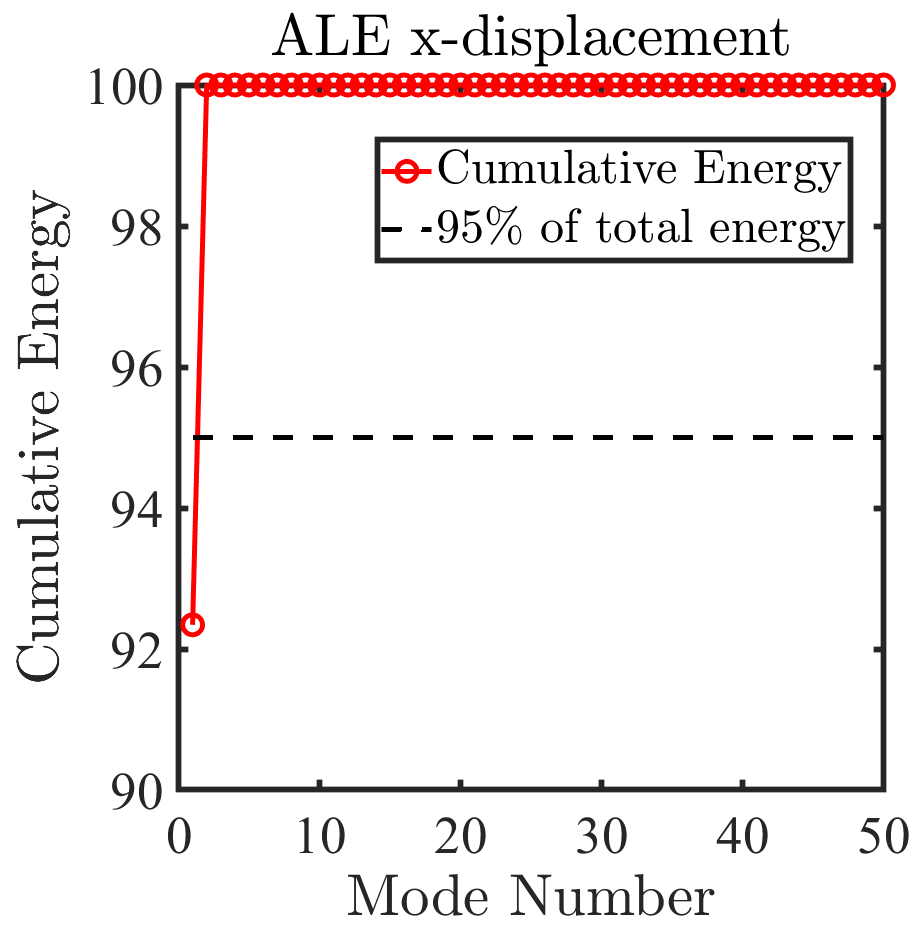}} 
\hspace{0.005\textwidth}
\subfloat[]{\includegraphics[width = 0.625\textwidth]{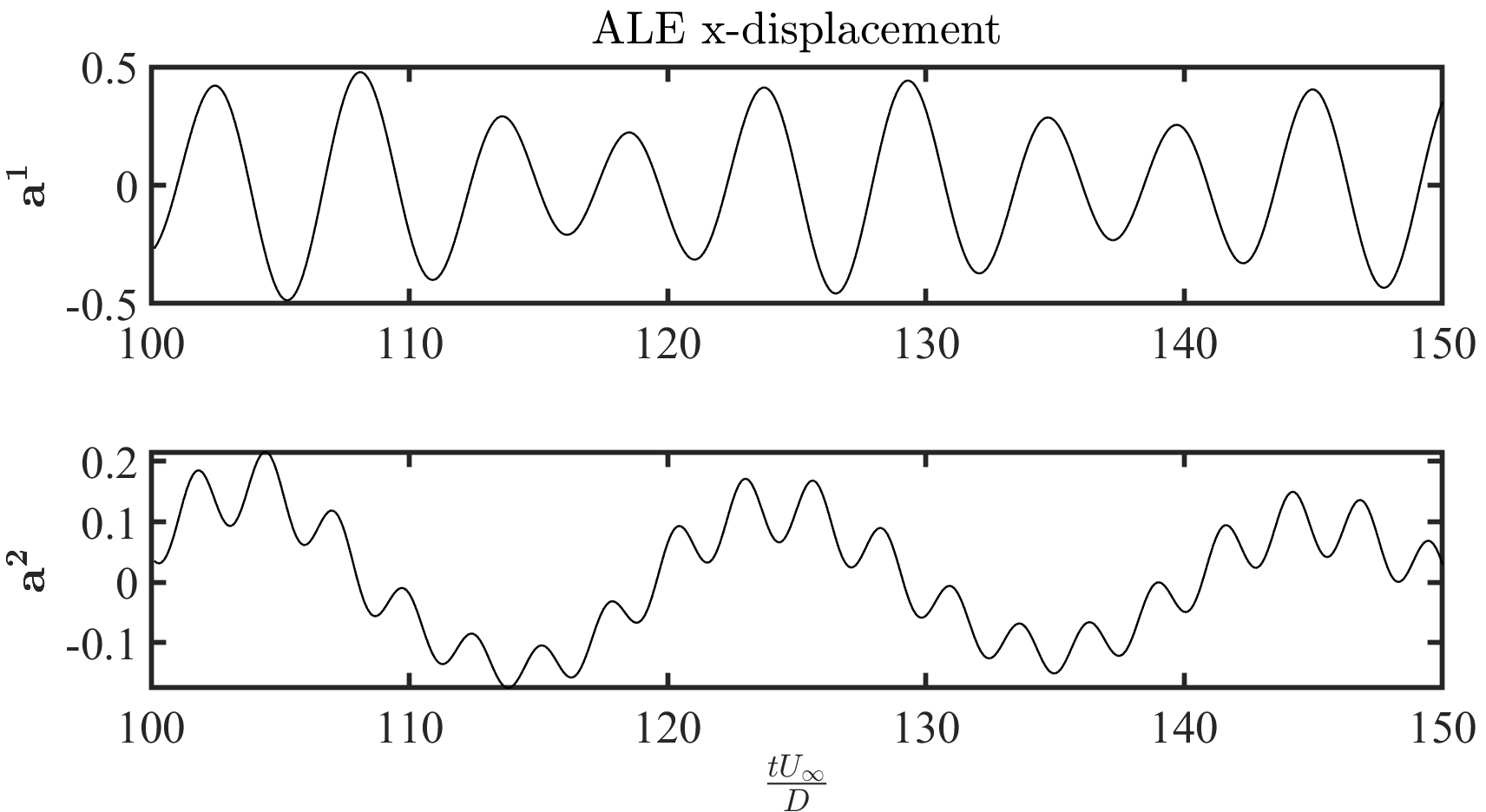}} \\ 
\subfloat[]{\includegraphics[width = 0.34\textwidth]{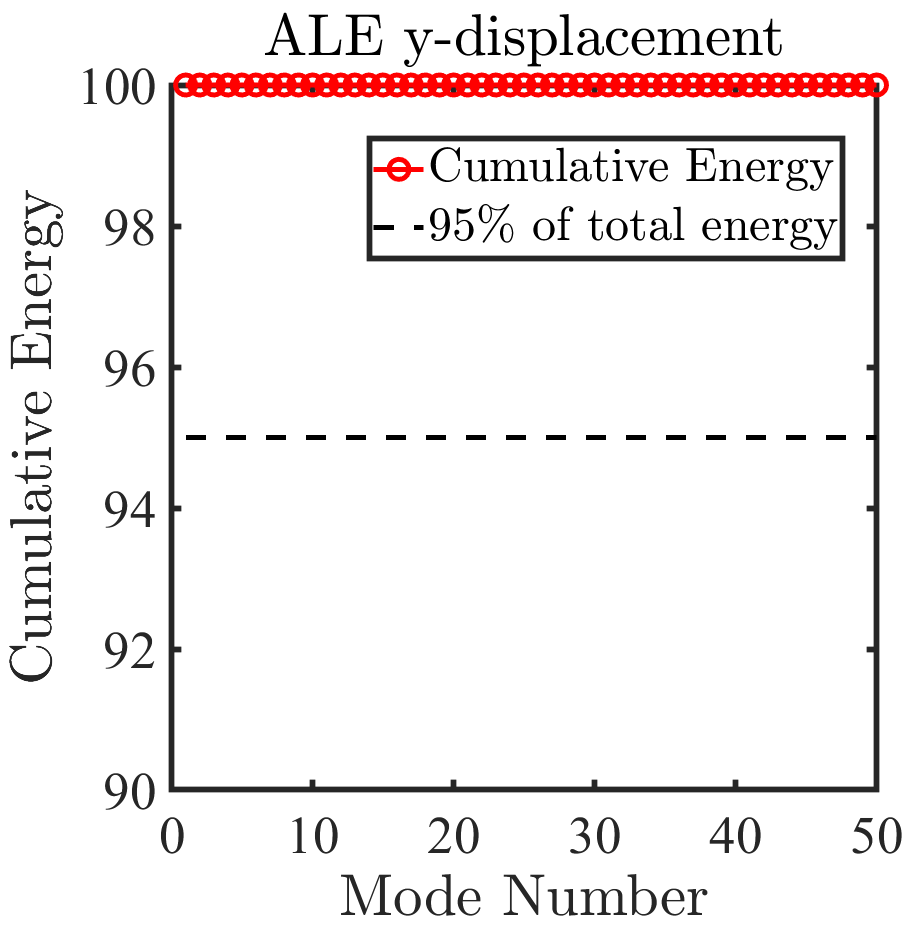}} 
\hspace{0.005\textwidth}
\subfloat[]{\includegraphics[width = 0.63\textwidth]{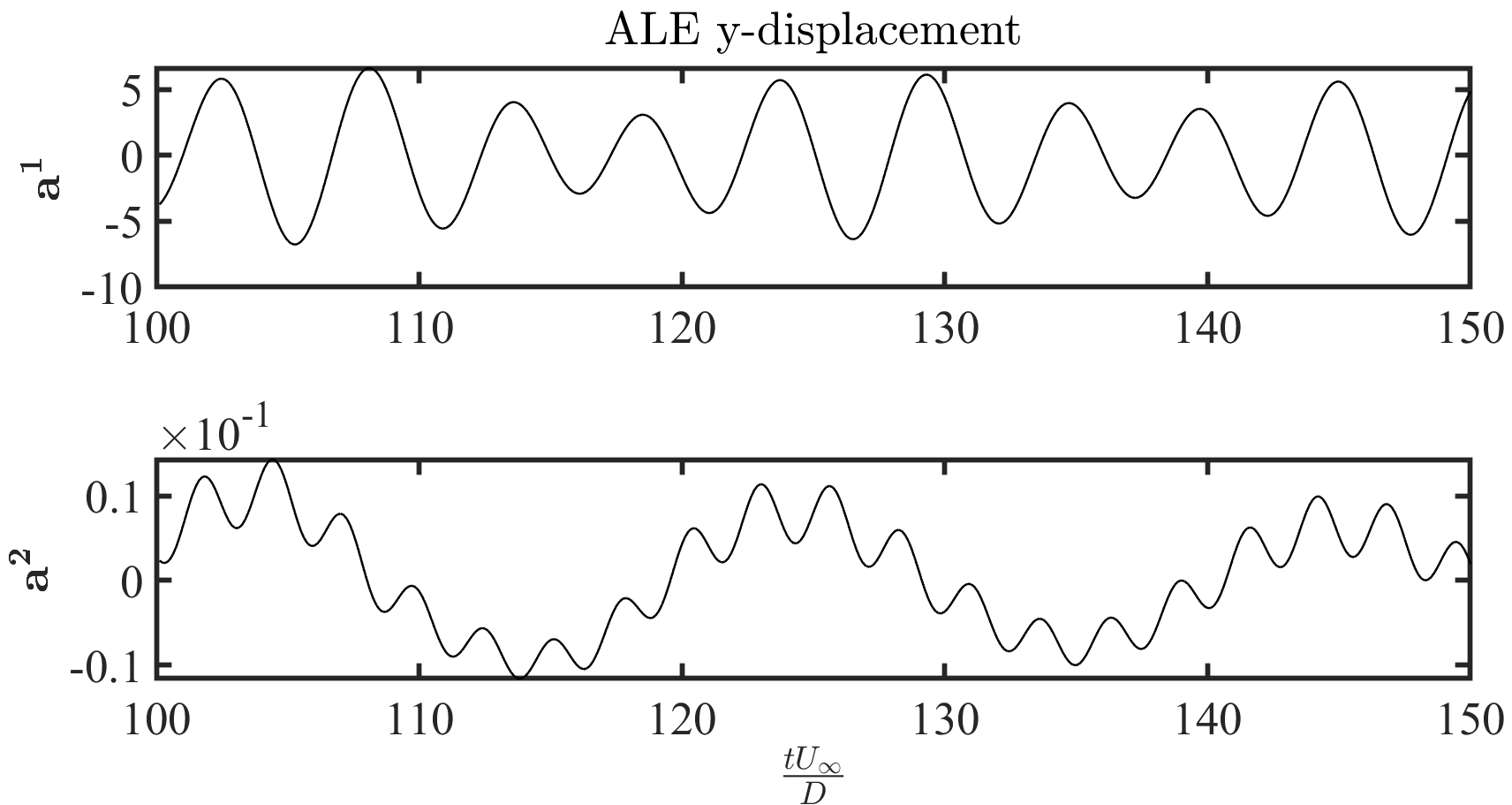}} 
\caption{Freely vibrating circular cylinder in a uniform flow ($m^{*}=10,U_r=7, Re=200$): Cumulative modal energies and the time history of $k=2$ modal coefficients shown from $100$ till $150\;tU_{\infty}/D$. (a)-(b) for the ALE x-displacements  $\textbf{Y}_{\textbf{x}}$, and (c)-(d) for the ALE y-displacements $\textbf{Y}_{\textbf{y}}$. }
\label{pod_aleU7}
\end{widepage}
\end{figure}

\begin{figure}
\begin{widepage}
\centering
\subfloat[]{\includegraphics[width = 0.6\textwidth]{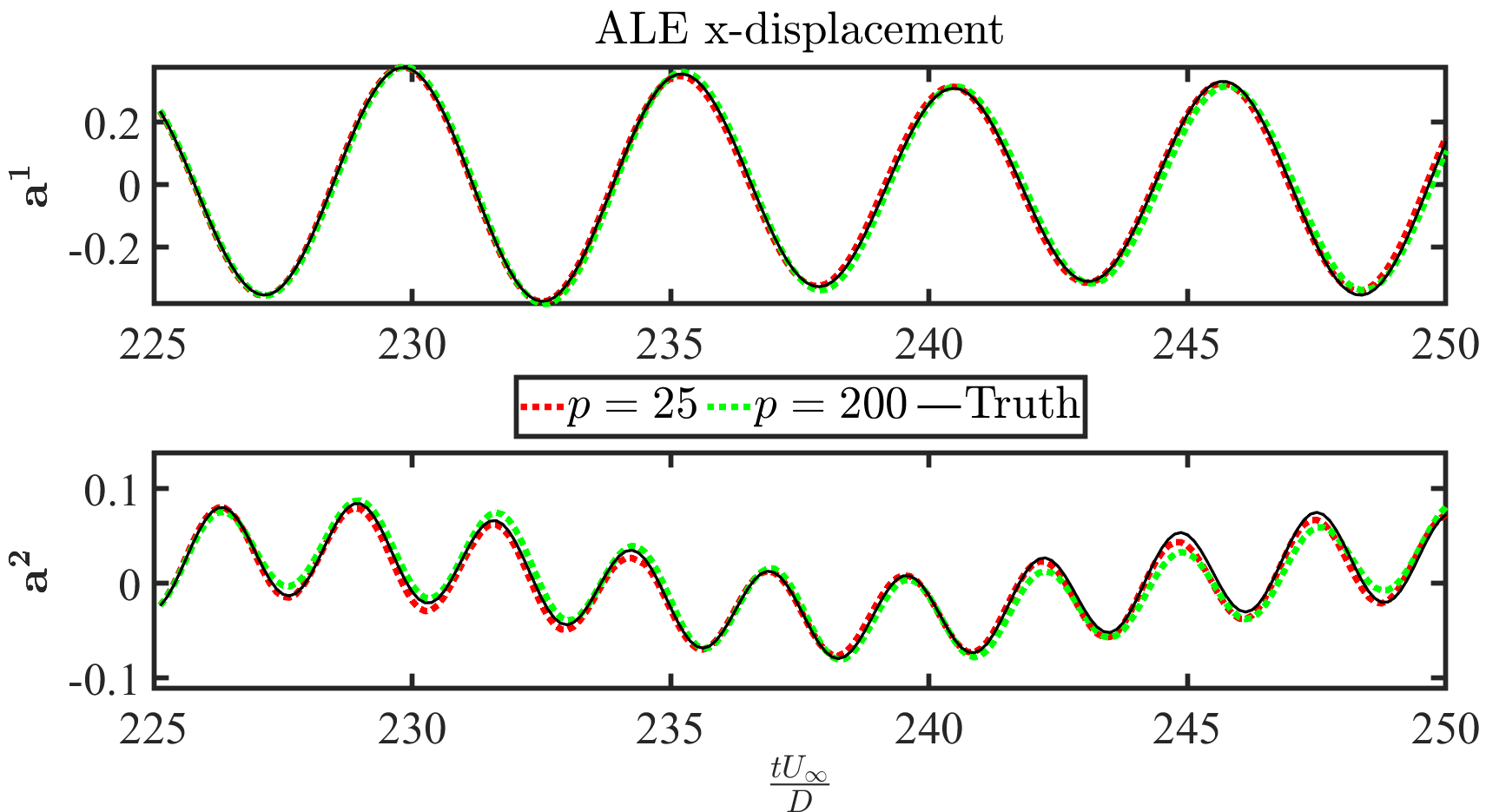}} 
\hspace{0.005\textwidth}
\subfloat[]{\includegraphics[width = 0.34\textwidth]{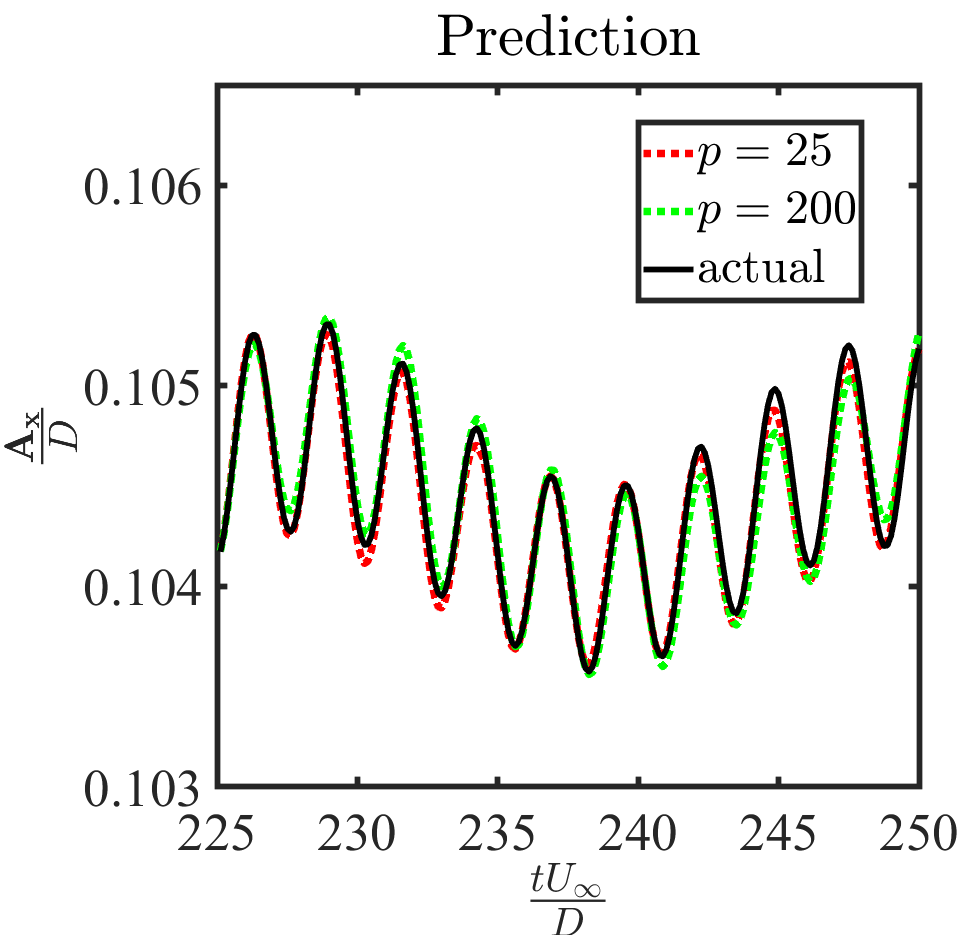}} \\
\subfloat[]{\includegraphics[width = 0.6\textwidth]{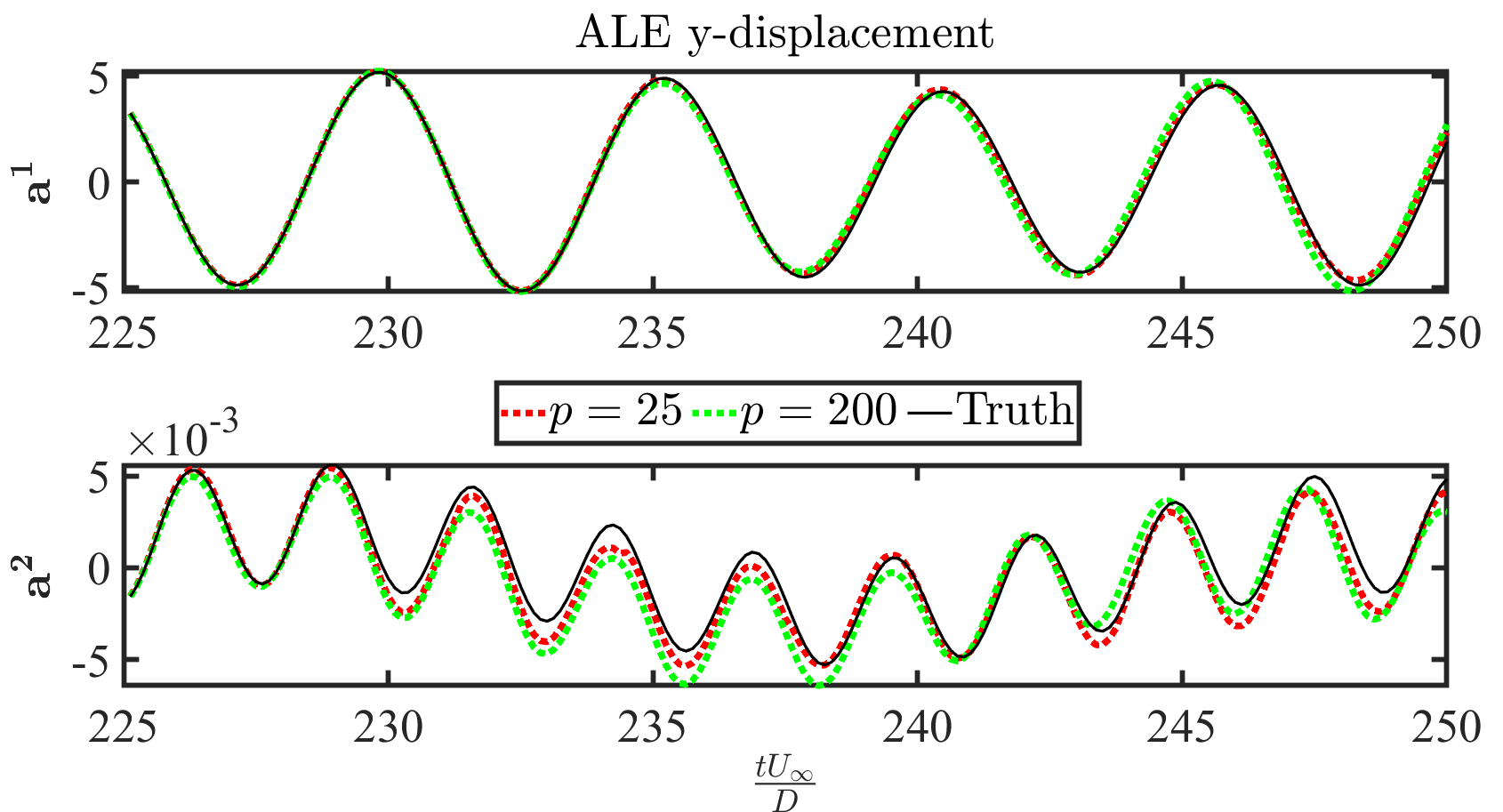}} 
\hspace{0.005\textwidth}
\subfloat[]{\includegraphics[width = 0.34\textwidth]{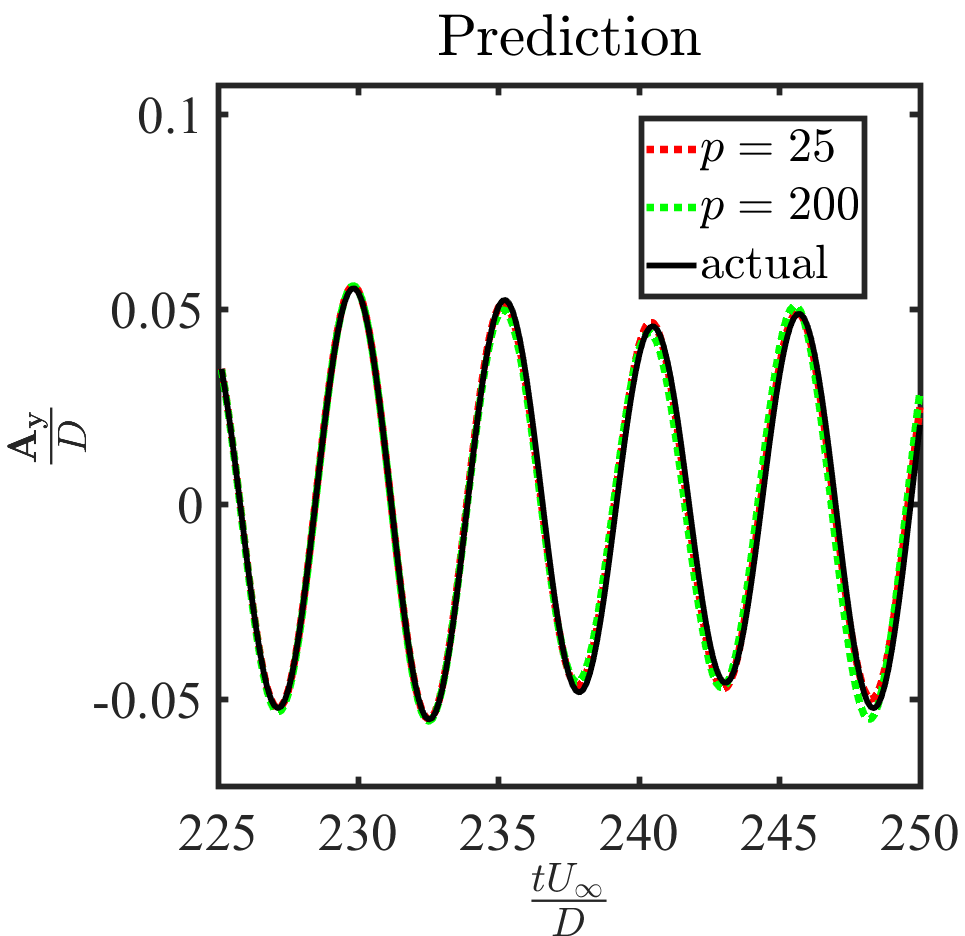}}
\caption{Freely vibrating circular cylinder in a uniform flow ($m^{*}=10,U_r=7, Re=200$): Prediction of the modal coefficients and the interface motion on the test data. (a) and (c) denote the predicted and actual time coefficients of the ALE x-displacements $\textbf{Y}_{\textbf{x}}$ and ALE y-displacements $\textbf{Y}_{\textbf{y}}$, respectively. (b) and (d) denote the corresponding  x and y position of the interface, normalised by diameter of the cylinder $D$. With $p=25$, the root mean squared error (RMSE) between the true and predicted is $5.58\times 10^{-4}$ and $2.22\times 10^{-3}$ for $\mbox{A}_{\mbox{x}} / D$ and $\mbox{A}_{\mbox{y}} /D$, respectively. With $p=200$, the RMSE between the true and predicted is $9.11\times 10^{-4}$ and $4.44\times 10^{-3}$ for $\mbox{A}_{\mbox{x}} / D$ and $\mbox{A}_{\mbox{y}} /D$, respectively.}
\label{pod_pred_aleU7}
\end{widepage}
\end{figure}

Fig.~\ref{pod_aleU7} (b) and (d) depict the dynamical evolution of the 2 POD modes for the ALE $\mathrm{x}$-displacements and $\mathrm{y}$-displacements, respectively, obtained from the POD analysis. As inherent in the high-dimensional VIV motion, we note that these time coefficients propagate as non-stationary variable amplitude responses with small frequency differences in every oscillation. The point cloud tracking, hence, reduces to effective learning and prediction of these dynamical coefficients accurately in time. 

Using a closed-loop LSTM-RNN with the same hyperparameters as in Table~\ref{tab:lstm_network}, we learn the POD time coefficients over the 1000 training time steps by optimising the evolver loss (Eq.~\ref{pod_rnn_loss}) as detailed in section \ref{pod_rnn}. With the training of 5 minutes on a graphics processing unit, the prediction of the modal coefficients is depicted in Fig.~\ref{pod_pred_aleU7} (a) and (c) for the ALE x-displacements and y-displacements, respectively, over the test data. We keep the multi-step prediction cycle of length $p=25$ (one input infers 25 test steps) and $p=200$ (one input infers all 200 test steps). We note that the predictions of these non-stationary time coefficients are in good agreement with the ground truth with slight improvements in $p=25$ predictions as compared to $p=200$. The predicted modal coefficients are reconstructed back to the point cloud using the offline database of the mean field and the POD modes to track the interface. Fig.~\ref{pod_pred_aleU7} (b) and (d) depict the comparison of the predicted and true values of the VIV motion in the x and y directions of the cylinder, respectively. The results indicate that the POD-RNN on the ALE motion can predict the position of the moving FSI interface with excellent accuracy for this parameter set.

The application of the snapshot-FTLR as a load recovery for this parameter set is showcased in Fig.~\ref{ftlr_aleU7} over the training forces. For the sake of uniformity, we have selected the same resolution of the snapshot DL-ROM grid $512 \times 512$ obtained via linear interpolation. Via Gaussian filtering of length 5 and the functional reconstruction mapping $\Psi$ (using Algorithm \ref{lab1}), the reconstructed force signals obtained show a good agreement for the drag and lift coefficients with the full-order values. As a result, we transfer the field point cloud to uniform and structured DL snapshots for the CRAN driver. For demonstration purposes, we only consider the pressure fields for testing the predictive abilities of the CRAN driver. 

\begin{figure}
\begin{widepage}
\centering
\subfloat[]{\includegraphics[width = 0.35\textwidth]{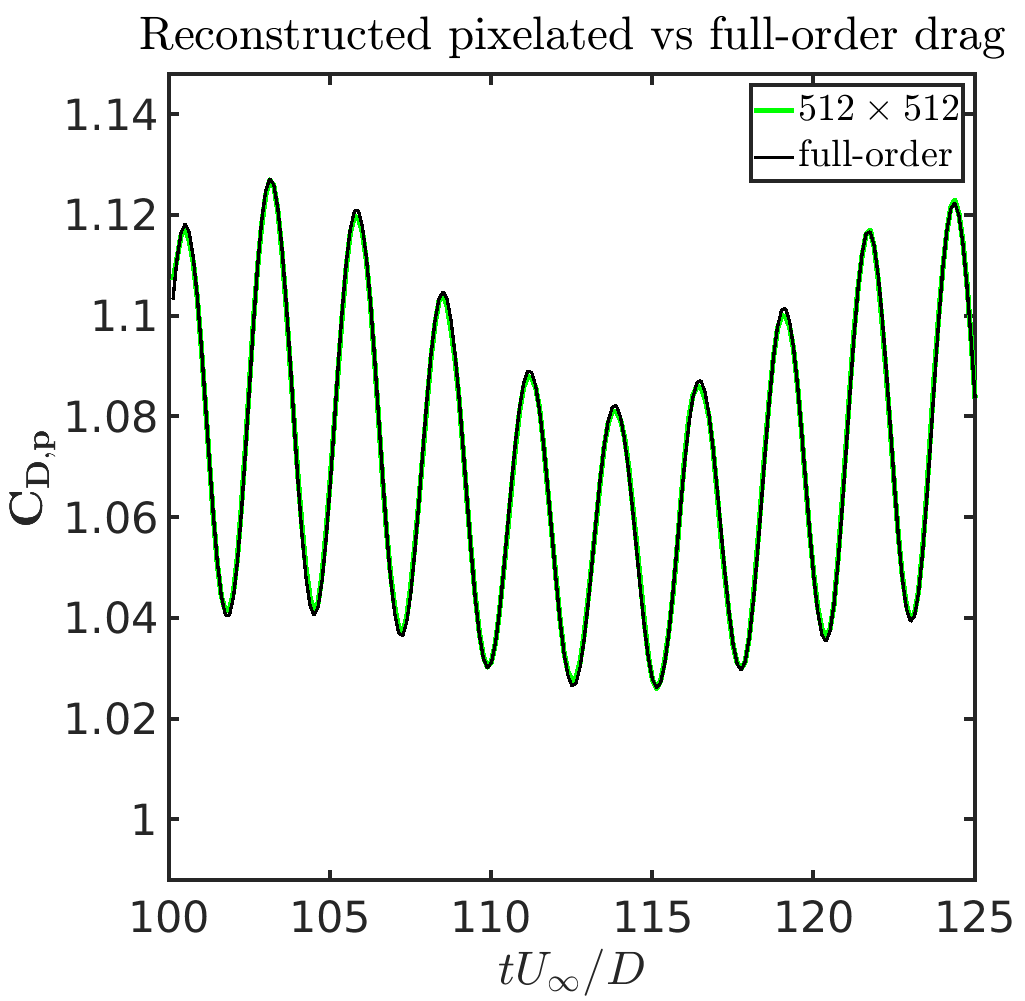}} 
\hspace{0.005\textwidth}
\subfloat[]{\includegraphics[width = 0.35\textwidth]{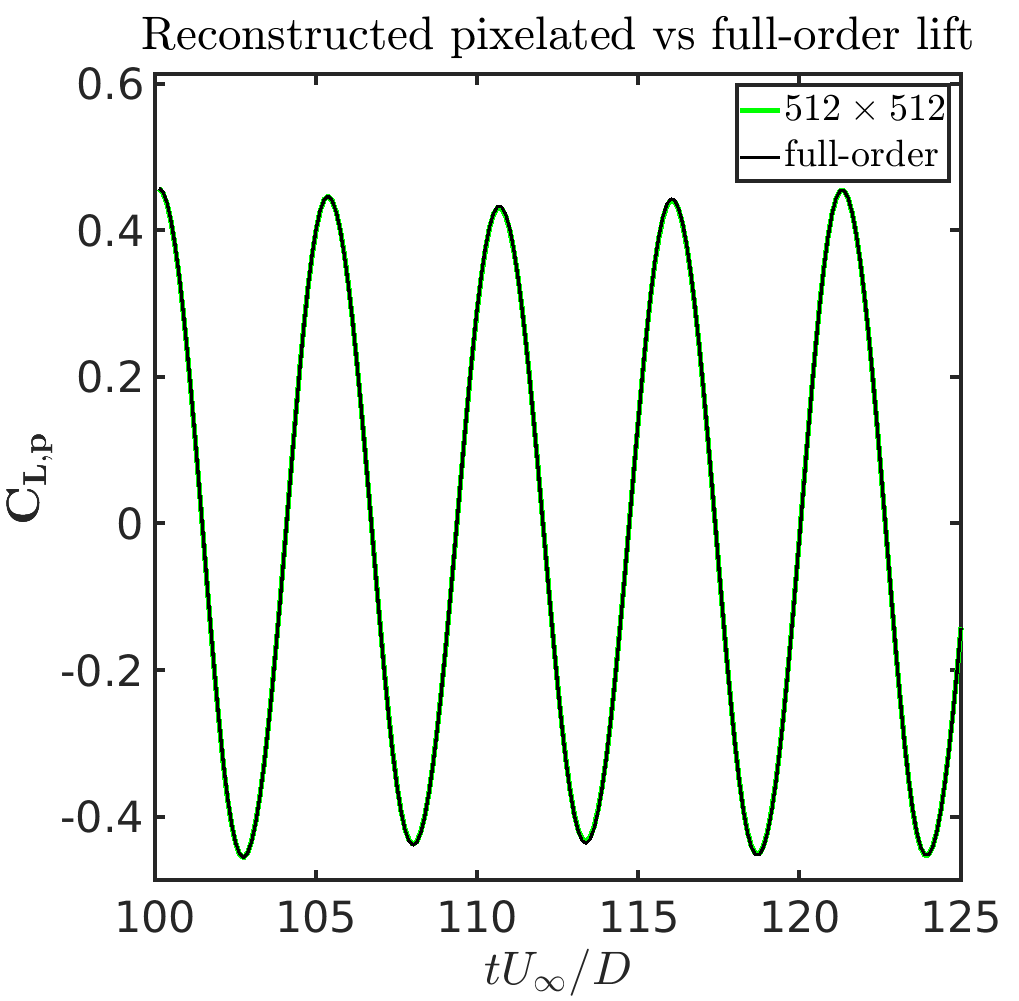}}
\caption{Freely vibrating circular cylinder in a uniform flow ($m^{*}=10,U_r=7, Re=200$): Recovered interface load behavior vs time shown from 100-125 $t U_{\infty} / D $ using the snapshot-FTLR for (a) the drag, and (b) the lift coefficients on the $512 \times 512$ snapshot DL-ROM grid with respect to full-order. } 
\label{ftlr_aleU7}
\end{widepage}
\end{figure}

\begin{figure}
\begin{widepage}
\centering
\subfloat[]
{\includegraphics[width = 0.33\textwidth]{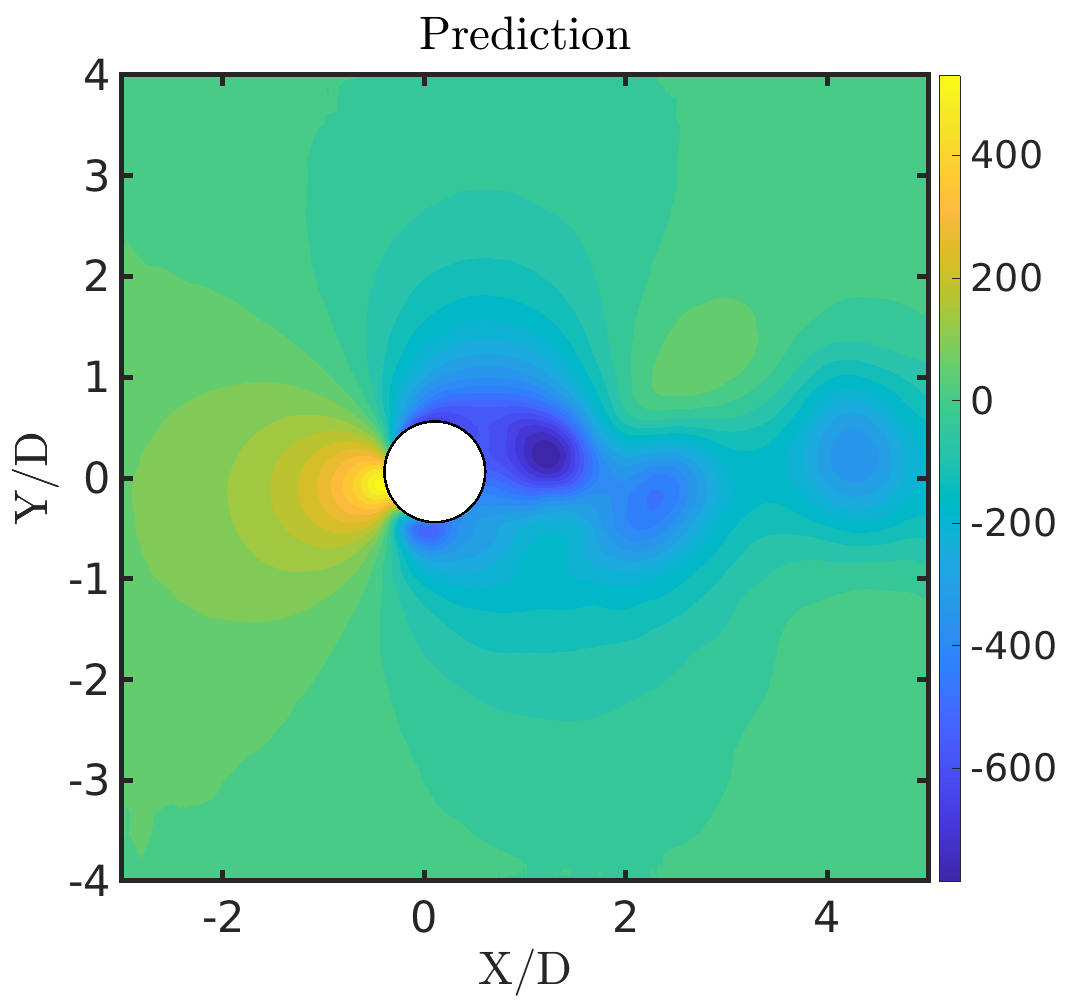}
\hspace{0.01\textwidth}
\includegraphics[width = 0.33\textwidth]{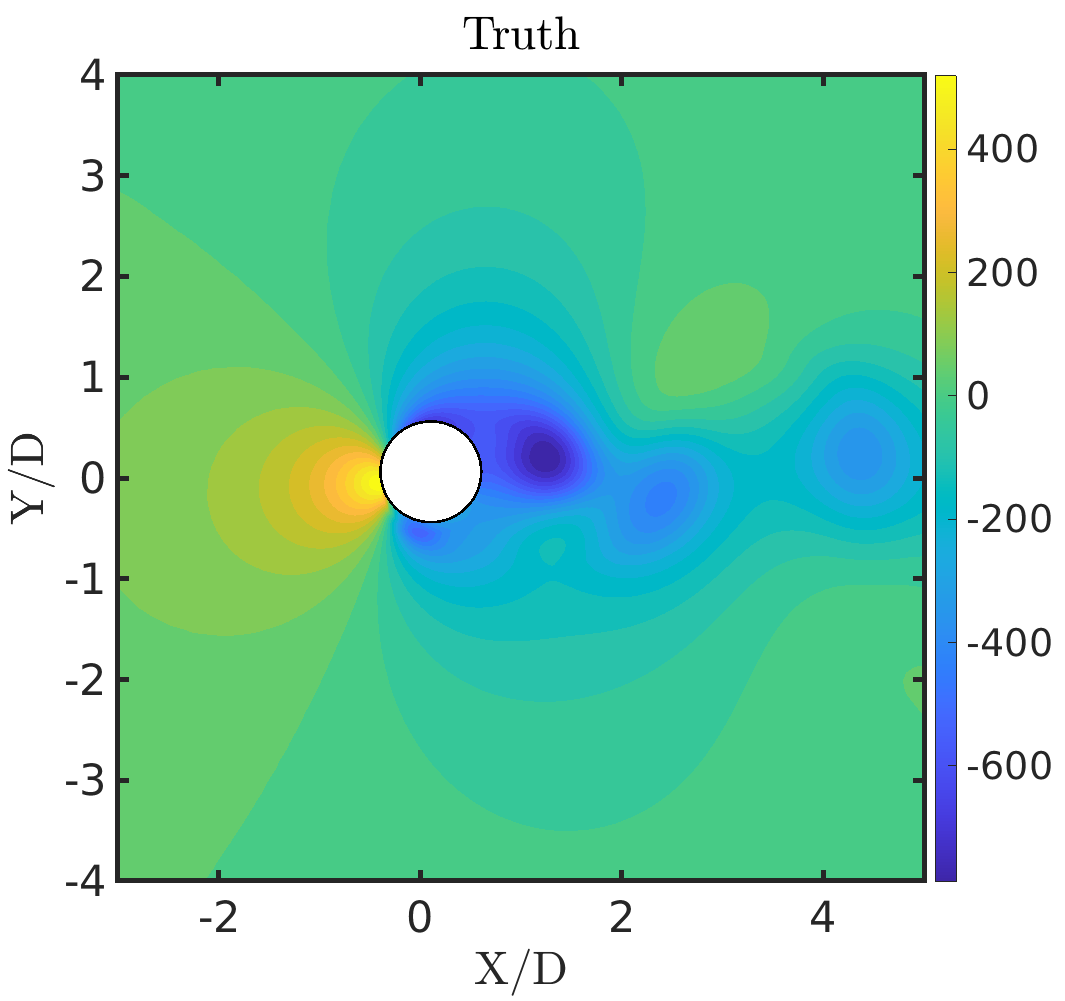}
\hspace{0.01\textwidth}
\includegraphics[width = 0.33\textwidth]{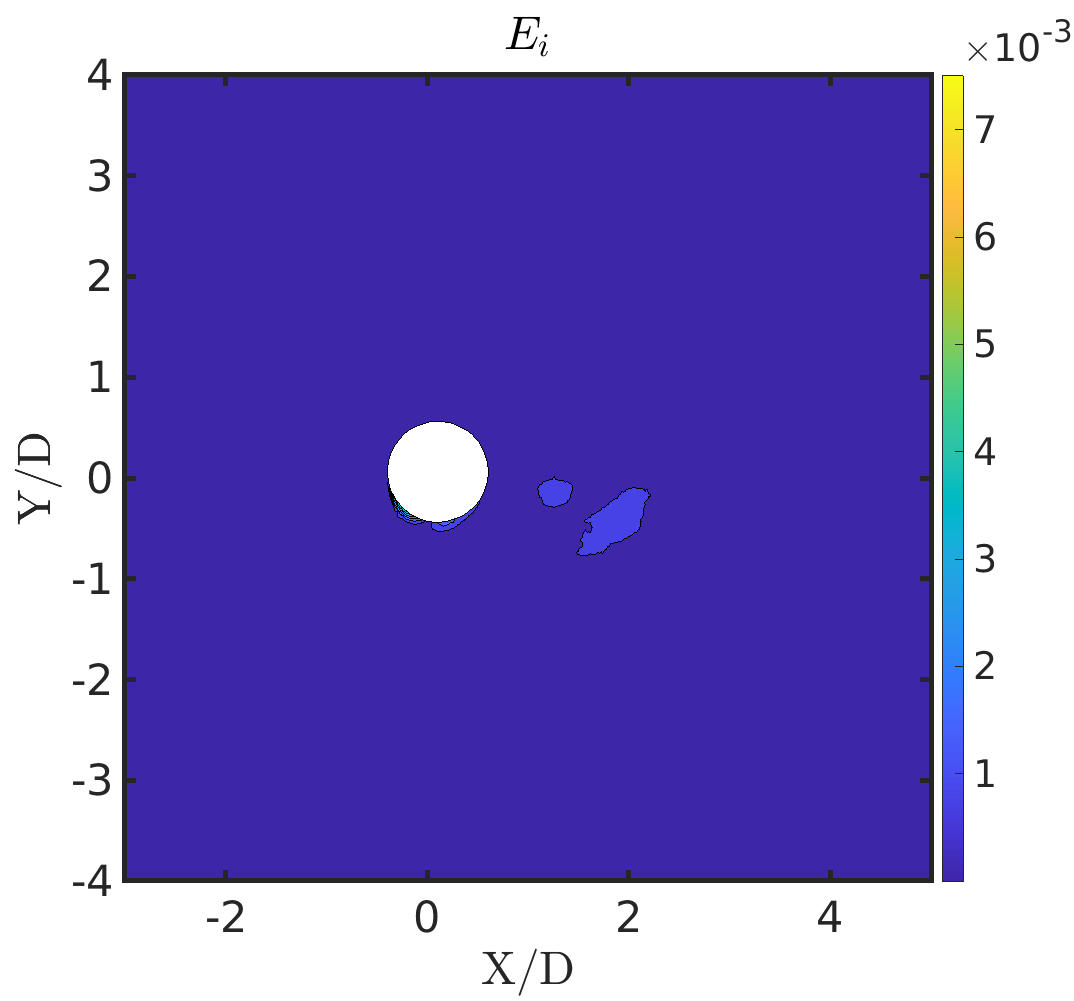}}
\\
\vspace{0.01\textwidth}
\subfloat[]{\includegraphics[width = 0.38\textwidth]{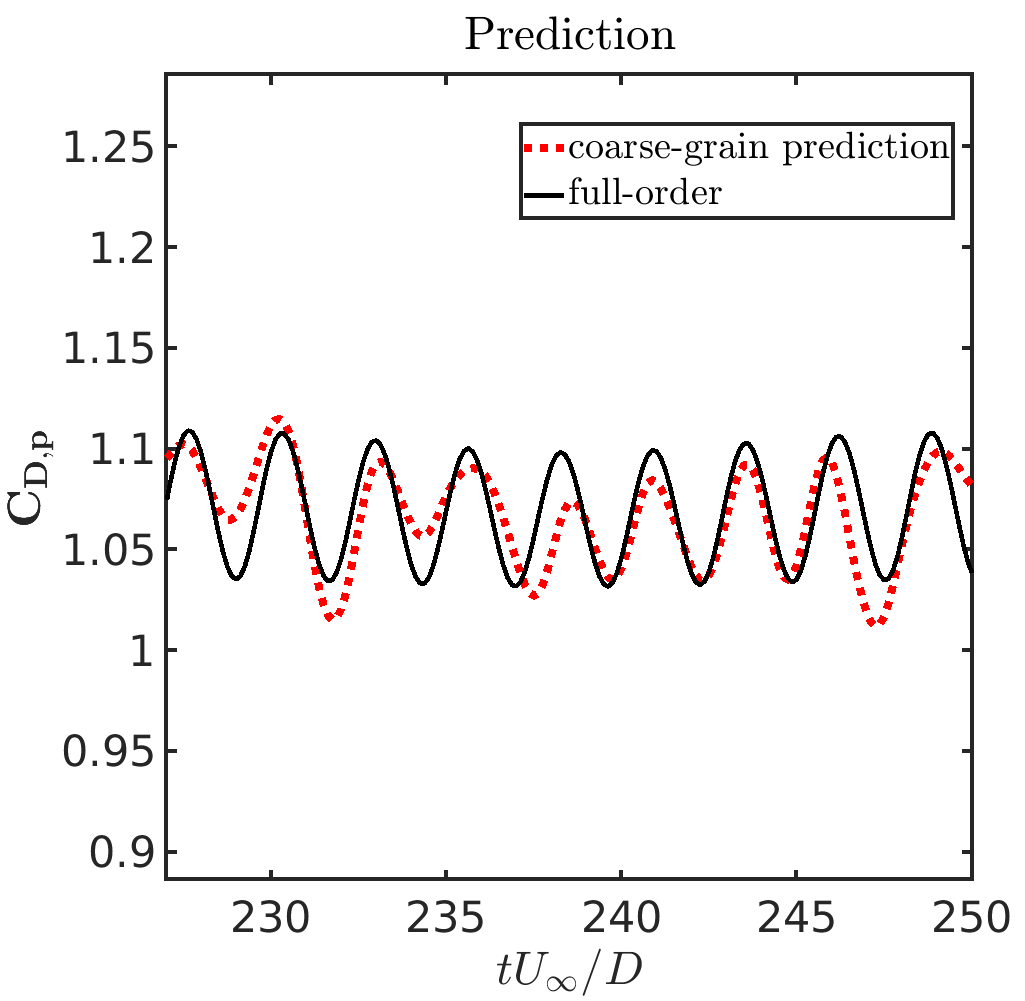}
\hspace{0.005\textwidth}
\includegraphics[width = 0.38\textwidth]{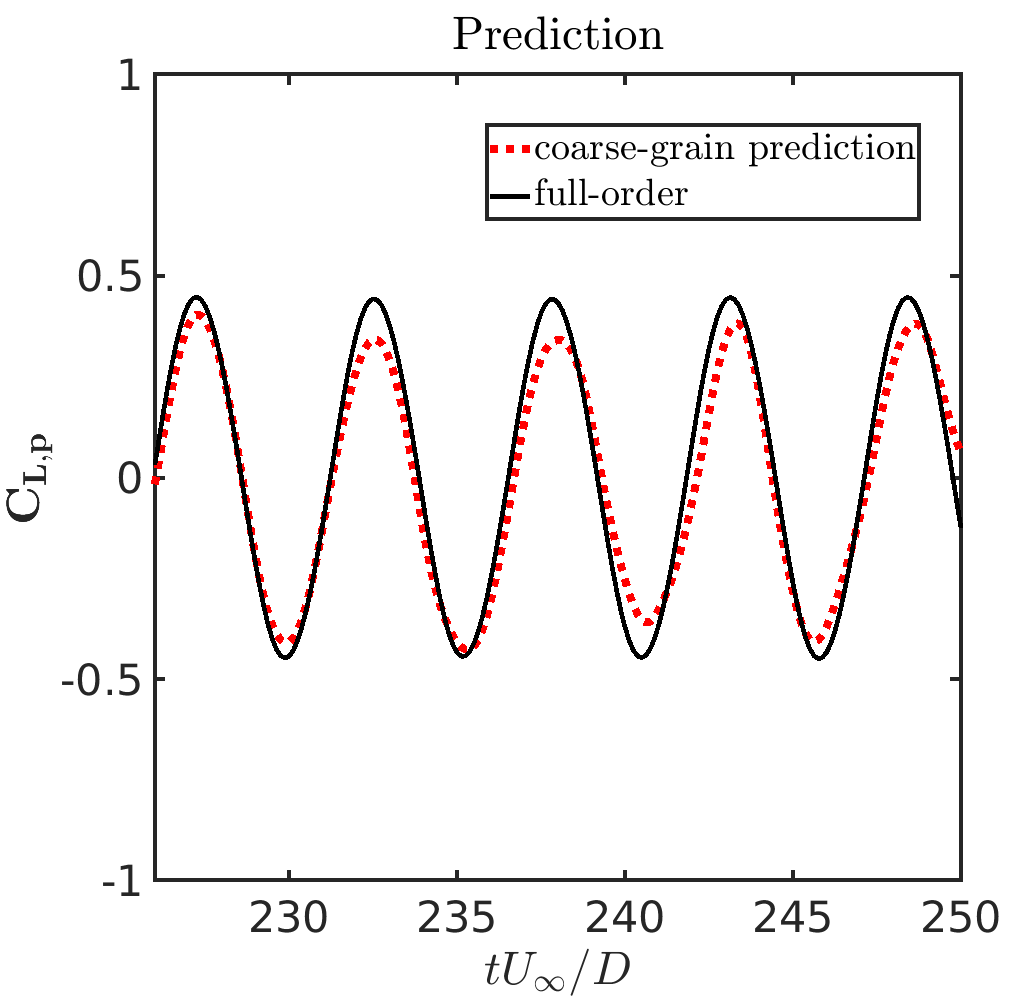}}
\end{widepage}
\caption{Freely vibrating circular cylinder in a uniform flow ($m^{*}=10,U_r=7, Re=200$): (a) Comparison of the predicted and the true fields along with the normalized reconstruction error $E^{i}$ at $\;tU_{\infty}/D = 250$ for the pressure field ($P$), and (b) interface drag and lift prediction on the test time steps from the predictive framework with multi-step predictive cycle $p=25$. The CRAN predictions are accurate for the first 25-30 time steps in the future, after which the predictions can diverge.}
\label{predictions_ur7}
\end{figure}


The generated pressure field information $\textbf{S} = \left\lbrace\textbf{S}^{1}\; \textbf{S}^{2}\; \dots\; \textbf{S}^{N}\right\rbrace \in \mathbb{R}^{N_{x} \times N_{y} \times N}$ using the snapshot-FTLR are decomposed into $n_{tr}=1000$ training data (100-225 $tU_{\infty}/D$) and $n_{ts}=200$ testing data (225-250 $tU_{\infty}/D$). As detailed in section \ref{cnn_rnn}, the training flow dataset is normalized and batch-wise arranged  to generate the scaled featured input with number of batches $N_s=40$ and the RNN time sequence $N_t=25$. To train the CRAN architecture on this flow dataset, we load the optimized $N_h=128$ CRAN parameters for the pressure model for the limit cycle case in section \ref{lco} and optimize further. This transfer of learning helps in optimizing the CRAN framework for the new parameter set in $N_{train}=50,000$ iterations, which took a training time of nearly 7 hours on single graphics processing unit with the final loss of $\approx 1.0\times 10^{-5}$. The test time steps depict the predictive performance of this network in terms of accuracy check. 

Fig.~\ref{predictions_ur7} (a) depict the comparison of the predicted and the true values of the pressure field at the last test time step $250\;tU_{\infty}/D$ obtained with a multi-step predictive cycle of $p=N_{t}=25$. From these predictions, we note that that the field predictions are in excellent agreement with the true field, with a reconstruction error $E^i$ in the order of $10^{-3}$. Similar to the limit cycle VIV oscillation, we note the coupled FSI prediction from the CRAN is accurate for the first 25-30 time steps in the future, after which the predictions can diverge. The feedback demonstrators combat the compounding errors and enforce the CRAN predictive trajectory devoid of divergence over the test data. Figs.~\ref{predictions_ur7} (b) depict the evolution of the integrated pressure forces $\mathrm{C}_{\mathrm{D},\mathrm{p}}$ and $\mathrm{C}_{\mathrm{L},\mathrm{p}}$ on the interface over the test data using Algorithm \ref{lab3}. From a direct integration of the predicted fields, we obtain a reasonable drag and lift prediction. The variation in the predicted amplitude from the force signals are minor for this case due to a smaller interface displacement in this coupled dataset. As a result, the noises in the forces are of a smaller magnitude.

\begin{table}
\footnotesize
\centering
\caption{Freely vibrating circular cylinder in a uniform flow: Summary of the training and testing times of the DL-ROM components along with force computation as compared to the full-order computational time. The testing of the POD-RNN, CRAN and Algorithm \ref{lab3} denote the run time needed to predict $p=25$ instances of displacements, fields and forces, respectively. The combined predictions denote the total run time for all these predictions. }
\begin{tabular}{P{1.3cm}|P{1.2cm}P{1.2cm}|P{1.2cm}P{1.2cm}P{1.2cm}|P{1.2cm}P{1.2cm}P{1.2cm}}
\toprule
\toprule
Parameter  &  Training$^{*}$ & Training$^{*}$  & Testing$^{*}$ & Testing$^{*}$ & Algo. \ref{lab3}$^{**}$ & Combined & FOM$^{**}$ & Speed-up\\
($Re=200$, $m^{*}=10$)  &  (POD-RNN) & (CRAN) & (POD-RNN) & (CRAN) & (Forces) & predictions  & predictions & factor \\
  &  (hours) & (hours) & (seconds) & (seconds) & (seconds) & (seconds)  & (seconds) & \\
\bottomrule
\bottomrule
$U_r = 5$  & $1.5 \times 10^{-1}$ & $6.0 \times 10^{1}$ & $2.8 \times 10^{-1}$ & $6.4 \times 10^{-1}$ & $8.9 \times 10^{-1}$ &  1.8 &  $4.3 \times 10^{2}$ & $2.4 \times 10^{2}$ \\
$U_r = 7$ & $1.5 \times 10^{-1}$ & 7.5 & $3.1 \times 10^{-1}$ & $6.5 \times 10^{-1}$ & $6.9 \times 10^{-1}$ & 1.7 & $4.3 \times 10^{2}$  & $2.5 \times 10^{2}$   \\
\hline
\end{tabular}\\
$^{*}$ Single GPU processor\\
$^{**}$ Single CPU processor\\
\label{tab:compsummary}
\end{table}

Finally, we provide an estimate of computational cost for the hybrid partitioned DL-ROM together with full-order simulations. In Table \ref{tab:compsummary}, we report the training times, the testing times and the total time needed to predict the solid displacements, the flow predictions and the forces (using Algorithm \ref{lab3}). We recall that the multi-level DL-ROM solution can offer remarkable speed-ups in computing as compared with the FOM, by nearly 250 times. While a canonical laminar flow of vibrating cylinder is considered, the proposed hybrid DL-ROM does not make any assumptions with regard
to geometry and boundary conditions.

}

\section{Conclusions} \label{conclusions}
We have presented a hybrid partitioned deep learning framework for the reduced-order modeling of moving interfaces and fluid-structure interaction. The proposed novel DL-ROM relies on the proper orthogonal decomposition combined with recurrent neural networks and convolutional recurrent autoencoder network. We decouple structural and fluid flow representations for easier and modular learning of two physical fields independently.
While POD-RNN provides an accurate extraction of the fluid-structure interface, the CRAN enables extraction of the fluid flow fields. We have successfully demonstrated the inference capability of the proposed DL-ROM framework by predicting the time series of the unsteady pressure field of an FSI set-up of an elastically-mounted circular cylinder.
Using coarse to fine-grained learning of fluid-structure interaction, a low-dimensional inference of the flow fields and the full-order mesh description with load recovery has been discussed. 

By analyzing a prototype FSI model for limit cycle and non-stationary structural motions, we have first shown that the POD-RNN component infers the point cloud dynamics by projecting the ALE displacements on the POD modes. A closed-loop LSTM-RNN effectively propagates the evolution of the POD time coefficients and tracks the full-order grid motion accurately. We then analyze an iterative low interface resolution DL-ROM grid search for the CNNs that preserves the full-order pressure stresses on the moving fluid-structure interface via snapshot-FTLR. 
We have shown that this snapshot-FTLR method selects the grid for the CRAN architecture but can exhibit spurious oscillations and data loss in the loads on a moving FSI system. These errors are shown to depend on the uniform grid resolution, and a detailed study has been performed to recover and filter the losses. A 16-layered trained CRAN network is shown to predict the flow dynamics involving a coupled FSI with accurate estimates of flow prediction. The CRAN network extrapolates the field accurately for 25-30 time steps from one input data, after which the predictions diverge owing to some noises in the CRAN reconstruction. This leads to instability in the force calculation on the moving interface. 
Once trained, the proposed DL-ROM technique can be implemented using only a fraction of computational resources to that of a full-order model for online predictions.

{ \color{black}

 Since the proposed hybrid DL-ROM framework is our first attempt for a moving fluid-structure interface, there are several possible extensions of the proposed approach. For example, the POD-RNN has been applied for the rigid body equations involving a simplified VIV motion. One could extend our framework to flexible structures with geometric and/or material nonlinearity via the CRAN driver. Moreover, other neural architectures such as graph neural networks \cite{pfaff2020learning} and transformers \cite{tay2020efficient} can be considered within the proposed partitioned DL-ROM framework. Instead of POD, hyperreduction techniques such as energy-conserving sampling and weighting \cite{An2008ECSW} can be used for evolving nonlinear structural dynamics.  In our current work, the DL-ROM methodology models VIV in a 2D non-body conformal uniform mesh. Although the snapshot-FTLR method allows a structured grid search for CNNs and preserves interface description, it can create a larger number of unknowns in the DL space if refined elsewhere. In future work, it is worth considering a 3D extension of the DL-ROM methodology and an adaptive refinement/coarsening of the non-interface cells in the DL-ROM grid to reduce the number of unknowns. 


}

\section*{Acknowledgement}
The authors would like to acknowledge the Natural Sciences and Engineering Research Council of Canada (NSERC) for the funding. This research was supported in part through computational resources and services provided by Advanced Research Computing at the University of British Columbia. {\color{black} We would also like to thank the reviewers for their constructive comments that has helped in improving the quality of the manuscript substantially.}

\bibliography{template}
\bibliographystyle{plain}

\end{document}